\documentclass[sn-mathphys,Numbered]{sn-jnl}

\usepackage{graphicx}%
\usepackage{multirow}%
\usepackage{amsmath,amssymb,amsfonts}%
\usepackage{amsthm}%
\usepackage{mathrsfs}%
\usepackage[title]{appendix}%
\usepackage{xcolor}%
\usepackage{textcomp}%
\usepackage{manyfoot}%
\usepackage{booktabs}%
\usepackage{algorithm}%
\usepackage{algorithmicx}%
\usepackage{algpseudocode}%
\usepackage{listings}%
 \usepackage{geometry}
 \usepackage[rightcaption]{sidecap}
 \usepackage{tocloft}

\begin{document}

\title[Article Title]{Fractional-statistics-induced entanglement from Andreev-like tunneling}


\author[1,2]{\fnm{Gu} \sur{Zhang}}
\email{zhanggu217@nju.edu.cn}

\author[3,4]{\fnm{Pierre} \sur{Glidic}}
\email{pierre.glidic@ftf.lth.se}

\author[3]{\fnm{Fr{\'e}d{\'e}ric} \sur{Pierre}}\email{frederic.pierre@c2n.upsaclay.fr}

\author[5]{\fnm{Igor} \sur{Gornyi}}\email{igor.gornyi@kit.edu}

\author[6]{\fnm{Yuval} \sur{Gefen}}\email{yuval.gefen@weizmann.ac.il}

\affil[1]{\mbox{National Laboratory of Solid State Microstructures, School of Physics}, \mbox{Jiangsu Physical Science Research Center}, \mbox{and Collaborative Innovation Center of Advanced Microstructures}, \mbox{Nanjing University, Nanjing 210093, China}}

\affil[2]{\mbox{Beijing Academy of Quantum Information Sciences, Beijing 100193, China}}

\affil[3]{\mbox{Universit{\'e} Paris-Saclay, CNRS}, \mbox{Centre de Nanosciences et de Nanotechnologies}, 91120 Palaiseau, France}

\affil[4]{\mbox{NanoLund and Solid State Physics, Lund University, Box 118, 22100 Lund, Sweden}}

\affil[5]{Institute for Quantum Materials and Technologies and Institut f\"ur Theorie der Kondensierten Materie, Karlsruhe Institute of Technology, 76131 Karlsruhe, Germany}

\affil[6]{Department of Condensed Matter Physics, Weizmann Institute of Science, Rehovot 761001, Israel}



\abstract{The role of anyonic statistics stands as a cornerstone in the landscape of topological quantum techniques. While recent years have brought forth encouraging and persuasive strides in detecting anyons, a significant facet remains unexplored, especially in view of connecting anyonic physics to quantum information platforms---whether and how entanglement can be generated by anyonic braiding.
Here, we demonstrate that even when two anyonic subsystems (represented by anyonic beams) are connected only by electron tunneling, entanglement between them, manifesting fractional statistics, is generated.
To demonstrate this physics, we rely on a platform where fractional quantum Hall edges are bridged by a quantum point contact that allows only transmission of fermions (so-called Andreev-like tunneling). This invokes the physics of two-beam collisions in an anyonic Hong-Ou-Mandel collider, accompanied by a process that we dub \textit{anyon-quasihole braiding}.
We define an entanglement pointer---a current-noise-based function tailored to quantify entanglement associated with {\it quasiparticle fractional statistics}.
Our work, which exposes, both in theory and in experiment, entanglement associated with anyonic statistics and braiding, prospectively paves the way to the exploration of entanglement induced by non-Abelian statistics.
}


\maketitle

\newpage
\tableofcontents

\newpage

\section{Introduction}\label{sec1}

One of the most fascinating classes of quasiparticles is known as anyons. Recent years have borne witness to an intensified spotlight on anyons within the condensed-matter community. The focal point of this scrutiny stems from the
fact that anyons exhibit fractional statistics, which touches on the very foundations of quantum mechanics. Furthermore, anyons may represent the promising toolbox for quantum information processing (see, e.g., Refs.~\cite{NayakRevModPhys08,NielsenChuangBook}).
These quasiparticles, defying conventional exchange statistics, are predicted to reside in topologically intricate states, e.g., those realized in the regime of fractional quantum Hall (FQH) effect~\cite{LaughlinPRL83, Arovas1984}.
In particular, anyonic quasiparticles are hosted by the edges of Laughlin quantum-Hall states. The landscape of anyons extends to encompass Majorana modes, foreseen to materialize at the edges of topological superconducting materials~\cite{Kitaev2001UFN, MongPRX14}.
Three decades have passed since the pioneering confirmation of the fractional charge of Laughlin quasiparticles~\cite{PSaminadayarPRL97, de-PicciottoPhysBConMatt98}. Inspired by earlier endeavors in the exploration of fractional statistics~(see e.g., Refs.~\cite{CaminoPRB05, OfekPNAS10,WillettPRL13}, most recently, highly persuasive signals of anyonic statistics have been directly and indirectly observed in Fabry–Perot~\cite{NakamuraNatPhys19, NakamuraNatPhys20, NakamuraNC22, Nakamura2023} and Hong-Ou-Mandel interferometers~\cite{BartolomeiScience20, PierrePRX23, LeeNature23, RuellePRX23}. 

This leap in the search for anyonic statistics has been accompanied by a series of landmark experiments that have unveiled a plethora of exotic anyonic features in FQH systems. Among these are the existence of charge neutral modes~\cite{RajarshiPRL19, DuttaScience22}, fractional Josephson relation~\cite{GlattliScience19}, and Andreev-like tunneling~\cite{SafiSchulzPRB95, Sandler1998,Hashisaka2021, PierrePRX23,cohen2022universal, PierreNC23} in anyonic systems~\cite{Comforti2002}. 
The agreement between the experimental findings and the theoretical predictions not only consolidates our understanding but also offers new horizons to fuse the physics of anyons with other foundational themes of quantum mechanics. Indeed, in addition to earlier theoretical ideas~\cite{SafiDevilardPRL01, KaneFisherPRB03, VishveshwaraPRL03, KimPRL05, LawPRB06, FeldmanPRB07, RosenowHalperinPRL07, CampagnanoPRL12, CampagnanoPRB13, CampagnanoPRB16, RosenowLevkivskyiHalperinPRL16, SimNC16}, most recently, there has been another surge of theoretical proposals~\cite{LeeSimNC22,  schillerPRL23, LeePRL19, RosenowSternPRL20, MartinDeltaT20, MorelPRB22, KyryloPRB22, GuPRB22,  JonckheerePRL23, JonckheerePRB23} on understanding and detecting anyonic features, and possibly harnessing them for quantum information processing platforms (see, e.g., Refs.~\cite{NielsenChuangBook, WildeBook, NayakRevModPhys08, AliceaFendleyParaReview16}).

Entanglement is another fundamental quantum-mechanical element and a prerequisite for the development of quantum technology platforms.
Despite its significance, experimentally quantifying entanglement remains a challenging endeavor. Recently,
Ref.~\cite{GuNC24} proposed to measure entanglement stemming from quantum statistics of quasiparticles by a certain combination of the current cross-correlation functions. 
The main message of that reference, addressing integer quantum Hall platforms, is that the statistics-induced entanglement targets genuine entanglement (manifest via collisions between indistinguishable quantum particles), without resorting (cf. Refs.~\cite{ChtchelkatchevPRB02, Samuelsson2004, Das2016}) to the explicit study of Bell's inequalities~\cite{Bell64} and, thus, establishes a possibility of directly accessing entanglement in transport experiments. 
Notably, extracting statistics-induced entanglement is far from being a trivial task, as statistical properties need not necessarily lead to entanglement, which is, for instance, the case for a fermionic product state.

When transitioning to anyonic systems, the quantification of entanglement becomes even more formidable.
This is, in particular, related to the lack of readily available fractional statistics in ``natural platforms,'' which hinders the development of intuition about the statistics-induced mechanisms of entanglement
(like bunching and antibunching for bosons and fermions, respectively).
Furthermore, the \textit{quasiparticle collisions} 
that can directly reveal anyons' statistics~\cite{CampagnanoPRL12} through entanglement are now commonly believed to be irrelevant 
to the noise measurements in anyonic Hong-Ou-Mandel colliders~\cite{BartolomeiScience20, LeeNature23, PierrePRX23}.
Indeed, when considering dilute anyonic beams, anyonic collisions are overshadowed by  \textit{time-domain braiding}~\cite{RosenowLevkivskyiHalperinPRL16,SimNC16,LeeSimNC22,schillerPRL23} of an incoming anyon with spontaneously generated quasiparticle-quasihole pairs.
Nevertheless, measurements of anyons' entanglement through their collisions hold immense potential for characterizing and manipulating anyonic states.
Despite the importance of anyonic statistics, the quest to generate, observe, and quantify anyonic statistics-induced entanglement remains a challenge to this day.

Here, we take on this challenge and investigate, both theoretically and experimentally, entanglement generated among particles of two subsystems,  which is underlain by anyonic statistics. We employ the so-called ``Andreev-like tunneling'' platform, where the central collider allows only fermions to tunnel between anyonic edge channels. Since the braiding phase of anyons with fermions is trivial, this setup does not support time-domain braiding at the collider.
We demonstrate that, nevertheless, anyonic statistics affects the correlations between the two subsystems. This is the result of braiding between a dilute-beam anyon and a quasihole, the latter being triggered by an Andreev tunneling event.  We refer to this process as an \textit{anyon-quasihole braiding}, and demonstrate that it gives rise to the dependence of collision-induced entanglement on the fractional braiding phase.

\begin{figure} 
\centering
\includegraphics[width=0.7 \linewidth]{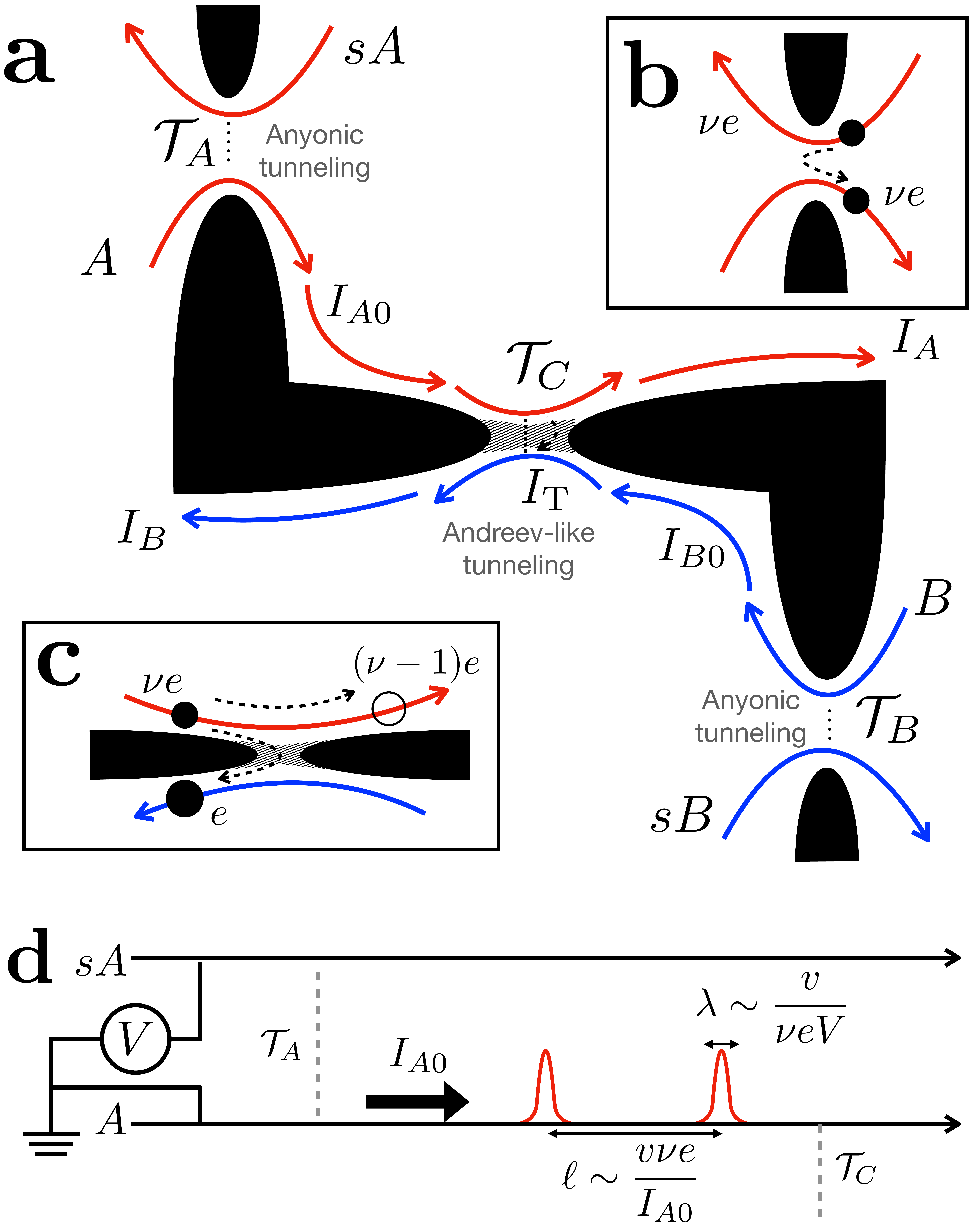}
  \caption{
  Schematic depiction of the model with Andreev-like tunneling between the fractional edges (cf. Fig.~S6 in the Supplementary Information). The quantum Hall bulk is represented by white regions separated by potential barriers (``fingers'' introduced by gates) shown in black; the gray areas correspond to barriers allowing for electron (but not anyon) tunneling.  \textbf{Panel~a:}~The entire setup involves two source arms ($sA$, $sB$) and two middle arms ($A$, $B$) in the FQH regime. They host chiral anyons that correspond to the bulk filling factor $\nu < 1/2$.
  Chiral edge-state transport modes are designated with the red and blue curved arrows for subsystems $\mathcal{A}$ (including $sA$ and $A$) and $\mathcal{B}$ ($sB$ and $B$), respectively.
  $I_A$ and $I_B$ represent the currents in middle arms ($A$ and $B$, respectively), past the central QPC.
  Before the central QPC, currents in arms $A$ and $B$ are represented by $I_{A0}$ and $I_{B0}$, respectively.
  Current $I_\text{T}$ tunnels through the central QPC connecting arms $A$ and $B$.
  \textbf{Panel~b:}~Anyons of charge $\nu e$ tunnel from the sources to corresponding middle arms $A$ and $B$ through diluters with transmissions $\mathcal{T}_A$ and $\mathcal{T}_B$, respectively.
  \textbf{Panel~c:}~Channels $A$ and $B$ communicate through the central QPC (central collider) with the transmission $\mathcal{T}_C$. The central QPC allows only electrons to tunnel, resulting in the ``reflection'' of an anyonic hole [with charge $(\nu - 1 ) e$, empty circle], which resembles Andreev reflection at the metal-superconductor interfaces.
  \textbf{Panel d}: Theoretical depiction of subsystem $\mathcal{A}$ that comprises channels $sA$ and $A$ (the upper half of panel \textbf{a}). Channel $A$ features the dilute current beam $I_{A0}$, coming from source $sA$ through a diluter with transmission probability $\mathcal{T}_A$. The schematics of subsystem $\mathcal{B}$ are similar.
    }
  \label{fig:model}
\end{figure}

\section{Entanglement pointer for Andreev-like tunneling}\label{sec2}

In this work, we combine anyonic statistics with quantum entanglement and define the \textit{entanglement pointer} to quantify the statistics-induced entanglement in a Hong-Ou-Mandel interferometer on FQH edges with filling factor $\nu$ (Fig.~\ref{fig:model}\textbf{a}).
Our platform contains three quantum point contacts (QPCs), including two diluters (Fig.~\ref{fig:model}\textbf{b}) and one central collider (Fig.~\ref{fig:model}\textbf{c}).
These QPCs bridge chiral channels propagating at different sample edges (indicated by red and blue arrows of Fig.~\ref{fig:model}\textbf{a}).
The setup is characterized by the experimentally measurable transmission probabilities $\mathcal{T}_A$, $\mathcal{T}_B$, and  $\mathcal{T}_C$ of the two diluters and the central QPC, respectively.

As far as the two diluter QPCs are concerned, they are characterized by a small rate of anyon tunneling through, generating dilute non-equilibrium beams in channels $A$ and $B$. These beams are characterized by two length scales (see Fig.~\ref{fig:model}\textbf{d}): (i) the typical width of non-equilibrium anyon pulses, $\lambda \sim v/\nu e V$, and (ii) the typical distance between two neighboring non-equilibrium anyon pulses, $\ell \sim v\nu e/I_{A 0}$ (and similarly for channel $B$), where $v$ is the velocity of edge excitations. 
When $I_{A 0} \ll (\nu e^2) V$, we obtain $\lambda \ll \ell$, such that incoming anyons can typically be considered as well-separated and, thus, independent quasiparticles.
This is the regime of diluted anyonic beams addressed here, characterized by the weak-tunneling condition for diluters, $\mathcal{T}_{A,B} \ll 1$.
For later convenience, we define
$I_+ = I_{A0} + I_{B0}$ as the sum of non-equilibrium currents $I_{A0}$ and $I_{B0}$ in arm $A$ and $B$ (see Fig.~\ref{fig:model}\textbf{a}), respectively, before arriving at the central collider. 

\begin{figure}
\centering
  \includegraphics[width=0.85\linewidth]{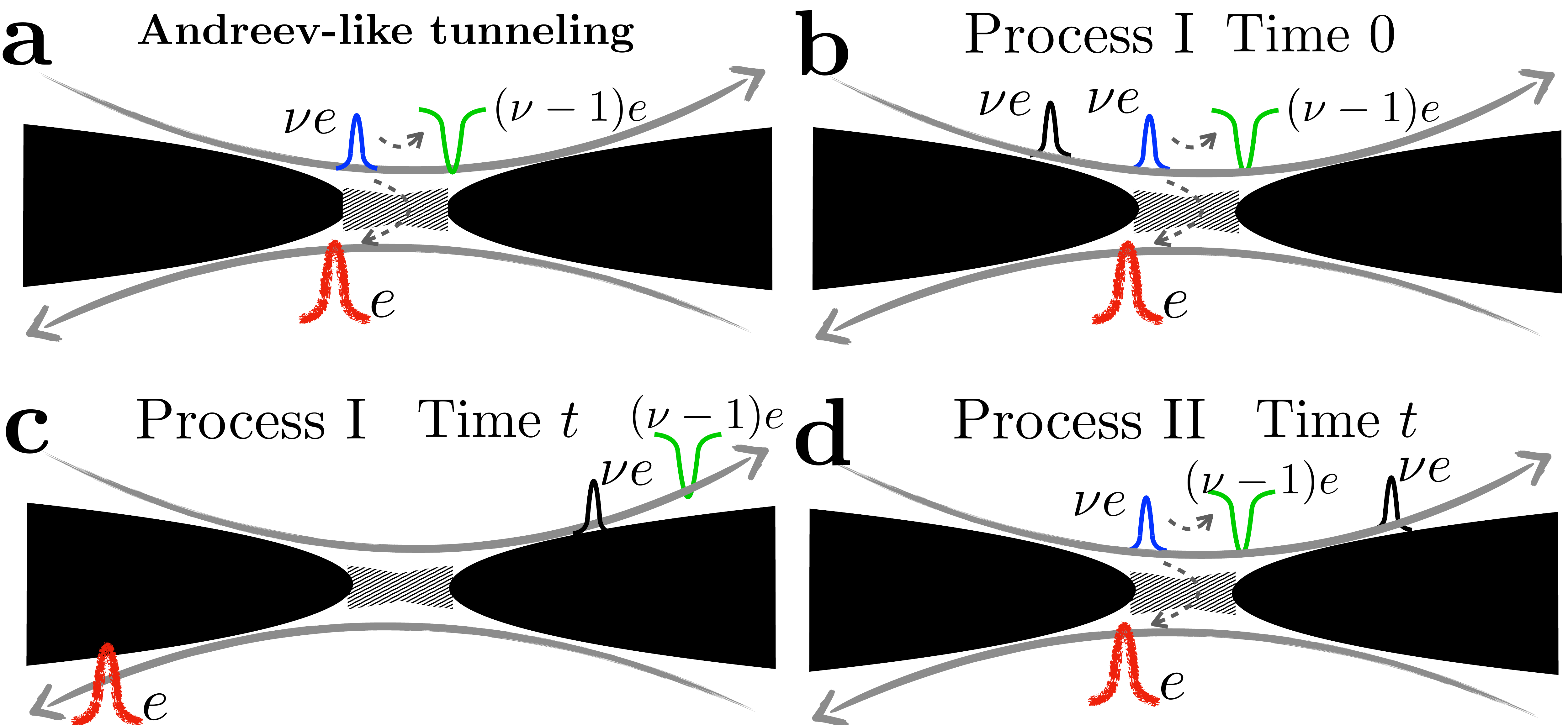}
  \caption{Anyon-quasihole braiding in Andreev-like tunneling processes. Gray arrows mark the chirality of the corresponding channels.
    \textbf{Panel a}: Leading-order Andreev-like tunneling, corresponding to Fig.~\ref{fig:model}\textbf{c}. Here, an anyon from the diluted beam (the blue pulse, of charge $\nu e$) arrives at the central collider, triggering the tunneling of an electron (the red pulse, of charge $e$) and the accompanied reflection of a hole [the green pulse, of charge $(\nu-1) e$]. \textbf{Panels b} and \textbf{c}: Illustrations of Process I, at times 0 and $t$, respectively. In \textbf{Panel b}, an Andreev-like tunneling occurs at time 0 at the collider. The positions of the particles involved at a later time $t$ are marked accordingly in \textbf{Panel c}. In both \textbf{b} and \textbf{c}, the non-equilibrium anyon (the black pulse) is located upstream (to the left) of the reflected hole (the green pulse). \textbf{Panel d}: Process II, in which the Andreev-like tunneling occurs at time $t$, when the black pulse has already passed the central QPC.     In comparison to \textbf{b} and \textbf{c}, here the non-equilibrium anyon (the black pulse) is located downstream (to the right) of the reflected hole (the green pulse). The interference of Processes I and II thus generates the \textit{anyon-quasihole braiding} between the non-equilibrium anyon (black) and the reflected quasihole (green). Note that this is not a vacuum-bubble braiding  (Ref.~\cite{SimNC16}) a.k.a. time-domain braiding (Refs.~\cite{RosenowLevkivskyiHalperinPRL16,LeeSimNC22,LeeNature23,schillerPRL23,LandscapePRL25}).
  }
\label{fig:time_domain_braiding_illustration}
\end{figure}

The model at hand is crucially distinct from more conventional anyonic colliders~\cite{RosenowLevkivskyiHalperinPRL16, BartolomeiScience20, LeeSimNC22,LeeNature23, PierrePRX23} in that its central QPC only allows the transmission of fermions~\cite{Comforti2002, Hashisaka2021, PierreNC23}.
This is experimentally realizable by electrostatically tuning the central QPC into the ``vacuum'' state (no FQH liquid), thus forbidding the existence and tunneling of anyons inside this QPC.
The dilute non-equilibrium currents in the middle arms are carried by anyons with charge $\nu e$ (Fig.~\ref{fig:model}\textbf{b}), where $\nu$ is the filling fraction. 
Since only electrons are allowed to tunnel across the central QPC, such a tunneling event must be accompanied by leaving behind a fractional hole of charge $-(1-\nu)e$; the latter continues to travel along the original middle edge (Fig.~\ref{fig:model}\textbf{c}). This ``reflection'' event is reminiscent of the reflection of a hole in an orthodox Andreev tunneling from a normal metal to a superconductor; hence, such an event is commonly dubbed ``quasiparticle Andreev reflection''~\cite{Sandler1998,JonckheerePRB23}. As distinct from the conventional normal metal-superconductor case, in an anyonic Andreev-like tunneling process, (i) both the incoming anyon and reflected ``hole'' carry fractional charges, and (ii) the absolute values of anyonic and hole charges differ.

It is known that for anyonic tunneling, time-domain braiding (or, alternatively, braiding with the topological vacuum bubbles~\cite{SimNC16}) can occur between an anyon-hole pair generated at the central QPC and anyons that bypass the collider~\cite{RosenowLevkivskyiHalperinPRL16, SimNC16, LeeSimNC22, schillerPRL23}.
Such a process is, however, absent for Andreev-like tunneling, where vacuum bubbles are made of fermions that cannot braid with anyons.
Instead, another mechanism of braiding is operative in Andreev-like setups, which requires the inclusion of higher-order tunneling processes at the diluters.  
As shown in Fig.~\ref{fig:time_domain_braiding_illustration} (where we take the single-source case as an example), the fractional statistics of a fractional-charge hole (the green pulse in Fig.~\ref{fig:time_domain_braiding_illustration}), which is left behind by the fermion tunneling, enables braiding of this quasihole with anyons supplied by the diluter (black pulses of Fig.~\ref{fig:time_domain_braiding_illustration}).
We term this type of braiding ``anyon-quasihole braiding''.
For comparison, time-domain braiding in an anyonic tunneling system is illustrated in Sec.~II in the Supplementary Information (SI).

\begin{figure}
\centering
\includegraphics[width=0.9 \linewidth]{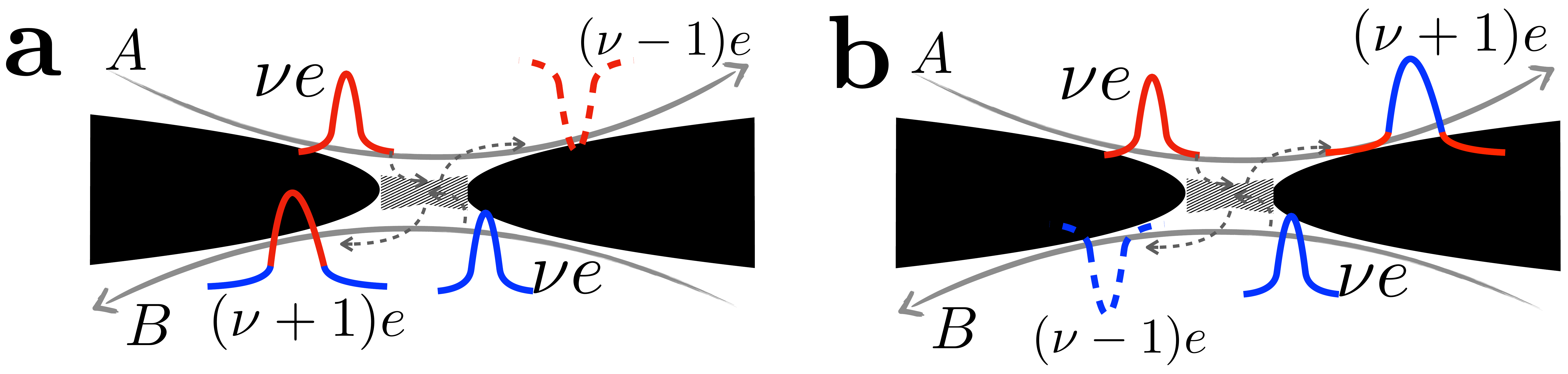}
  \caption{
  Generation of cross-correlation by the collision of two anyons (the red and blue pulses) arriving at the collider from uncorrelated sources. There are two possible processes. \textbf{Panel a}: A pulse of charge $(\nu + 1 )e$ is generated in channel $B$, leaving an outgoing fractional-charge hole with charge $(\nu -1 )e$ in channel $A$. \textbf{Panel b}: An alternative process, where a pulse of charge $(\nu +1) e$ is created in channel $A$, leaving a quasihole with charge $(\nu -1 )e$ in Channel $B$.
  The resulting cross-correlation is intrinsically related to the entanglement generated  between two subsystems ($\mathcal{A}$ and $\mathcal{B}$), which is captured by the entanglement pointer.
  The Andreev two-anyon collision processes are further ``decorated'' by anyon-quasihole braiding (which involves additional anyons supplied by the diluters, cf. Fig.~\ref{fig:time_domain_braiding_illustration}). To keep the figure simple, the latter is not shown.
    }
  \label{fig:schematics_and_entanglement}
\end{figure}

Anyon-quasihole braiding significantly influences the generation of entanglement between the two parts of the system---subsystems $\mathcal{A}$ and $\mathcal{B}$ (see Fig.~\ref{fig:model}\textbf{a}). 
To characterize this statistics-induced entanglement we introduce the {\it entanglement pointer}~(cf. Ref.~\cite{GuNC24}) for Andreev-like tunneling:
\begin{equation}
    \mathcal{P}_\text{Andreev} \equiv -S_\text{T} (\mathcal{T}_A, \mathcal{T}_B) + S_\text{T} (\mathcal{T}_A,0) + S_\text{T} (0,\mathcal{T}_B)  .
    \label{eq:paa}
\end{equation}
Here,
$S_\text{T} (\mathcal{T}_A,\mathcal{T}_B)$
refers to the noise of the tunneling current between the two subsystems, which is a function of the transmission probabilities of the two diluters.
The entanglement pointer effectively subtracts out those contributions to the tunneling noise that are present when only one of the two sources is biased (which is equivalent to setting one of the two transmissions to zero), thus highlighting the effects of correlations formed between the two diluted anyonic beams. 
Indeed, the contribution to the noise that results from collisions between particles from the two beams is absent in the sum of two single-source processes; the corresponding difference is captured in Eq.~\eqref{eq:paa}.
By construction, $\mathcal{P}_\text{Andreev}$ naturally quantifies entanglement generated by two-particle collisions (see Fig.~\ref{fig:schematics_and_entanglement} for the illustration of corresponding correlations), through which quasiparticle statistics is manifest (in analogy with bunching and antibunching for bosons and fermions, respectively).
Although the entanglement pointer is defined relying on the tunneling-current noise, it can also be measured with the cross-correlation noise, see Eq.~\eqref{eq:noise_two_particle} below and the ensuing discussion.

\section{Tunneling-current noise}\label{sec3}

As discussed above, anyonic statistics is manifest in Andreev-like tunneling through the central collider and the corresponding tunneling-current noise.
Remarkably, although the tunneling particles are fermions, the associated transport still reveals the anyonic nature of the edge quasiparticles. This is the consequence of anyon-quasihole braiding between reflected fractional-charge qusihole (green pulse in Fig.~\ref{fig:time_domain_braiding_illustration}) and an anyon from diluted beams (black pulse in Fig.~\ref{fig:time_domain_braiding_illustration}, generated by tunneling through diluters).
More specifically, for Andreev-like transmission through the central QPC at $\nu<1/2$, the expression for the 
tunneling noise, when the two sources are biased by the same voltage $V$, can be decomposed as follows: $S_\text{T} = S_\text{T}^\text{single} + S_\text{T}^\text{collision}$, where (for simplicity, here and in what follows, we set $\hbar=v=1$)
\begin{equation}
\begin{aligned}
S_\text{T}^\text{single} &\! = \text{Re}\left\{\mathcal{T}_C\, \mathcal{T}_A \,\frac{\nu e^3 V}{2} \frac{ (2\pi \nu)^{1-\nu_\text{s}} e^{i\pi ( \nu_\text{s} -  \nu \nu_\text{s}+1 )}}{ \pi\nu\sin (\pi\nu_\text{s}) + 2 f_1(\nu) \mathcal{T}_A}   
 \left[ 2i\pi \nu -  \mathcal{T}_A\left( 1 - e^{-2i\pi\nu} \right)  \right]^{\nu_\text{s}-1} \right\}  + \left\{A\to B\right\} ,\\
&\text{with}\ \ f_1(\nu) \!\equiv (\nu_\text{s} - 1) \sin (\pi\nu) \big\{\sin \left[\pi (\nu_\text{s} -\nu)\right] + \sin (\pi\nu) \big\},
\end{aligned}
\label{eq:S1AB}
\end{equation}
is the sum of single-source noises resulting from separately activating sources $sA$ and $sB$, and 
\begin{equation}
\begin{aligned}
 S_\text{T}^\text{collision}& =  \text{Re} \left\{\mathcal{T}_C\, e^3 V\frac{ \sqrt{\mathcal{T}_A \mathcal{T}_B}\, f_2(\nu) \cos\left(\pi \nu_\text{d}/{2} \right) }{\pi\nu\sin\left( \pi \nu_\text{s} \right) + 2 f_1 (\nu) \sqrt{\mathcal{T}_A \mathcal{T}_B}  }\,     \left[\mathcal{T}_A \left( 1 - e^{-2i\pi\nu} \right) + \mathcal{T}_B \left( 1 - e^{2i\pi\nu} \right)\right]^{\nu_\text{d} - 1} \right\},\\
 \text{with}&\ \ f_2 (\nu) \equiv \frac{4\pi^3\,  (2\pi\nu)^{1-\nu_\text{d}}\, \Gamma(1 - \nu_\text{d})}{\sin(2\pi\nu)\, \Gamma (1 - 2\nu)\, \Gamma\left(1-\nu_\text{s} \right)} ,
\end{aligned}
\label{eq:andreev noises}
\end{equation}
is the double-source ``collision contribution'' (see Sec.~IIIB of the SI). In Eqs.~\eqref{eq:S1AB} and \eqref{eq:andreev noises}, $\nu_\text{s} \equiv 2/\nu + 2 \nu - 2$ and $\nu_\text{d} \equiv 2/\nu + 4\nu - 4$ reflect the scaling features of Andreev-like tunneling, for the single-source and the double-source (collision-induced) contributions, respectively.
Crucially, 
a phase factor $\exp(\pm2i\pi\nu)$ appearing in Eqs.~\eqref{eq:S1AB} and \eqref{eq:andreev noises} is generated by the braiding of two Laughlin quasiparticles (i.e., by the anyon-quasihole braiding) that have the statistical phase $\pi \nu$. This factor, multiplying the diluter transparency, is, however, concealed in the single-source noise in the strongly dilute limit ($\mathcal{T}_{A,B}\ll 1$) by the constant term $-2i\pi \nu$ in the square brackets of Eq.~\eqref{eq:S1AB}. On the contrary, this statistical factor appears already in the leading term of Eq.~\eqref{eq:andreev noises}, rendering the collision contribution to the noise particularly handy for extracting the information on the quasiparticle statistics.

\begin{figure}
  \includegraphics[width=1\linewidth]{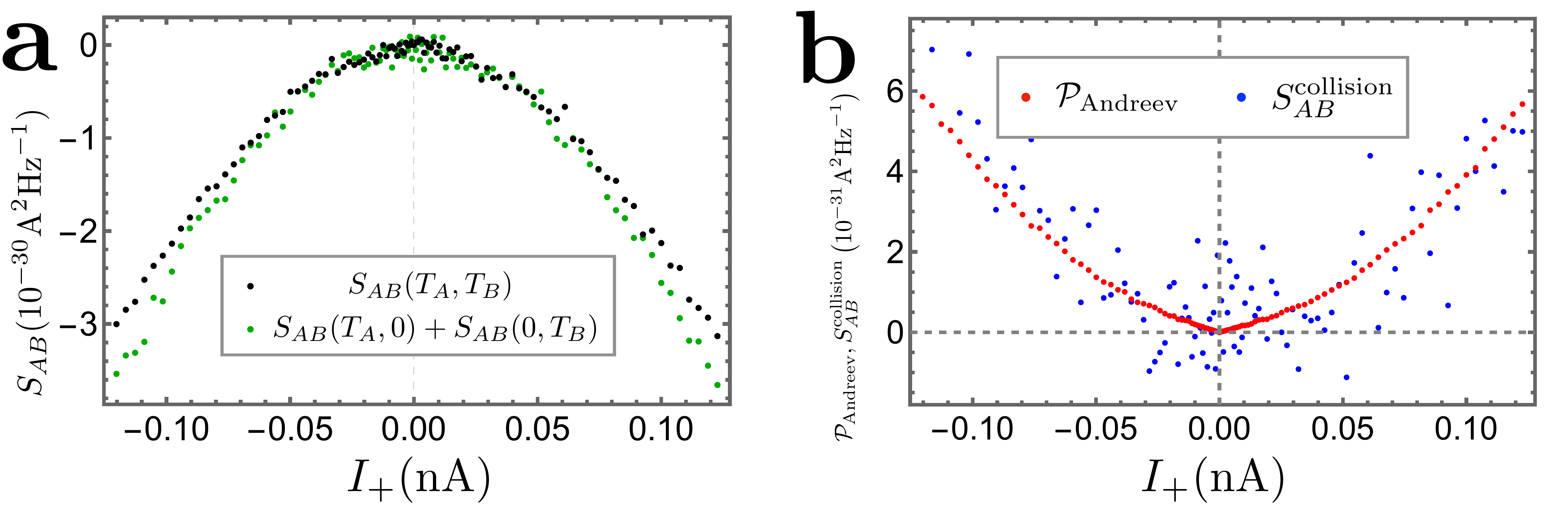}
  \caption{Cross-correlation noise: theory vs. experiment.
\textbf{Panel~a:} Experimental data for the cross-correlation $S_\text{AB}$, for the single-source (green points, corresponding to the sum of two single-source signals) and double-source (black points) scenarios, respectively.
Here, the $x$-axis represents $I_+$ of the double-source situation.
\textbf{Panel~b:} The theory-experiment comparison. The experimental data $\mathcal{P}_\text{Andreev}$ refers to the collision contribution to the cross-correlation, obtained following the definition of Eq.~\eqref{eq:paa}. For comparison, one can alternatively obtain $S_\text{AB}^\text{collision}$, following Sec.~VIII of the SI, indirectly from transmissions at diluters and the central collider.
The two approaches compare well for most values of $I_+$.
For small values of $I_+$, the weight of thermal fluctuations becomes more significant, leading to a larger deviation.
Further experimental information, concerning the transmission probability through the central collider ($\mathcal{T}_C$) and that through the diluters ($\mathcal{T}_A$ and $\mathcal{T}_B$) is provided in Fig.~\ref{fig:tc} and  Fig.~S7 of the SI, respectively.
}
  \label{fig:experimental_data}
\end{figure}

According to Eq.~\eqref{eq:paa}, the entanglement pointer is determined by the value of $S_\text{T}^\text{collision}$: 
\begin{equation}
    \mathcal{P}_\text{Andreev} =  -S_\text{T}^\text{collision}.
    \label{eq:paa_nu}
\end{equation}
Note that $S_\text{T}^\text{collision}$ vanishes when $\nu = 1$, indicating that the Andreev entanglement pointer, $\mathcal{P}_\text{Andreev}$, represents a quantity that is unique for anyons.
As an important piece of the message, when $\nu = 1/3$, the extra noise induced by collisions between two Laughlin quasiparticles is negative, i.e., $S_\text{T}^\text{collision} (\mathcal{T}_A,\mathcal{T}_B) < 0$.
This indicates that the simultaneous arrival of anyons reduces the probability of Andreev-like reflection at the central collider, as supported by the experimental data (cf. Fig.~\ref{fig:experimental_data}).

The entanglement pointer, $\mathcal{P}_\text{Andreev}$, has three advantages over the total tunneling-current noise $S_\text{T} (\mathcal{T}_A, \mathcal{T}_B)$ or current cross-correlations.
Firstly, it rids of single-beam contributions to the current correlations, which are not a manifestation of genuine statistics-induced entanglement.
Secondly, $\mathcal{P}_\text{Andreev}$ reflects the statistics-induced extra Andreev-like tunneling for two-anyon collisions.
It provides an alternative option (other than the braiding phase~\cite{NakamuraNatPhys19} and two-particle bunching or anti-bunching preferences~\cite{CampagnanoPRL12}) to disclose anyonic statistics.
Thirdly, $\mathcal{P}_\text{Andreev}$ is resilient against intra-edge interactions, in edges that host multiple edge channels.
In the setup we consider here, the interaction occurs between the edges coupled by the central QPC.
Since the region where the two edges come close to each other has a rather small spatial extension, the effect of such inter-edge interaction is weak, leading to small corrections to both cross-correlation and the entanglement pointer (see SI Sec.~VB).
The situation will be, however, different, when considering systems with complex edges that contain multiple edge channels.
Indeed, following our discussions at the end of Sec.~VC in the SI, interactions in such setups may lead to a significant correction to the cross-correlation due to the so-called charge fractionalization.
This correction, which may even exceed the interaction-free noise, is avoided by the subtraction of the single-source noises, when evaluating the entanglement pointer.

\section{Physical interpretation of the entanglement pointer}\label{sec4}

The essence of an entanglement pointer can be illustrated by resorting to single-particle (Fig.~\ref{fig:time_domain_braiding_illustration}) and two-particle  
(Fig.~\ref{fig:schematics_and_entanglement}) scattering formalism revealing the statistical properties of anyons in the course of two-particle collisions,
via bunching or anti-bunching preferences~\cite{BlanterButtikerPhysRep00,CampagnanoPRL12}.
The situation is more involved for the model under consideration, as particles that are allowed to tunnel at the central collider (fermions) are of distinct statistics that differs from that of the colliding particles (anyons).
As we have emphasized above, although only fermions can tunnel through the central collider, anyonic statistics still manifests itself by influencing
the probability of Andreev-like tunneling events, when two anyons arrive at the collider simultaneously.

The probability of a two-anyon scattering event is proportional to $\mathcal{T}_A \mathcal{T}_B$; 
Andreev-like tunneling then produces fractional charges on both arms, as shown in Fig.~\ref{fig:schematics_and_entanglement}.
These processes establish the entanglement between two subsystems $\mathcal{A}$ and $\mathcal{B}$, that are initially independent from each other otherwise.
Noteworthily, here, the entanglement is induced by the statistics of colliding anyons, not interactions at the collider.
After including both single-particle and two-particle scattering events, we obtain (SI Sec.~VI) the differential noises at a given voltage $V$:
\begin{equation}
    \begin{aligned}
        s_\text{T} & = (s_\text{T})_\text{single} + (s_\text{T})_\text{collision} \!=\! (\mathcal{T}_A \! + \! \mathcal{T}_B) \mathcal{T}_C \!-\! ( \mathcal{T}_A^2 \!+\! \mathcal{T}_B^2 ) \mathcal{T}_C^2 \! + \! \mathcal{T}_A \mathcal{T}_B P_\text{Andreev}^\text{stat},\\
        s_\text{AB} & = (s_\text{AB})_\text{single} + (s_\text{AB})_\text{collision} 
        =\! -(1\!-\!\nu) \mathcal{T}_C (\mathcal{T}_A \! + \!\mathcal{T}_B) \!- \!\mathcal{T}_C ( \nu \!-\! \mathcal{T}_C ) (\mathcal{T}_A^2 \!+\! \mathcal{T}_B^2 ) - \mathcal{T}_A \mathcal{T}_B P_\text{Andreev}^\text{stat},
    \end{aligned}
    \label{eq:noise_two_particle}
\end{equation}
where the subscripts ``single'' and ``collision'' indicate contributions from single-particle and two-particle scattering events, respectively.
Here, ${s_\text{T}=\partial_{e I_+/2} S_\text{T}}$ and ${s_\text{AB}=\partial_{e I_+/2} S_\text{AB}}$ are the differential noises, and ${S_\text{AB}=\int dt\langle \delta \hat{I}_A (t) \delta \hat{I}_B (0) \rangle}$ is the irreducible zero-frequency cross-correlation with $\delta \hat{I}_{A,B} \equiv \hat{I}_{A,B} -  I_{A,B}$ the fluctuation of the current operator $\hat{I}_{A,B}$. 
The factor $P_\text{Andreev}^\text{stat}$
refers to extra Andreev-like tunneling induced by anyonic statistics in the course of two-anyon collisions.
It would be equal to zero if anyons from subsystem $\mathcal{A}$ were distinguishable from those in $\mathcal{B}$. In this case, the noise would be equal to the sum of two single-source ones. 
By comparing with Eqs.~\eqref{eq:S1AB}, \eqref{eq:andreev noises}, and \eqref{eq:paa_nu}, $P_\text{Andreev}^\text{stat}$ can be expressed via the microscopic parameters [see Eq.~(S100) of the SI Sec.~VI and more details in SI Secs.~I and IV]; 
furthermore,  $\mathcal{T}_{A,B} = \partial_V I_{A0,B0}h/(e^2\nu)$ are directly related to the conductance of the corresponding diluter.
As another feature of Andreev-like tunnelings, $S_\text{T}$ in Eq.~\eqref{eq:noise_two_particle}
does not explicitly depend on $\nu$, since the central QPC allows only charge $e$ particles to tunnel.

Equation~\eqref{eq:noise_two_particle} exhibits several features of Andreev-like tunneling in an anyonic model.
Firstly, in the strongly dilute limit, $s_\text{AB} \approx (\nu-1)s_\text{T}$, when considering only the leading contributions to the noise, i.e., the terms linear in both $\mathcal{T}_A$ (or $\mathcal{T}_B$) and $\mathcal{T}_C$.
Both $(s_\text{T})_\text{single}$ and $(s_\text{AB})_\text{single}$ correspond to $S_\text{T}^\text{single}$ [Eq.~\eqref{eq:S1AB}] and are subtracted out following our definition of the entanglement pointer, Eq.~\eqref{eq:paa}.
In both functions, the double-source collision contributions, i.e., the bilinear terms $\pm \mathcal{T}_A \mathcal{T}_B P_\text{Andreev}^\text{stat}$, involve $P_\text{Andreev}^\text{stat}$. This is the contribution to the entanglement pointer generated by statistics, when two anyons collide at the central QPC.
Most importantly, bilinear terms $\propto \mathcal{T}_A \mathcal{T}_B$ of both functions in Eq.~\eqref{eq:noise_two_particle} have the same magnitude, i.e., $(s_\text{T})_\text{collision} = -(s_\text{AB})_\text{collision}$. Consequently, the experimental measurement of $\mathcal{P}_\text{Andreev}$, though defined with tunneling current noise, can be performed by measuring the cross-correlation of currents in the drains, which is more easily accessible in real experiments: 
\begin{align} 
\mathcal{P}_\text{Andreev}
=& -\frac{e \mathcal{T}_A \mathcal{T}_B}{2} \int dI_+ \, P_\text{Andreev}^\text{stat} (I_+)
= S_\text{AB} (\mathcal{T}_A, \mathcal{T}_B) - S_\text{AB} (\mathcal{T}_A,0) - S_\text{AB} (0,\mathcal{T}_B).
\label{eq6}
\end{align}

\section{Comparison with experiment}\label{sec5}

We now compare our theoretical predictions with the experimental data (cf. Refs.~\cite{PierreNC23} and \cite{Glidic2024}), see Fig.~\ref{fig:experimental_data}.
Panel \textbf{a} shows the raw data for the double-source noise $S_\text{AB}(\mathcal{T}_A,\mathcal{T}_B)$ and for the sum of single-source noises $S_\text{AB}(\mathcal{T}_A,0)+S_\text{AB}(0,\mathcal{T}_B)$. For the single-source data, the $x$-axis 
represents $I_{A0} (\mathcal{T}_A , 0) + I_{B0} (0,\mathcal{T}_B)$, i.e., the sum of non-equilibrium currents in the two single-source settings (with either source $sA$ or source $sB$ biased).
Firstly, as shown in panel \textbf{a}, the double-source cross-correlation, $S_\text{AB} (\mathcal{T}_A, \mathcal{T}_B )$, is smaller in magnitude compared to the sum, $S_\text{AB} (\mathcal{T}_A, 0 ) + S_\text{AB} (0, \mathcal{T}_B )$, of two single-source cross-correlations.
This fact agrees with the negativity of $S_\text{T}^\text{collision}$ (tunneling-current noise induced by two-anyon collision) of Eq.~\eqref{eq:andreev noises} for $\nu = 1/3$.
To verify our theoretical result, we compare, in Fig.~\ref{fig:experimental_data}\textbf{b}, the values of $\mathcal{P}_\text{andreev}$ and $S_\text{AB}^\text{collision}$.
Here, the former is obtained directly from the measured noises by virtue of Eq.~\eqref{eq6}, while the latter, defined as $S_\text{AB}^\text{collision} (\mathcal{T}_A, \mathcal{T}_B) \equiv S_\text{AB} (\mathcal{T}_A, \mathcal{T}_B) - S_\text{AB} (0, \mathcal{T}_B) - S_\text{AB} (\mathcal{T}_A, 0) $, is calculated from the measured dependence of the tunneling current on the incoming currents using the following relation:
\begin{equation}
\begin{aligned}
    & S_\text{AB}^\text{collision} 
    \!=\! \frac{e I_+ \tan (\pi\nu)}{2(\nu_\text{d} - 1) \tan\left( \pi \nu_\text{d}/2 \right)   }
    \left\{ \left( \frac{\partial}{\partial I_{A0}} \!-\! \frac{\partial}{\partial I_{B0}}  \right) \Big[ I_\text{T} (\mathcal{T}_A, 0) \!+\!  I_\text{T} (0,\mathcal{T}_B) \!-\! I_\text{T} (\mathcal{T}_A, \mathcal{T}_B) \Big]  \right\}\Bigg|_{I_- = 0}.
\end{aligned}
\label{eq:sab_with_it_conductance}
\end{equation}
Derivation of Eq.~\eqref{eq:sab_with_it_conductance} 
relies on explicit expressions for the noise, Eqs.~\eqref{eq:S1AB} and \eqref{eq:andreev noises}, as well as expressions \eqref{eq:explicit_current_expression} for the tunneling currents presented in Methods (see details in Sec.~VIII of the SI).
Figure~\ref{fig:experimental_data}\textbf{b} demonstrates remarkable agreement between the theory and experiment for $\mathcal{P}_\text{Andreev}$.
This indicates the validity of the qualitative picture based on the phenomenon of anyon-quasihole braiding, which influences Andreev-like tunneling as described in Sec.~\ref{sec4}.

\section{Conclusions}\label{sec8}

In this work, we have 
studied, both theoretically and experimentally, the generation of entanglement associated with the fractional quasiparticle statistics
in an anyonic (with filling factor ${\nu<1/2}$) Hong-Ou-Mandel interferometer that exhibits
Andreev-like tunneling through the central QPC.
We defined the entanglement pointer through the
associated noise functions that
are obtained by considering anyon-triggered fermion tunneling at the central QPC, which is accompanied by ``anyon-quasihole braiding'' of Andreev-reflected anyonic charges with anyons from non-equilibrium beams.
The Andreev-like tunneling in an anyonic collider is ``halfway'' 
between
the integer case of Ref.~\cite{GuNC24} (where both tunneling and dynamics along the arms are fermionic) 
and
a purely anyonic collider (both tunneling and dynamics are anyonic).
The latter case will be addressed elsewhere, with insights from the present work indicating
that quasiparticle collisions do matter in the collider geometry.
The identification of the peculiar braiding mechanism in the Andreev-like platform studied here suggests a variety of potential unexpected phenomena determined by anyonic statistics.

The Hong-Ou-Mandel setup in the Andreev-like tunneling regime provides us with a convenient platform for a direct inspection and study of real anyonic collisions, and, especially, the resulting generation of entanglement of initially unentangled anyons in transport experiments (cf. Fig.~\ref{fig:schematics_and_entanglement}).
Our analysis indicates that the exchange of electrons (carrying ``trivial'' fermionic statistics) between two anyonic subsystems suffices to render those subsystems aware of their mutual \textit{anyonic statistics}, generating non-trivial statistical entanglement between the subsystems.
The theory predictions are verified in
the experiment; the measured data agree remarkably well with the theoretically calculated one, for both the current cross-correlations $S_\text{AB}$ and the entanglement pointer $\mathcal{P}_\text{Andreev}$.
We have thus demonstrated the crucial role of two-particle collisions in establishing fractional-statistics-induced entanglement in anyonic colliders.

An interesting issue here concerns the role of electrostatic interactions: To what extent is our entanglement pointer sensitive to such interactions? One identifies two types of interactions: intra-edge interactions among several chiral modes in a given edge and inter-edge interactions around the collider. To leading order, the contribution of the former to the current-current correlations is subtracted when calculating the entanglement pointer. The magnitude of the latter is parametrically small, given the relatively small size of the collider and typically weak interaction between the edges (cf. Ref.~\cite{GuNC24}).

Our theory demonstrates that the idea of entanglement pointer, introduced in Ref.~\cite{GuNC24} for fermions and bosons, can be non-trivially extended to the anyonic case, capturing the effect of braiding of Abelian quasiparticles, which manifests fractional statistics. 
Prospectively, our work unveils the relevance of statistics-induced entanglement to even more sophisticated settings.
In particular, our work motivates further studies of Andreev-like tunneling beyond Laughlin quasiparticles, employing either particle-like or hole-like FQH fractions, as well as more exotic quasiparticles (like, e.g., ``neutralons'' and non-Abelian anyons), under non-equilibrium conditions.
Although Andreev-like reflection has been investigated in various setups that comprise non-Abelian edge states~\cite{ImuraSSC98, OhashiJPS22,MaX2022}, 
the highly intriguing challenge from this perspective would be to 
realize strongly diluted beams, 
facilitating the braiding of non-Abelian anyons
with Andreev-reflected fractional quasiparticles.
Given the present analysis, we expect that such braiding will generate entanglement induced by non-Abelian statistics. It is feasible (see Sec. IX of the SI) to extend our framework, which quantifies entanglement induced by Laughlin quasiparticle statistics, to non-Abelian systems. 
Generating and quantifying the statistics-induced entanglement through transport experiments will allow both the identification of non-Abelian states 
(cf. Refs.~\cite{Park2020, park2024fingerprints} and references therein) and the manipulation of entanglement in topological quantum platforms.
In particular, this may shed more light on the complex structure of non-Abelian edges---distinguishing between the candidate states---through the entanglement content obtained from transport noise measurements.

Finally, the statistics-induced entanglement reported here is evidently a topological phenomenon, since the anyonic fractional statistics is a manifestation of topology.
Crucially, our Andreev entanglement pointer, given by Eqs.~(\ref{eq:paa_nu}) and (\ref{eq:andreev noises}),  explicitly vanishes at $\nu = 1$, which is a feature shared by the topological entanglement entropy \cite{KitaevPreskillPRL06}.
It stands to reason to envision that employing our framework can provide direct access to topological entanglement entropy, which is expected to open up further avenues in experimentally studying systems with topological order.

\section{Methods}\label{sec10}

\subsection{Theoretical model}
\label{theo-methods}

We consider the anyonic setup shown in Fig.~\ref{fig:model}, which consists of two source arms ($sA$, $sB$) and two middle ones ($A$, $B$). 
The system is viewed as comprising two subsystems, $\mathcal{A}$ (including $sA$ and $A$) and $\mathcal{B}$ ($sB$ and $B$).
The system Hamiltonian contains the three parts: $H = H_\text{arms} + H_\text{diluter} + H_\text{T}$.
The arms, carrying quasiparticles of charge $\nu e$, can be described by the bosonized edge Hamiltonian $H_\text{arms} = v \sum_\alpha  \int dx [\partial_x \phi_\alpha (x)]^2/4\pi $, with $\phi_\alpha$ the bosonic field labeled by $\alpha = sA, sB, A, B$, see Fig.~\ref{fig:model}\textbf{d}.
The dynamical bosonic phase obeys the standard commutation relation $[\partial_x \phi_\alpha (x), \phi_\beta (x') ] = i \pi \delta_{\alpha\beta} \delta (x-x')$.

Fractional charges tunnel from sources to middle arms through the FQH bulk at two QPCs. These two diluter QPCs are described by the Hamiltonians written in terms of the anyon field operators $\psi_\alpha$: $H_\text{diluter} = \zeta_A \psi^\dagger_A \psi_{sA} + \zeta_B \psi^\dagger_B \psi_{sB} + \text{H.c.}$.
Via bosonization, tunneling operators can be written through $\psi_\alpha  = \exp(i\sqrt{\nu}\phi_\alpha)/(2\pi a)$, with $a$ 
an ultraviolet cutoff (the smallest length scale in the problem). Strictly speaking, tunneling operators contain Klein factors that guarantee proper commutation relations for distant tunneling operators, which is important in systems that support circulating currents (e.g., in a Mach-Zehnder interferometer, see, e.g., Refs.~\cite{JonckheerePRB05,LawPRB06}).
The Klein factors are, however, irrelevant to the HOM setup, where currents cannot travel back and forth (cf.~ Ref.~\cite{GuyonPRB02}), so that we do not introduce them here.
The tunneling amplitudes $\zeta_A$ and $\zeta_B$
define the ``bare'' tunneling probabilities at the diluters, $\mathcal{T}^{(0)}_A=|\zeta_A|^2$ and $\mathcal{T}^{(0)}_B=|\zeta_B|^2$.
The experimentally accessible transmission probabilities of diluters $\mathcal{T}_A$ and $\mathcal{T}_B$ 
are proportional to the corresponding bare probabilities:  $\mathcal{T}_A \propto \mathcal{T}_A^{(0)}$ and $\mathcal{T}_B \propto \mathcal{T}_B^{(0)}$.
We assume strong dilution, $\mathcal{T}^{(0)}_A, \mathcal{T}^{(0)}_B \ll 1$. 
In this work, the same voltage bias $V$ is assumed in both sources, and the single-source scenario is realized by pinching off either diluter.

The middle arms $A$ and $B$ communicate at the central QPC characterized by the bare transmission probability $\mathcal{T}_C^{(0)}$,
which is related to the experimentally accessible transmission probabilities following $\mathcal{T}_C^{(0)} \propto \mathcal{T}_C/\sqrt{\mathcal{T}_A \mathcal{T}_B}$.
The central QPC is placed at a distance $L$ from two diluters, in the downstream transport direction [Fig.~\ref{fig:model}\textbf{a}].
At variance with the two diluters, where the two depletion gates (the black area in Fig.~\ref{fig:model}\textbf{b}) are well separated in space, the central QPC is in the opposite limit, where the gates are almost ``touching'' each other  
(Fig.~\ref{fig:model}\textbf{c}). Following self-duality of tunneling through FQH QPCs (see, e.g., Refs.~\cite{FendleyLudwigSaleurPRL95, ShopenGefenMeirPRL05, WeissBook12}), only fermionic tunneling is allowed in this limit.
Physically, there is no bulk states with filling factor $\nu$ between the two arms (red and blue in Fig.~\ref{fig:model}), and, hence, between the subsystems $\mathcal{A}$ and $\mathcal{B}$, inside the central QPC.
The tunneling at the central QPC is therefore described by the Hamiltonian $H_\text{T} = \zeta_C \Psi^\dagger_A \Psi_B + \text{H.c.}$, with $\zeta_C\propto \left[\mathcal{T}_C^{(0)}\right]^{1/2}$ and $\Psi_\alpha = 
\exp(i\phi_\alpha/\sqrt{\nu})/\sqrt{2\pi a} $ is the fermionic field operator. This bosonized expression contains $\sqrt{1/\nu}$ instead of $\sqrt{\nu}$ encountered above, which is a hallmark of electron tunneling in anyonic systems.

The building blocks of the entanglement pointer are current correlators. In the Andreev-tunneling limit at the collider, $\mathcal{T}_C^{(0)} \ll 1$, the noise of the current operator $\hat{I}_\text{T} = i \zeta_C \Psi_B^\dagger \Psi_A + \text{H.c.}$ is given by
\begin{equation}
    S_\text{T} = e^2 \mathcal{T}^{(0)}_C \int dt \Big\langle \left\{ \Psi_B^\dagger (0) \Psi_A (0), \Psi_A^\dagger (t) \Psi_B (t) \right\} \Big\rangle_{\mathcal{T}^{(0)}_C = 0},
    \label{eq:st}
\end{equation}
with $\left\{\ ,\ \right\}$ denoting an anticommutator.
Evaluation of $S_\text{T}$, which yields  Eqs.~\eqref{eq:S1AB} and \eqref{eq:andreev noises}, involves correlators $\langle \Psi_A^\dagger(t) \Psi_A (0) \rangle $ and $\langle \Psi_B^\dagger(t) \Psi_B (0) \rangle $ at the position of the central QPC (see Sec.~I of the SI).
These correlation functions are greatly influenced by statistics of the quasiparticles involved, thus generating dependence on statistics in the observables---tunneling current and noise.

The evaluation of Eq.~\eqref{eq:sab_with_it_conductance} requires explicit expression for the tunneling current, 
\begin{equation}
    I_\text{T} = e \mathcal{T}^{(0)}_C \int dt \Big\langle \left[ \Psi_B^\dagger (0) \Psi_A (0), \Psi_A^\dagger (t) \Psi_B (t) \right] \Big\rangle_{\mathcal{T}^{(0)}_C = 0},
\end{equation}
with $[\ ,\ ]$ denoting a commutator. It is obtained using the correlation functions similar to those in Eq.~\eqref{eq:st} (see Sec.~IIIB of the SI), yielding $I_\text{T}=I_\text{T}^\text{single}+I_\text{T}^\text{collision}$, where 
\begin{align}
    I_\text{T}^\text{single} \! = &\text{Re}\left\{\mathcal{T}_C\, \mathcal{T}_A \,\frac{\nu e^2 V}{2}\frac{ (2\pi \nu)^{1-\nu_\text{s}}\, e^{i\pi(\nu_\text{s}-\nu \nu_s+1)} }{ \pi\nu\sin (\pi\nu_\text{s}) + 2 f_1(\nu) \mathcal{T}_A}     
 \left[ 2i\pi \nu -  \mathcal{T}_A\left( 1 - e^{-2i\pi\nu} \right)  \right]^{\nu_\text{s}-1} \right\}
    - \left\{A\to B\right\},
    \label{eq:explicit_current_expression-s}
    \\
    I_\text{T}^\text{collision}&\! = \text{Im} \left\{\mathcal{T}_C\, e^2 V \frac{ \sqrt{\mathcal{T}_A \mathcal{T}_B}\,  f_2(\nu) \sin \left(\pi \nu_\text{d}/{2} \right)}{\pi\nu\sin\left( \pi \nu_\text{s} \right) + 2 f_1 (\nu) \sqrt{\mathcal{T}_A \mathcal{T}_B}  } 
     \left[
    \mathcal{T}_A\left( 1 - e^{-2i\pi\nu} \right) + \mathcal{T}_B \left( 1 - e^{2i\pi\nu} \right)
    \right]^{\nu_\text{d} - 1} \right\},
\label{eq:explicit_current_expression}
\end{align}
with functions $f_{1,2}(\nu)$ defined in Eqs.~\eqref{eq:S1AB} and \eqref{eq:andreev noises}.
Compared to those expressions for the contributions to the tunneling noise $S_\text{T}$, the main difference of the tunneling currents from the corresponding noises is (apart from a trivial overall prefactor $1/e$) in their parity: ``$- \left\{A\to B\right\}$'' replaces ``$+\left\{A\to B\right\}$'' in the single-source term \eqref{eq:explicit_current_expression-s} and ``$\text{Im}$'' replaces ``$\text{Re}$'' in the collision term \eqref{eq:explicit_current_expression}.

\textit{Note added:} While preparing the first version of our manuscript, we noticed Ref.~\cite{JonckheerePRB23}, which concerned a single-source platform
and did not address the effects of collisions and anyon-quasihole braiding.

\begin{figure}
\centering
  \includegraphics[width=0.98\linewidth]{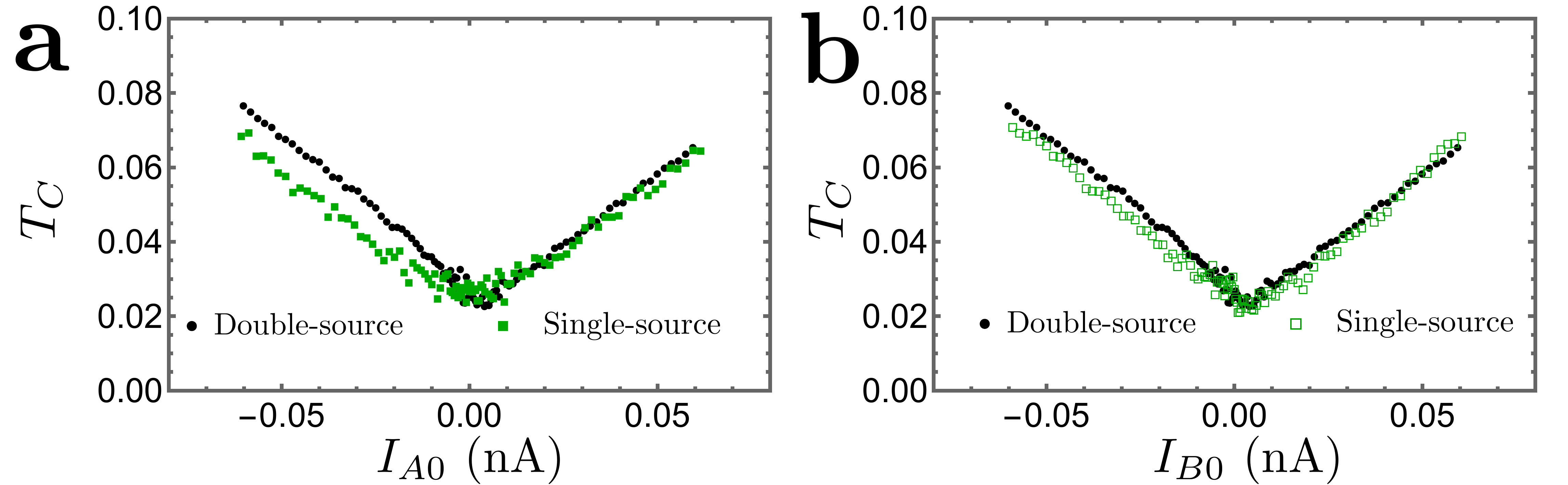}
  \caption{
  Extracted $\mathcal{T}_C$ in the double-source and single-source situations. This figure shows difference between the measured $\mathcal{T}_C$ for the single-source and double-source settings, which likely results from the effect of bias voltages on the overall electrostatic landscape of the setup.
  \textbf{Panel a}: Values of $\mathcal{T}_C$, for the double-source (black dots) and the single-source cases with $I_{A0}$ finite (solid green squares).
  \textbf{Panel b}: The same double-source transmission $\mathcal{T}_C$ (black dots) is compared to that of the single-source case with $I_{B0}$ finite (empty green squares).
  Corresponding values of $\mathcal{T}_A$ and $\mathcal{T}_B$ (arranging approximately between $0.02$ and $0.08$) are provided by the supplementary figure, Fig.~S7.
  }
  \label{fig:tc}
\end{figure}

\subsection{Experiment}

The measurements are realized at $T\approx35$\,mK on a 2DEG set to $\nu=1/3$.
The device includes two nominally identical source QPCs positioned symmetrically with respect to a central QPC (see the SI and Ref.~\cite{PierreNC23}).
Gate voltages allow us to tune the QPCs in the configuration where the Andreev tunneling of quasiparticles takes place. 
The source QPCs are set in the 
anyonic-tunneling regime (Fig.~\ref{fig:model}\textbf{b})
and exhibit a shot-noise Fano factor corresponding to a fractional charge $e^*\approx e/3$, whereas the central QPC is tuned in the
Andreev-like tunneling regime (Fig.~\ref{fig:model}\textbf{c})
with the tunneling charge $e^*\approx e$,
as deduced from shot noise~\cite{PierreNC23}.
An experimental challenge is to be able to obtain reliably the entanglement pointer. 
Indeed, $\mathcal{P}_\text{Andreev}$ is a small difference between larger quantities measured separately, which increases the sensitivity to experimental artifacts such as drifts in time between compared configurations or unwanted small capacitive cross-talks.
The difficulty is further enhanced by the difference between the measured $\mathcal{T}_C$ for the single-source and double-source settings (see Fig.~\ref{fig:tc}), which results, apparently, from the different electrostatic landscapes.
As further detailed in Sec.~VIII of the SI, the data set used to extract the entanglement pointer was obtained following a specific protocol reducing such artifacts.
In particular, there are no changes in the device gates voltages, and the time between compared configurations is minimized. Further details on the experiment can be found in Sec.~VII of the SI.

\backmatter

\section*{Supplementary Information}

In the Supplementary Information, we provide extra material on (i) derivation of correlation functions, tunneling current, and tunneling noise in the Andreev-like tunneling limit;
(ii) comparison between correlation functions of two opposite limiting cases;
(iii) evaluation of relevant integrals for tunneling current and its corresponding noise;
(iv) finite-temperature expressions for observables; (v) influence of interaction on correlation functions and noises; (vi) derivation on single-particle and two-particle expressions for different types of noise; (vii) experimental details, (viii) theoretical analysis of experimental data, and (ix) prospective applications to non-Abelian states.

\section*{Acknowledgments}

We are grateful to Gabriele Campagnano, Domenico Giuliano, Moty Heiblum, Thierry Martin, Bernd Rosenow, In{\`e}s Safi, and Kyrylo Snizhko for fruitful discussions.
We thank O. Maillet, C. Piquard, A. Aassime, and A. Anthore for their contribution to the experiment.
IG and YG acknowledge the support from the DFG grant No. MI$658/10$-$2$ and German-Israeli Foundation (GIF) grant No. I-1505-303.10/2019. 
YG acknowledges support from the Helmholtz International Fellow Award, by the DFG
Grant RO $2247/11$-$1$, by CRC 183 (project C01), the US-Israel Binational Science Foundation, and by the Minerva Foundation. FP acknowledges the support of the European Research Council (ERC-2020-SyG-951451) and of the French RENATECH network.

\section*{Declarations}

\begin{itemize}

 \item Availability of data and materials: Raw data of this work can be accessed via Zenodo:\\ https://doi.org/10.5281/zenodo.10434474.
 \item Code availability: Relevant Mathematica notebook can be accessed via Zenodo:\\ https://doi.org/10.5281/zenodo.10434474.
 \item Authors' contributions: GZ, IG, and YG carried out the theoretical analysis.
 PG obtained the experimental data and performed the low-level data analysis under the supervision of FP. GZ designed and performed the data-theory comparison, with critical inputs from FP. All authors participated in the scientific discussions, contributed to the preparation of this work and to the writing of the paper, and proofread the manuscript.
\end{itemize}



\newpage

\clearpage

\appendix

\renewcommand{\bibnumfmt}[1]{[S#1]}
\renewcommand{\citenumfont}[1]{S#1}
\global\long\def\theequation{S\arabic{equation}}
\global\long\def\thefigure{S\arabic{figure}}
\renewcommand{\thesection}{\Roman{section}}
\renewcommand{\thesubsection}{\Alph{subsection}}
\setcounter{equation}{0}
\setcounter{figure}{0}

\begin{center}
\textbf{\Large Supplementary Information}\\
 \vspace{15pt}
\end{center}
\vspace{10pt}

Here, we provide additional information on (i) the derivation of correlation functions, tunneling current, and tunneling noise in the Andreev-like tunneling limit;
(ii) comparison between correlation functions of two opposite limiting cases;
(iii) the evaluation of relevant integrals for tunneling current and its corresponding noise;
(iv) the finite-temperature expressions for observables; (v) the influence of interaction on correlation functions and noises; (vi) the derivation on single-particle and two-particle expressions for different types of noise; (vii) experimental details, (viii) theoretical analysis of experimental data, and (ix) prospective applications to non-Abelian states.

For simplicity, we set $v = \hbar = k_B  =1$ throughout the derivations (whenever it does not create confusion).

\section{{\,\,\,\,\,\,\,}Time-dependent correlation functions at zero temperature}
\label{sec:correlation_derivations}

In this section, we derive correlation functions $\langle \Psi_A^\dagger(L,t) \Psi_A (L,0) \rangle $ and $\langle \Psi_B^\dagger(L,t) \Psi_B (L,0) \rangle$.
To leading order in tunneling at the central QPC $\mathcal{T}^{(0)}_C$, these two correlation functions are needed [following Eqs.~(8) and (9) of the main text] to obtain both the tunneling current [Eqs.~(10) and (11) of the main text], and the current noise [Eqs.~(2) and (3) of the main text].

\subsection{Leading-order correlations}
\label{sec:da1}

We begin with expansions of the correlation functions to leading order in dilution $\mathcal{T}^{(0)}_{A,B}$ at the corresponding diluter.
For concreteness, we focus on the correlation function of operators in edge $A$, i.e., $\langle \Psi_A^\dagger(L,t) \Psi_A (L,0) \rangle $, where $\Psi_\alpha$ are the \textit{fermionic} field operators (see Methods) at the central collider ($x=L$).
The first-order term in expansion of the correlator in  $\mathcal{T}^{(0)}_{A}$ is represented as a double time integral, where the integrand contains a product of two expectation values: one in channel $A$, combining fermion operators $\Psi_A$ with anyon operators  $\psi_A$ at the diluter ($x=0$), and the other in the source channel $sA$ where only anyon operators at the diluter are involved:
\begin{equation}
\begin{aligned}
D_{A1} \! = \! & \!-\!  \mathcal{T}^{(0)}_A \!\sum_{\eta_1\eta_2}\! \eta_1\eta_2\! \iint \!ds_1 ds_2 e^{-i\nu e V(s_1 - s_2)} \big\langle \Psi^\dagger_A (L,t^-) \Psi_A (L,0^+) \psi^\dagger_A (0, s_1^{\eta_1}) \psi_A (0, s_2^{\eta_2}) \big\rangle \big\langle \psi_{sA} (0, s_1^{\eta_1}) \psi_{sA}^\dagger (0, s_2^{\eta_2}) \big\rangle \\
& = -\frac{\mathcal{T}^{(0)}_A}{(2\pi \tau_0)^3} \sum_{\eta_1\eta_2} \eta_1\eta_2 \iint ds_1 ds_2 \,  e^{-i\nu e V(s_1 - s_2)} \,\frac{\tau_0^{{1}/{\nu} + 2\nu}\,}{(\tau_0 + it)^{1/\nu} [\tau_0 + i (s_1 - s_2) \chi_{\eta_1\eta_2} (s_1 - s_2)]^{2\nu}}  \\
&\qquad \qquad \qquad \qquad   \qquad\times \frac{[\tau_0 + i (t - s_1 - L) \chi_{-\eta_1} (t-s_1)] [\tau_0 + i (-s_2 - L) \chi_{+\eta_2} (-s_2)]}{[\tau_0 + i (t - s_2 - L) \chi_{-\eta_2} (t-s_2)] [\tau_0 + i (-s_1 - L) \chi_{+\eta_1} (-s_1)]}.
\end{aligned}
\label{eq:leading1}
\end{equation}
Here, $\tau_0$ is the ultraviolet time cutoff (related to the length cutoff $a$ as $\tau_0=a/v$), ``A1'' indicates the expansion to the leading order in $\mathcal{T}^{(0)}_A$, $s_1$ and $s_2$ are the times when anyons tunnel from $sA$ to $A$, and  $\eta_1$ and $\eta_2$ the corresponding Keldysh indexes.
For brevity, we address these anyons, supplied from the sources through the diluters to the main arms, as ``non-equilibrium anyons,'' underscoring the non-equilibrium nature of dilute beams in channels $A$ and $B$.
The function $\chi_{\eta\eta'}(t-t')$ reflects the relative positions of $t^\eta$ and $t^{\eta'}$: it equals 1 if $t^\eta$ is in front of $(t')^{\eta'}$ along the Keldysh contour, equals $-1$ for the opposite situation, and equals zero if $t = t'$ and $\eta = \eta'$.
The voltage bias is included in the phase factor $\exp[-i\nu e V (s_1 - s_2)]$ by the standard transformation, as employed in, e.g., Ref.~\cite{KaneFisherPRB92S}.
With the identity
\begin{equation}
\begin{aligned}
& \frac{1}{(i\tau_0 - t) [i\tau_0 \chi_{\eta_1\eta_2} (s_1 - s_2) - (s_1 - s_2)]}\frac{[i\tau_0 \chi_{-\eta_1} (t-s_1) - (t-s_1-L)] [i\tau_0 \chi_{+\eta_2} (-s_2) - (-s_2 - L)]}{[i\tau_0 \chi_{-\eta_2} (t-s_2) - (t-s_2-L)] [i\tau_0 \chi_{+\eta_1} (-s_1) - (-s_1 - L)/v]}\\
&=  \frac{1}{(i\tau_0 - t) [i \tau_0 \chi_{\eta_1\eta_2} (s_1 - s_2) - (s_1 - s_2)]} +\frac{1}{ [i\tau_0 \chi_{-\eta_2} (t-s_2) - (t-s_2-L)] [i\tau_0 \chi_{+\eta_1} (-s_1) - (-s_1 - L)]},
\end{aligned}
\label{eq:identify}
\end{equation}
Eq.~\eqref{eq:leading1} can be simplified into
\begin{equation}
\begin{aligned}
&D_{A1}=\!- \!\frac{\mathcal{T}^{(0)}_A}{(2\pi \tau_0)^3} 
\frac{(i \tau_0)^{1/\nu}}{(i\tau_0 \! -\! t)^{1/\nu}}
\sum_{\eta_1\eta_2} \eta_1\eta_2 \iint ds_1 ds_2\, e^{-i\nu e V(s_1 \!-\! s_2)} \, \frac{[i \tau_0\chi_{\eta_1\eta_2} (s_1 \!-\! s_2)]^{2\nu} }{ [i\tau_0  \chi_{\eta_1\eta_2} (s_1 \!-\! s_2) - (s_1 \!-\! s_2)]^{2\nu}} \\
& \qquad \qquad \qquad \qquad  \times \frac{\chi_{-\eta_2} (t - s_2) \chi_{+\eta_1} (-s_1)}{\chi_{-\eta_1} (t - s_1) \chi_{+\eta_2} (-s_2)}\, \frac{[i\tau_0 \chi_{-\eta_1} (t-s_1) - (t - s_1 - L) ] [i \tau_0 \chi_{+\eta_2} (-s_2) - (-s_2 - L) ]}{[i \tau_0 \chi_{-\eta_2} (t-s_2)-  (t - s_2 - L) ] [i \tau_0 \chi_{+\eta_1} (-s_1) - (-s_1 - L) ]} \\
 & = - \frac{\mathcal{T}^{(0)}_A}{(2\pi \tau_0)^3} \sum_{\eta_1\eta_2} \eta_1\eta_2 \iint ds_1 ds_2 \frac{e^{-i\nu e V(s_1 - s_2)} (i \tau_0)^{1/\nu}[i \tau_0\chi_{\eta_1\eta_2} (s_1 \!-\! s_2)]^{2\nu} }{(i\tau_0  - t)^{1/\nu - 1} [i\tau_0  \chi_{\eta_1\eta_2} (s_1 - s_2) - (s_1 - s_2)]^{2\nu - 1}}\, 
 \frac{\chi_{-\eta_2} (t - s_2) \chi_{+\eta_1} (-s_1)}{\chi_{-\eta_1} (t - s_1) \chi_{+\eta_2} (-s_2)}
\\
& \times \left\{ \frac{1}{(i\tau_0 - t) [i\tau_0 \chi_{\eta_1\eta_2} (s_1 - s_2) - (s_1 - s_2)]} + \frac{1}{ [i\tau_0 \chi_{-\eta_2} (t-s_2) - (t-s_2-L)] [i\tau_0 \chi_{+\eta_1} (-s_1) - (-s_1 - L)]} \right\},
\end{aligned}
\label{eq:leading2}
\end{equation}
where
the first and second terms within the curly brackets correspond to two sets of singularities that are related to the integration variables, $s_1$ and $s_2$: (i) $s_1 \to s_2$, and (ii) $s_2 \to t-L$, $s_1\to -L$.
In the limit of Andreev-like tunneling, the first pair of singularities leads to a vanishing contribution.
Indeed, after taking the first term within the curly brackets, Eq.~\eqref{eq:leading2} contains a factor,
\begin{equation}
    \sum_{\eta_1\eta_2} \eta_1\eta_2 \frac{e^{-i\nu e V (s_1 - s_2)}}{[\tau_0  +i \chi_{\eta_1\eta_2}(s_1 - s_2) (s_1 - s_2)]^{2\nu}} = 0,
    \label{eq:vanishing_summation}
\end{equation}
that vanishes after summing over Keldysh indexes $\eta_1$ and $\eta_2$.
We stress that this is in great contrast to the situation of the opposite tunneling limit, i.e., the limit of anyonic tunneling through the central collider  (see, e.g., Refs.~\cite{RosenowLevkivskyiHalperinPRL16S, SimNC16S, LeeSimNC22S, LeeNature23S, schillerPRL23S, LandscapePRL25S}). In this limit, the correlation functions are similar to Eq.~(\ref{eq:leading1}) but comprise only the anyonic $\psi$ operators, so that the $s_1\to s_2$ singularity is then  of major importance, as the time-domain braiding process correlates tunneling events at different time moments.
In the Andreev-like tunneling limit, this time-domain braiding process is absent, as fermions do not produce a nontrivial braiding phase when braiding with non-equilibrium anyons.
As will be shown shortly in Secs.~\ref{sec:higher_order_dilution} and \ref{sec:proceeses_(ii)_(iii)}, in the Andreev-like tunneling limit, another type of braiding (i.e., the anyon-quasihole braiding introduced in the main text) will occur.
This anyon-quasihole braiding process is absent in $D_{A1}$, which is linear in $\mathcal{T}_A^{(0)}$, but appears in the terms of higher order in the diluter transparency. We compare time-domain braiding (relevant to anyonic tunneling) and anyon-quasihole braiding (relevant to Andreev-like tunneling) in Sec.~\ref{sec:andreev_comparison}.

Following the above discussion, we focus on the second term within the curly brackets of Eq.~\eqref{eq:leading2}, which contains poles at $s_1 \to -L$ and $s_2 \to t- L$.
By taking these two poles, the time arguments of the fermionic tunneling operators coincide with the time moment when a non-equilibrium anyon has arrived at the central collider.
Physically, it indicates that tunneling at the central collider is Andreev-like, as it is triggered by the non-equilibrium anyon.
Mathematically, after choosing this pair of singularities, the rest of the integrand has no other singularities. Indeed, since $\nu < 1/2$ in this work, the factor $[i\tau_0  \chi_{\eta_1\eta_2} (s_1 - s_2) - (s_1 - s_2)]^{1-2\nu}$ becomes non-singular.

Now, we are ready to evaluate the integral Eq.~\eqref{eq:leading2}.
The integrals over $s_1$ and $s_2$ can be straightforwardly obtained via the residue theorem, leading to:
\begin{equation}
    \begin{aligned}
        \int d(t - s_2) \frac{e^{-i\nu eV (t - s_2)}}{i \tau_0 \eta_2 - (t-s_2) } \int ds_1 \frac{e^{-i\nu e V s_1}}{ i\tau_0 \eta_1 + s_1} = (\eta_2 - 1) (\eta_1 + 1) \pi^2,
    \end{aligned}
\end{equation}
which indicates that only the option $\eta_1 =1 $ and $\eta_2 = -1$ yields a finite result.
This choice of Keldysh indexes is related to the fact that $V > 0$ preselects the allowed contour (i.e., upper or lower half of the complex plane).
With this integral, we obtain
\begin{equation}
D_{A1} = \mathcal{T}^{(0)}_A e^{i\nu e V t} \frac{ 1 }{2\pi } \frac{\tau_0^{1/{\nu}+2\nu-3}}{(\tau_0 + i t)^{2 \nu + 1/\nu - 2}}.
\label{eq:da1_expression}
\end{equation}
We combine $D_{A1}$ with the equilibrium contribution given by \begin{equation}
D_{A0} = \frac{1}{2\pi} \frac{\tau_0^{{1}/{\nu}-1}}{ (\tau_0 + it)^{1/\nu}},
\end{equation}
and use the leading-order expression for the corresponding non-equilibrium current 
\begin{equation}
I_{A0} = \mathcal{T}^{(0)}_A \nu e\tau_0^{2\nu-2} \sin(2\pi\nu) \Gamma (1-2\nu) (\nu e V)^{2\nu-1}/2\pi^2,
\label{eq:ia0_expression}
\end{equation}
where $\Gamma(x)$ is the gamma function, to arrive at 
\begin{equation}
\begin{aligned}
D_{A0} + D_{A1} & = \frac{\tau_0^{\frac{1}{\nu}-1}}{2\pi (\tau_0 + it)^{1/\nu}} \left[ 1 +  c(\nu) \frac{I_{A0}}{(\nu e V)^{2\nu - 1}} ( it)^{2-2\nu} e^{i\nu e V t} \right],\\
\text{with}&\ \ c (\nu) = \frac{2\pi^2}{\sin (2\pi\nu) \Gamma (1-2\nu) \nu}.
\end{aligned}
\label{eq:da0_da1}
\end{equation}

It is instructive to compare the $\nu\to 1$ limit of Eq.~\eqref{eq:da0_da1}, where $\lim_{\nu\to 1} c(\nu) = 2\pi$, with that of a non-interacting fermionic system: the latter has the correlation function
\begin{equation}
\langle{\Psi}^\dagger_A(t) {\Psi}_A (0) \rangle_\text{fermion} = \frac{1}{2\pi (\tau_0 + it)} \left( 1 + e^{i e V t} \mathcal{T}^{(0)}_A - \mathcal{T}^{(0)}_A \right).
\label{eq.S10}
\end{equation}
After taking $\nu \to 1$, and $I_{A0}/e^2 V = 2 \pi \mathcal{T}^{(0)}_A$, we notice that Eq.~\eqref{eq:da0_da1} perfectly captures the first two terms of the non-interacting fermionic result, missing,  however, the last term.
This missing term corresponds to the simple pole $s_1\to s_2$, which emerges at $\nu=1$ from the factor $[i\tau_0  \chi_{\eta_1\eta_2} (s_1 - s_2) - (s_1 - s_2)]^{1-2\nu}$, obtained after choosing the singularities in the second term in the curly brackets of Eq.~\eqref{eq:leading2}.
This term is absent in Eq.~\eqref{eq:da0_da1}, as the (to be integrated) function becomes regular when $\nu < 1/2$ (i.e., for Laughlin quasiparticles). We also note that Eq.~(\ref{eq:da0_da1}) is well-defined and nontrivial also for $\nu=1/2$, as the singularity in $\Gamma(1-2\nu)$ is canceled at $\nu=1/2$ by $\sin(2\pi \nu)$.

\subsection{Contributions to the correlation function from next-to-leading-order tunneling through diluters}
\label{sec:higher_order_dilution}

Now, we proceed to analyze processes of higher order in dilution, where the Andreev-like tunneling is influenced by multiple pairs of non-equilibrium anyons.
As discussed in the main text, these non-equilibrium anyonic pairs can be separated into two categories: the pair that triggers Andreev-like tunneling and
the rest non-equilibrium pair(s) of anyons that do not tunnel at the central QPC. Most interestingly, the latter anyons are responsible for nontrivial ``anyon-quasihole braiding'', by braiding with the fractional-charge hole generated by Andreev-like tunneling.
In the remaining part of Sec.~\ref{sec:correlation_derivations}, we provide detailed derivations of these higher-order processes.
We begin with the fourth-order expansion in the Hamiltonian at the upper diluter, which is of the second order in $\mathcal{T}^{(0)}_A$:
\begin{equation}
\begin{aligned}
D_{A2} \! \equiv  &  \frac{\big(4\mathcal{T}^{(0)}_A\big)^2 }{4!}\!\!\! \sum_{\eta_1\eta_2\eta_3\eta_4} \!\!\!\eta_1\eta_2 \eta_3\eta_4 \!\int ds_1 ds_2 ds_3 ds_4\, e^{-i\nu e V(s_1 - s_2 + s_3 - s_4)} \big\langle \psi_{sA} (0,s_1^{\eta_1}) \psi^\dagger_{sA} (0,s_2^{\eta_2}) \psi_{sA} (0,s_3^{\eta_3}) \psi^\dagger_{sA} (0,s_4^{\eta_4}) \big\rangle \\
&\qquad \qquad \qquad \qquad \qquad \qquad \times  \big\langle \Psi^\dagger_A (L,t^-) \Psi_A (L,0^+) \psi^\dagger_A (0, s_1^{\eta_1}) \psi_A (0, s_2^{\eta_2})  \psi_A^\dagger (0, s_3^{\eta_3}) \psi_A (0, s_4^{\eta_4}) \big\rangle\\
= & \frac{\big(\mathcal{T}^{(0)}_A\big)^2}{48 \pi^5 \tau_0^5}\! \sum_{\eta_1\eta_2\eta_3\eta_4}\! \eta_1\eta_2 \eta_3\eta_4 \int\! ds_1 ds_2 ds_3 ds_4 e^{-i\nu e V(s_1 - s_2 + s_3 - s_4)}
\\
 & \qquad  \qquad \qquad \times
\big\langle e^{i\sqrt{\nu} \phi_{sA}(0,s_1^{\eta_1})} e^{-i\sqrt{\nu} \phi_{sA}(0,s_2^{\eta_2})} e^{i\sqrt{\nu} \phi_{sA}(0,s_3^{\eta_3})} e^{-i\sqrt{\nu} \phi_{sA}(0,s_4^{\eta_4})} \big\rangle 
\\
& \qquad  \qquad \qquad
\times  \big\langle e^{-\frac{i}{\sqrt{\nu}} \phi_A (L,t^-) } e^{\frac{i}{\sqrt{\nu}} \phi_A (L,0^+) } e^{-i\sqrt{\nu} \phi_{A}(0,s_1^{\eta_1})} e^{i\sqrt{\nu} \phi_{A}(0,s_2^{\eta_2})} e^{-i\sqrt{\nu} \phi_{A}(0,s_3^{\eta_3})} e^{i\sqrt{\nu} \phi_{sA}(0,s_4^{\eta_4})} \big\rangle.
\end{aligned}
\label{eq:da2}
\end{equation}
This expression contains vertex operators of non-equilibrium anyons in channel $A$ at the position of the diluter ($x=0$), with four time arguments: $s_1,s_2,s_3$ and $s_4$. Again, in Eq.~\eqref{eq:da2}, the bias $V$ is incorporated through the phase factor, $e^{-i\nu e V (s_1 - s_2 + s_3 - s_4)}$. The vertex correlation functions is then evaluated with equilibrium bosonic correlators, leading to
\begin{equation}
\begin{aligned}
    D_{A2}=  \frac{\big(\mathcal{T}^{(0)}_A\big)^2}{48 \pi^5 \tau_0^5} \frac{\tau_0^{4\nu + 1/\nu}}{(\tau_0 + i t)^{1/\nu}}&  \sum_{\eta_1\eta_2\eta_3\eta_4}\! \eta_1\eta_2 \eta_3\eta_4 \int\! ds_1 ds_2 ds_3 ds_4\, e^{-i\nu e V(s_1 - s_2 + s_3 - s_4)} \\
    \times & \frac{1}{[\tau_0 + i (s_1 - s_2)\chi_{\eta_1\eta_2} (s_1 - s_2)]^{2\nu}} \frac{1}{[\tau_0 + i (s_3 - s_4)\chi_{\eta_3\eta_4} (s_3 - s_4)]^{2\nu}}\\
    \times & \frac{[\tau_0 + i (s_1 - s_3)\chi_{\eta_1\eta_3} (s_1 - s_3)]^{2\nu} [\tau_0 + i (s_2 - s_4)\chi_{\eta_2\eta_4} (s_2 - s_4)]^{2\nu}}{[\tau_0 + i (s_1 - s_4)\chi_{\eta_1\eta_4} (s_1 - s_4)]^{2\nu} [\tau_0 + i (s_2 - s_3)\chi_{\eta_2\eta_3} (s_2 - s_3)]^{2\nu}}\\
    \times & \frac{[\tau_0 + i (t-L - s_1)\chi_{-\eta_1} (t-s_1)][\tau_0 + i (-L-s_2) \chi_{+\eta_2} (-s_2)]}{[\tau_0 + i (t-L - s_2)\chi_{-\eta_2} (t-s_2)][\tau_0 + i (-L-s_1) \chi_{+\eta_1} (-s_1)]}
    \\
    \times & 
    \frac{[\tau_0 + i (t-L - s_3)\chi_{-\eta_3} (t-s_3)][\tau_0 + i (-L-s_4) \chi_{+\eta_4} (-s_4)]}{[\tau_0 + i (t-L - s_4)\chi_{-\eta_4} (t-s_4)][\tau_0 + i (-L-s_3) \chi_{+\eta_3} (-s_3)]},
\end{aligned}
\label{eq:da2_initial}
\end{equation}
where the second and third lines (characterized by fractional powers $2\nu$) describe correlations among operators of non-equilibrium anyons. The last two lines, on the other hand, represent correlations between non-equilibrium anyons and fermions that tunnel at the central collider.

The integrand of Eq.~\eqref{eq:da2_initial} contains several singularities (zeros in the denominator) that lead to dominant contributions to the integral.
Different choices of these singularities produce then distinct integral outcomes.
For later convenience, we introduce the terminology ``\textit{contract}'' (not to be confused with contractions in  Wick's theorem for fermions and bosons, although the term introduced for anyons is intentionally similar to ``contraction'')
to describe the chosen singularities.
For instance, when we ``contract'' non-equilibrium operators at time moments $s_1$ and $s_2$, we refer to focusing on the origin of the branch cut $s_1 = s_2$.
More specifically, to single out the contribution of this branching point, the condition $|s_1 - s_2| \ll \lambda/v = 1/\nu e V$ is required, with $\lambda$ describing the width of a typical non-equilibrium anyonic pulse (see Fig.~1\textbf{d} of the main text).
Physically, when two operators are chosen to ``contract'' with each other, the corresponding wave functions overlap significantly and are thus correlated.
The correlator given by Eq.~\eqref{eq:da2_initial}  involves the three types of contract options: 
\begin{equation}
\text{(i)}\ s_2= s_1\ \text{and} \ s_4 = s_3;
\quad
\text{(ii)}\ s_2, s_4 = t-L\ \text{and}\ s_1, s_3 = -L; 
\quad
\text{(iii)}\ s_1 = -L,\, s_2= t-L,\ \text{and}\ s_3 = s_4.
\label{options}
\end{equation}
All possible permutations within these sets should also be included: for instance, in option (i), one can alternatively take $s_4\to s_1$ and $s_3 \to s_2$. The contributions of these permutations to the correlation function are included by a proper constant prefactor.

Let us now analyze these contract options.
Firstly, contract option (i) yields a vanishing contribution to $D_{A2}$, similarly to the vanishing of the leading-order integral, Eq.~\eqref{eq:leading2}, when taking $s_2 \to s_1$.
Indeed, when we take $s_2\to s_1$ and $s_4 \to s_3$ in Eq.~\eqref{eq:da2_initial}, the product of the factors in its last three lines become simply unity. As a consequence, following a similar identity of Eq.~\eqref{eq:vanishing_summation}, this contract option yields zero upon summation over Keldysh indices.
Physically, this vanishing originates from the lack of time-domain braiding in a system with Andreev-like tunneling. In the next subsection, we will analyze the contributions to $D_{A2}$ from contract options (ii) and (iii).

\subsection{Non-vanishing next-to-leading-order contributions}
\label{sec:proceeses_(ii)_(iii)}

\begin{SCfigure}
  \includegraphics[width=0.4\linewidth]{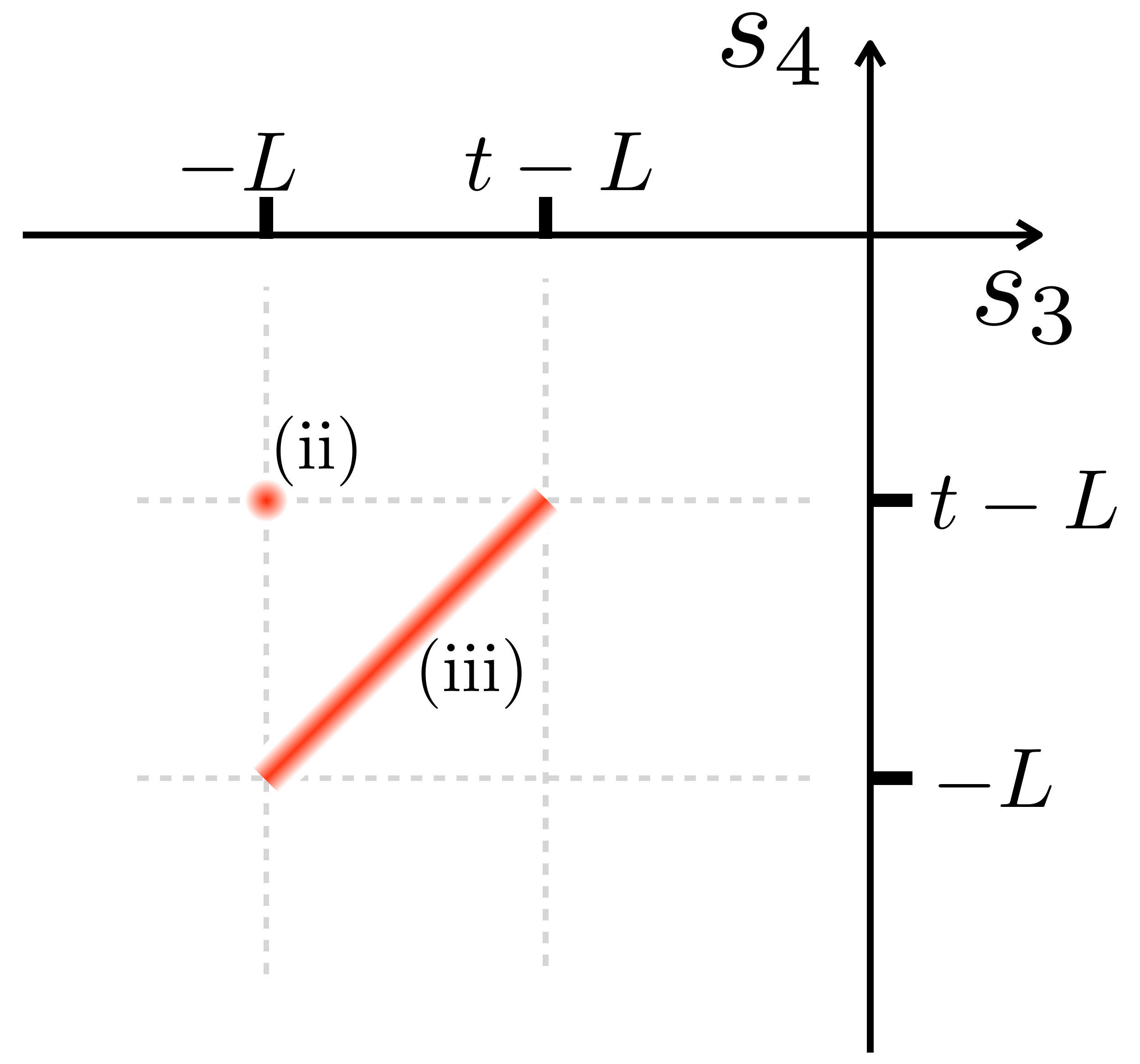}\hspace{0.5cm}
  \caption{Two types of contract options, (ii) and (iii), for the integral over $s_3$ and $s_4$ in Eq.~\eqref{eq:da2_ii_2}. The regions determining the integral are marked in red. Notice that one needs to consider permutations among all four time moments, $s_1, s_2, s_3, s_4$, appearing in Eq.~(\ref{eq:da2_initial}); the picture illustrates one of the possible choices of variables, where we have, without loss of generality, chosen operators with time arguments $s_1$ and $s_2$ as those that trigger Andreev-like tunneling. By doing so, $s_1 $ and $s_2$ are set to $-L$ and $t-L$, respectively.}
  \label{fig:da2_s3_s4}
\end{SCfigure}

As the poles $s_1 \to -L$ and $s_2 \to t-L$ are included in both contract options (ii) and (iii), we simplify Eq.~\eqref{eq:da2_initial} by first performing integrals over these two poles, following details provided in Sec.~\ref{sec:da1}.
After this integration, Eq.~\eqref{eq:da2_initial} simplifies into
\begin{equation}
\begin{aligned}
    D_{A2}^\text{(ii),(iii)}= & \left[\mathcal{T}^{(0)}_A\right]^2 e^{i\nu e V t} \frac{ 1 }{12\pi^3 } \frac{\tau_0^{1/{\nu}+4\nu-5}}{(\tau_0 + i t)^{2 \nu + 1/\nu - 2}} \sum_{\eta_3\eta_4} \eta_3\eta_4 \iint ds_3 ds_4 \frac{e^{-i\nu e V (s_3 - s_4)}}{[\tau_0 + i (s_3 - s_4)\chi_{\eta_3\eta_4} (s_3 - s_4)]^{2\nu}} \\
    \times &\, \frac{[\tau_0 + i (-L - s_3)\chi_{+\eta_3} (-L - s_3)]^{2\nu} [\tau_0 + i (t-L - s_4)\chi_{-\eta_4} (t-L - s_4)]^{2\nu}}{[\tau_0 + i (-L - s_4)\chi_{+\eta_4} (-L - s_4)]^{2\nu} [\tau_0 + i (t-L - s_3)\chi_{-\eta_3} (t-L - s_3)]^{2\nu}}\\
    \times & \,\frac{[\tau_0 + i (t-L - s_3)\chi_{-\eta_3} (t-s_3)][\tau_0 + i (-L-s_4) \chi_{+\eta_4} (-s_4)]}{[\tau_0 + i (t-L - s_4)\chi_{-\eta_4} (t-s_4)][\tau_0 + i (-L-s_3) \chi_{+\eta_3} (-s_3)]},
\end{aligned}
\label{eq:da2_ii}
\end{equation}
where the residue theorem replaced $s_1 $ and $s_2$ in the original integrand with $-L$ and $t-L$, respectively.
In Eq.~\eqref{eq:da2_ii}, the superscript ``(ii), (iii)'' indicates that both contract options (ii) and (iii) are included.
Using the identity [cf. Eq.~\eqref{eq:identify}]
\begin{equation}
\begin{aligned}
& \frac{1}{(i\tau_0 - t) [i\tau_0 \chi_{\eta_3\eta_4} (s_3 - s_4) - (s_3 - s_4)]}\frac{[i\tau_0 \chi_{-\eta_3} (t-s_3) - (t-s_3-L)] [i\tau_0 \chi_{+\eta_4} (-s_4) - (-s_4 - L)]}{[i\tau_0 \chi_{-\eta_4} (t-s_4) - (t-s_4-L)] [i\tau_0 \chi_{+\eta_3} (-s_3) - (-s_3 - L)]}\\
&=  \frac{1}{(i\tau_0 - t) [i\tau_0 \chi_{\eta_3\eta_4} (s_3 - s_4) - (s_3 - s_4)]} +\frac{1}{ [i\tau_0 \chi_{-\eta_4} (t-s_4) - (t-s_4-L)] [i\tau_0 \chi_{+\eta_3} (-s_3) - (-s_3 - L)]},
\end{aligned}
\label{eq:identify_2}
\end{equation}
we rewrite Eq.~\eqref{eq:da2_ii} as
\begin{equation}
\begin{aligned}
   &  D_{A2}^\text{(ii),(iii)}= \frac{\left[\mathcal{T}^{(0)}_A\right]^2}{12\pi^3}  \frac{\tau_0^{1/{\nu}+4\nu-5} e^{i\nu e V t}}{(\tau_0 + i t)^{2 \nu + 1/\nu - 3}} \sum_{\eta_3\eta_4} \eta_3\eta_4 \iint ds_3 ds_4  \frac{ 
 e^{-i\nu e V (s_3 - s_4)} \chi_{\eta_3\eta_4} (s_3-s_4) }{[\tau_0 + i (s_3 - s_4)\chi_{\eta_3\eta_4} (s_3 - s_4)]^{2\nu - 1}} \\
 & \times \frac{\chi_{-\eta_4} (t - s_4) \chi_{+\eta_3} (-s_3)}{\chi_{-\eta_3} (t - s_4) \chi_{+\eta_4} (-s_4)} \frac{[\tau_0 + i (-L - s_3)\chi_{+\eta_3} (-L - s_3)]^{2\nu} [\tau_0 + i (t-L - s_4)\chi_{-\eta_4} (t-L - s_4)]^{2\nu}}{[\tau_0 + i (-L - s_4)\chi_{+\eta_4} (-L - s_4)]^{2\nu} [\tau_0 + i (t-L - s_3)\chi_{-\eta_3} (t-L - s_3)]^{2\nu}}  \\
    &\times \left\{ \frac{1}{ [i\tau_0 \chi_{-\eta_4} (t\!-\!s_4) \!-\! (t\!-\!s_4\!-\!L)] [i\tau_0 \chi_{+\eta_3} (-s_3) \!+\! (s_3 \!+\! L)]} + \frac{1}{(i\tau_0 \!-\! t) [i \tau_0 \chi_{\eta_3\eta_4} (s_3 \!-\! s_4) \!-\! (s_3 \!-\! s_4)]}  \right\},
\end{aligned}
\label{eq:da2_ii_2}
\end{equation}
where in the last line, the first and second terms in the curly brackets correspond to the contract options (ii) and (iii), respectively (see Fig.~\ref{fig:da2_s3_s4}).
Now, we evaluate the terms stemming from these contract options separately.

\textbf{Contract option (ii).} We begin with the contract option (ii), where we take the first term within the curly brackets of the last line of Eq.~\eqref{eq:da2_ii_2} to perform the pole integration.
In this case, integrals over $s_3$ and $s_4$ are determined by singularities located at $s_3 \to -L$ and $s_4 \to t-L$. The integral in Eq.~\eqref{eq:da2_ii_2} then yields
\begin{equation}
\begin{aligned}
     D_{A2}^\text{(ii)} & = \frac{\left[\mathcal{T}^{(0)}_A\right]^2}{12\pi^3}  \frac{\tau_0^{4\nu + 1/{\nu}-5} e^{2 i\nu e V t}}{(\tau_0 + i t)^{4 \nu + 1/\nu - 3}} \sum_{\eta_3\eta_4} \eta_3 \frac{1}{ (\tau_0 + i \eta_3 t)^{4\nu - 1}} \iint  ds_3 ds_4 
  \frac{ e^{-i\nu e V (t - s_4)}}{ [\tau_0  \!+\! i(t-s_4) \eta_4]^{1-2\nu} } \frac{e^{-i\nu e V s_3} }{[\tau_0 \!+\! i (-s_3 ) \eta_3 ]^{1-2\nu}} \\
  & = \frac{\left[\mathcal{T}_A^{(0)}\right]^2}{3\pi^3} \frac{\tau_0^{4\nu + 1/{\nu}-5} (\nu e V)^{-4\nu}}{(\tau_0 + i t)^{8 \nu + 1/\nu - 4}} e^{2 i\nu eV t} \cos^2\left[ \frac{\pi }{2} (1 - 2\nu) \right] \Gamma^2 (2\nu)\\
  & =  \frac{\Gamma^2 (2\nu)\pi}{3 \Gamma^2 (1-2\nu) \cos^2 (\pi\nu)} (\nu e V)^{2-8\nu} \frac{I_{A0}^2}{\nu^2 e^2} \frac{\tau_0^{\frac{1}{\nu}-1}}{(\tau_0 + it)^{\frac{1}{\nu} + 8\nu - 4}},
\end{aligned}
\label{eq:da2_ii_3}
\end{equation}
where $\eta_3 = 1$ and $\eta_4 = -1$ have been taken ( otherwise the integral vanishes).
In the last line of Eq.~\eqref{eq:da2_ii_3}, we have rewritten $\mathcal{T}_A^{(0)}$ through the non-equilibrium current $I_{A0}$.

\textbf{Contract option (iii).}
Alternatively, we can take the second term of the last line within the curly brackets of Eq.~\eqref{eq:da2_ii_2}.
With this option, the time arguments of the two operators of non-equilibrium anyons ($s_3$ and $s_4$) are equal to each other.
This also occurred for the contract option (i), where both $|s_1- s_2|$ and $|s_3- s_4|$ are assumed to be within $1/\nu e V$, the width (in time) of a typical non-equilibrium anyon pulse, but option (i) yielded zero to the correlation function after summation over Keldysh indexes, because of the absence of braiding phase.
The situation is, however, different for option (iii), where a finite braiding phase is introduced by the anyon-quasihole braiding process, thus generating a finite result. In what follows, this fact will be explicitly presented with detailed derivations.

To begin with, with contract option (iii), the product of anyonic correlators (the ``tangling factor'' in terminology of Ref.~\cite{GuOneHalf24S}) in  Eq.~\eqref{eq:da2_ii_2} simplifies to a phase factor
\begin{equation}
\frac{[\tau_0 + i (-L - s_3)\chi_{+\eta_3} (-L - s_3)]^{2\nu} [\tau_0 + i (t-L - s_4)\chi_{-\eta_4} (t-L - s_4)]^{2\nu}}{[\tau_0 + i (-L - s_4)\chi_{+\eta_4} (-L - s_4)]^{2\nu} [\tau_0 + i (t-L - s_3)\chi_{-\eta_3} (t-L - s_3)]^{2\nu}} = \exp[ i\pi \nu (\eta_4 -\eta_3) ],
\label{eq:non_trivial_phase}
\end{equation}
for $t>0$ and $-L<s_3, s_4 < t-L$ (the case of $t<0$ will be discussed separately).
The braiding phase defined by  Eq.~\eqref{eq:non_trivial_phase} is similar to that for the time-domain braiding in an anyonic tunneling system (see e.g., Refs.~\cite{RosenowLevkivskyiHalperinPRL16S, SimNC16S,LeeSimNC22S, schillerPRL23S, GuOneHalf24S, LandscapePRL25S}).
However, we stress that the braiding phase in Eq.~\eqref{eq:non_trivial_phase}, stemming from anyon-quasihole braiding in the present work, has a totally different origin compared to the phase appearing in time-domain braiding.
Indeed, in the Andreev-tunneling case, the phase is generated by the braiding between the reflected fractional-charge quasihole and extra non-equilibrium anyons---in sharp contrast to the time-domain braiding phase that resorts to anyonic pairs that tunnel at the central collider (see Sec.~\ref{sec:andreev_comparison} for detailed comparisons).
With the braiding phase factor \eqref{eq:non_trivial_phase}, the contract option (iii) yields
\begin{equation}
\begin{aligned}
    &\sum_{\eta_3\eta_4} \eta_3\eta_4 \int_{0}^t ds_3 \int_{0}^t ds_4 \frac{e^{-i\nu e V (s_3 - s_4)} e^{ i\pi \nu (\eta_4 -\eta_3) }}{[\tau_0 + i (s_3 - s_4) \chi_{\eta_3\eta_4} (s_3 - s_4)]^{2 \nu}} \Bigg|_{t>0} \\
    &\simeq  2 i \sin (2\pi\nu) \left( 1 \!-\! e^{-2i\pi\nu} \right) \left\{ \!- i \nu e V t\left[\Gamma(1-2\nu) \!-\! \Gamma(1-2\nu, -i\nu e V t) \right] \!-\! \Gamma (2 \!-\! 2 \nu) + \Gamma (2 \!-\! 2\nu, -i \nu e V t) \right\} (\nu e V)^{2\nu -2}
\end{aligned}
\label{eq:tdb_integral}
\end{equation}
for positive $t$. Here, $\Gamma(x,y)$ is the incomplete gamma function, and only the leading order in $\tau_0$ is kept for $t\gg \tau_0$.
For $t \ll \tau_0$, the integral is actually proportional to $t^2/\tau_0^{2\nu}$.
In Eq.~\eqref{eq:tdb_integral}, the integration is first carried out over the sum of the time arguments $s_3 + s_4$ in the range $|s_3 - s_4| < s_3 + s_4 < 2 t - |s_3 - s_4|$ and over the difference $s_3 - s_4$ in the range $-t<s_3 - s_4 < t$ (see Fig.~\ref{fig:near_zero_t}).
Depending on the parameter $\nu e V t$,  the asymtotics in the limiting cases of large and small times read:
\begin{equation}
\begin{aligned}
&\sum_{\eta_3\eta_4} \eta_3\eta_4 \int_{0}^t ds_3 \int_{0}^t ds_4 \frac{e^{-i\nu e V (s_3 - s_4)} e^{ i\pi \nu (\eta_4 -\eta_3) }}{[\tau_0 + i (s_3 - s_4) \chi_{\eta_3\eta_4} (s_3 - s_4)]^{2 \nu}} \\
   &\qquad \approx 2 \sin (2\pi\nu)\times    \begin{cases}
     \left( 1 - e^{-2i\pi\nu} \right) \Gamma (1 - 2\nu) \,(\nu e V)^{2\nu - 1}\,t,&\quad t\gg 1/\nu e V,\\
    \dfrac{\sin(\pi\nu) }{(1 - \nu) (1 - 2\nu)} \,\dfrac{t^{2}}{(t^2 + \mu\tau_0^2)^{\nu}},              & \quad  0<t\ll 1/\nu e V.
\end{cases}
\end{aligned}
\label{eq:limit_positive}
\end{equation}
Here, in the short-time limit, we have replaced $t^{-2\nu}$ with $(t^2 + \mu\tau_0^2)^{-\nu}$,
where
$
    \mu =  \cos^{1/\nu}(\pi\nu)[(1-\nu) (1 - 2\nu )]^{-1/\nu},
$
in order to reflect the fact that the integral equals $4 t^2 \tau_0^{-2\nu} \sin^2(\pi\nu)$ in the $t \ll \tau_0$ limit.

\begin{figure}
  \includegraphics[width=0.99\linewidth]{fig_near_zero_t}
  \caption{Illustration of the domains for integrals 
    in Eqs.~\eqref{eq:tdb_integral} and \eqref{eq:tdb_integral_negative} at $t>0$ (\textbf{Panels a} and \textbf{b}) and $t < 0$ (\textbf{Panels c} and \textbf{d}), respectively.
    Both integrals can be performed in two steps.
    When $t > 0$, these two steps involve the integral over $s_3 + s_4$ from $|s_3 - s_4|$ to $2 t - |s_3 - s_4|$, followed by the integral over $s_3 - s_4$ from $-t$ to $t$.
    When $t < 0$, one instead first takes the integral over $s_3 + s_4$ from $2t + |s_3 - s_4|$ to $ - |s_3 - s_4|$, which is followed by the integration over $s_3 - s_4$ from $t$ to $-t$.
    }
  \label{fig:near_zero_t}
\end{figure}

Before moving on to contributions of higher orders in $\mathcal{T}_{A,B}$ to the correlation function, we show the result for negative times, $t < 0$.
This calculation differs from that for $t>0$ in two respects. First, the braiding phase in Eq.~\eqref{eq:non_trivial_phase} changes its sign in (after noticing that $s_1 \to -L$ and $s_2 \to t - L$ for two involved poles).
Second, the integration range for the integral over $s_3 + s_4$ changes to  $2 t  + |s_3 - s_4| < s_3 + s_4 < -|s_3 - s_4|$, as shown in \textbf{panels c} and \textbf{d} of Fig.~\ref{fig:near_zero_t}. This is different from the range for positive $t$ (\textbf{panels a} and \textbf{b} of Fig.~\ref{fig:near_zero_t}), where $|s_3 - s_4| < s_3 + s_4 < 2 t - |s_3 - s_4|$.
With these two modifications, we obtain for $t<0$:
\begin{equation}
\begin{aligned}
    &\sum_{\eta_3\eta_4} \eta_3\eta_4 \int_t^0 ds_3 \int_t^0 ds_4 \frac{e^{-i\nu e V (s_3 - s_4)} e^{ i\pi \nu (\eta_3 -\eta_4) }}{[\tau_0 + i (s_3 - s_4) \chi_{\eta_3\eta_4} (s_3 - s_4)]^{2 \nu}} \Bigg|_{t<0} =  \sum_{\eta_3\eta_4} \eta_3\eta_4 \int_{t}^{-t} ds \frac{e^{-i\nu e V s} e^{i\pi\nu (\eta_3 - \eta_4)}}{[\tau_0 + i s \chi_{\eta_3\eta_4}(s)]^{2\nu}} (-|s| - t) \\
    &\simeq   2 i \sin (2\pi\nu)\left(1 - e^{2i\pi\nu} \right) \left\{ i \nu e V t\left[\Gamma(1-2\nu)\! -\! \Gamma(1-2\nu, - i\nu e V t) \right] + \Gamma (2 - 2 \nu)\! -\! \Gamma (2 - 2\nu, -i \nu e V t) \right\} (\nu e V)^{2\nu -2},
\end{aligned}
\label{eq:tdb_integral_negative}
\end{equation}
with the following asymptotics:
\begin{equation}
\begin{aligned}
&\sum_{\eta_3\eta_4} \eta_3\eta_4 \int_{-t}^0 ds_3 \int_{-t}^0 ds_4 \frac{e^{-i\nu e V (s_3 - s_4)} e^{ i\pi \nu (\eta_3 -\eta_4) }}{[\tau_0 + i (s_3 - s_4) \chi_{\eta_3\eta_4} (s_3 - s_4)]^{2 \nu}} \Bigg|_{t<0} \\
&\approx 2 \sin (2\pi\nu)\times    
\begin{cases}
     \left( 1 - e^{2i\pi\nu} \right) \Gamma (1 - 2\nu) \,(\nu e V)^{2\nu - 1}\,|t|,&\quad |t|\gg 1/\nu e V,\\
    \dfrac{\sin(\pi\nu) }{(1 - \nu) (1 - 2\nu)} \,\dfrac{t^{2}}{(t^2 + \mu\tau_0^2)^{\nu}},              & \quad  |t|\ll 1/\nu e V.
\end{cases}
\end{aligned}
\label{eq:limit_negative}
\end{equation}
Here,  in the small time limit, we have again replaced $|t|^{-2\nu}$ with $(t^2 + \mu\tau_0^2)^{-\nu}$ to incorporate the result in the $|t| \ll \tau_0$ limit.
We see that Eq.~\eqref{eq:limit_negative} differs from Eq.~\eqref{eq:limit_positive} (expressed through $|t|$) only by the complex conjugation in the long-time asymptotics.

Combining Eqs.~\eqref{eq:tdb_integral} and \eqref{eq:tdb_integral_negative}, we arrive at \begin{equation}
\begin{aligned}
    D_{A2}^\text{(iii)} = \frac{\tau_0^{\frac{1}{\nu}-1} c(\nu)}{2\pi (\tau_0 + it)^{1/\nu}} \frac{I_{A0}/e}{(\nu e V)^{2\nu - 1}} ( it)^{2-2\nu} e^{ i\nu e V t} \left[ - \zeta_+ (\nu, \nu e Vt) \frac{I_{A0}}{\nu e}\, t \right],
\end{aligned}
\label{eq:S14}
\end{equation}
where we have introduced the function 
\begin{equation}
\begin{aligned}
    \zeta_\pm (\nu, y) &\equiv \Big\{ \left[ 1 - \cos(2\pi\nu) \right] \text{sgn}(y)  \pm i \sin (2\pi\nu) \Big\} \, 
    \frac{   y\left[\Gamma(1-2\nu)\! -\! \Gamma(1-2\nu, -iy) \right] -i\,[ \Gamma (2 - 2 \nu)\! -\! \Gamma (2 - 2\nu,- i y)] }{y\,  \Gamma(1 - 2 \nu)}\\
    &\approx  2 \sin (\pi\nu)   
\begin{cases}
    \mp i\,\text{sgn} (y) \exp\left[ \mp i\pi\nu\,\text{sgn} (y)   \right],&\quad |y|\gg 1,\\
     \pm \dfrac{y^{1-2\nu}}{\Gamma (3 - 2\nu) }  \exp\left\{i\pi\nu \left[ 1 \mp \text{sgn} ( y)  \right] \right\} ,              & \quad  |y|\ll 1,
\end{cases}
\end{aligned}
    \label{eq:zeta_definition}
\end{equation}
with the subscripts ``$+$'' and ``$-$'' referring to the evaluation of $\langle \Psi_\alpha^\dagger (L,t) \Psi_\alpha (L,0)\rangle $ and $\langle \Psi_\alpha (L,t) \Psi_\alpha^\dagger (L,0)\rangle $, respectively (here, $\alpha = A, B$).
Strictly speaking, Eq.~\eqref{eq:S14} is valid when $|t| \gg \tau_0$.
Indeed, for $|t|\ll \tau_0$, according to the second lines of Eqs.~\eqref{eq:limit_positive} and \eqref{eq:limit_negative}, the singularity related to the jump $\text{sgn}(y)$ in $\zeta_\pm$ disappears.
As another feature, as follows from the first lines of Eqs.~\eqref{eq:limit_positive} and \eqref{eq:limit_negative}, for $|t| \gg 1/\nu e V$, the term in the square brackets of Eq.~\eqref{eq:S14} is similar to the one reported in Refs.~\cite{RosenowLevkivskyiHalperinPRL16S, LeeSimNC22S}, where time-domain braiding was considered for a system with anyonic tunneling at the central collider (see Sec.~\ref{sec:andreev_comparison} for a more detailed discussion).

Before ending this section, we compare the contributions of contract options (ii) and (iii) to $D_{A2}$ in the relevant long-time limit of $D_{A2}^\text{(iii)}$:
\begin{equation}
   \frac{D_{A2}^\text{(ii)}}{D_{A2}^\text{(iii)}} \sim \frac{(\tau_0 \nu e V)^{1-6\nu}}{(t/\tau_0)} \xrightarrow{\nu = 1/3} \frac{3}{t e V}.
    \label{eq:da2_ratio}
\end{equation}
As will be shown shortly in Sec.~\ref{sec:resummation}, after including processes of higher-order in diluter transmissions (involving multiple non-equilibrium anyons), the characteristic time scale in $D_A$ is set by $t \sim \nu e/I_{A0}$, such that the factor of Eq.~\eqref{eq:da2_ratio} becomes
\begin{equation}
    D_{A2}^\text{(ii)}/D_{A2}^\text{(iii)} \sim \mathcal{T}_A,
\end{equation}
which is a small quantity in the strongly diluted limit, $\mathcal{T}_A \ll 1$.
Because we focus on this limit, in the derivations that follow, only the contribution from the contract option (iii) will be taken into consideration.

\subsection{Resummation of higher-order contributions to the correlation function involving multiple non-equilibrium anyons}
\label{sec:resummation}

In Sec.~\ref{sec:proceeses_(ii)_(iii)}, we have evaluated the leading and next-to-leading contributions to the correlation functions determining the tunneling current and its noise.
As has been shown, the dominant contribution at order $[\mathcal{T}_A^{(0)}]^2$ comes from the process corresponding to contract option (iii) [Eq.~\eqref{options}], where one of two non-equilibrium anyons triggers the Andreev-like tunneling.
The fractional-charge hole, generated in the course of Andreev tunneling, braids with the other non-equilibrium anyon (anyon-quasihole braiding process, see Fig.~2 of the main text and Fig.~\ref{fig:disconnected_diagrams}) below.

In this section, we perform resummation of contributions to the correlation function resulting from higher-order processes involving multiple non-equilibrium anyons.
Generalizing Eqs.~(\ref{eq:leading1}) and (\ref{eq:da2_initial}), we consider the expansion of Eq.~(8) of the main text to the $2n$th order in diluter transmissions.
Without loss of generality, we assume that Andreev-like tunneling is triggered by the anyon operators taken at times $s_1$ and $s_2$. For the rest quasiparticle operators, we arrange the time such that the (annihilation) operator at time $s_{2i-1}$ contracts with the creation operator at time $S_{2i}$, where $2\leq i\leq n$.
Notice that this contract option can be considered as a natural extension of contract option (iii) defined in Eq.~\eqref{options} of Sec.~\ref{sec:proceeses_(ii)_(iii)}, which was quadratic in $\mathcal{T}_A^{(0)}$, to correlations due to higher-order processes $\propto \left[\mathcal{T}_A^{(0)}\right]^{n}$.
The correlation function then contains the following product:
\begin{equation}
    \begin{aligned}
        & \frac{\tau_0^{2\nu + 1/\nu}}{(\tau_0 + i t)^{1/\nu - 1}[\tau_0 + i(s_1 - s_2 ) \chi_{\eta_1\eta_2} (s_1 - s_2)]^{2\nu - 1}} \frac{[\tau_0 + i (t - s_1 - L) \chi_{-\eta_1} (t-s_1)] [\tau_0 + i ( - s_2 - L) \chi_{+\eta_2} (-s_2)]}{(\tau_0 + it) [\tau_0 + i(s_1 - s_2 ) \chi_{\eta_1\eta_2} (s_1 - s_2)]}\\
        & \times  \frac{1}{[\tau_0 + i (t-s_2 - L) \chi_{-\eta_2} (t-s_2)] [\tau_0 + i (-s_1 - L) \chi_{+\eta_1} (-s_1)] } \\
        & \times  \prod_{j=2}^n \iint ds_{2j-1} ds_{2j} \exp[i\pi\nu (\eta_{2j} - \eta_{2j - 1})] \frac{\tau_0^{2\nu }}{[\tau_0 + i (s_{2j-1} - s_{2j}) \chi_{\eta_{2j-1}\eta_{2j}} (s_{2j-1} - s_{2j})]^{2 \nu}},
    \end{aligned}
    \label{eq:2n_contraction}
\end{equation}
where the first two lines are induced by the Andreev-like tunneling triggered by non-equilibrium anyons that tunneled through the diluter at times $s_1$ and $s_2$.
The last line in Eq.~\eqref{eq:2n_contraction} describes ``dressing'' this process with extra $n-1$ pairs of ``self-contracted'' non-equilibrium anyons,
where the anyon-quasihole braiding phase, akin to Eq.~\eqref{eq:non_trivial_phase}, has already been included. Importantly, following Eq.~\eqref{eq:2n_contraction}, with multiple ($n-1$) pairs of self-contracted operators, the contribution of these pairs [the last line of Eq.~\eqref{eq:2n_contraction}] equals the product of $n-1$ copies of the single-pair result. This fact is the prerequisite of resummation performed in, e.g., Ref.~\cite{LeeSimNC22S}.

Now, we consider the combinatorics of the corresponding processes (cf. Ref.~\cite{GuOneHalf24S}). We have $2n (2n - 1)$ ways to choose two operators (one creation and one annihilation) that trigger Andreev-like tunnelings.
Next, we need to pair up the rest $2n-2$ operators of non-equilibrium anyons into self-contracted pairs, yielding
\begin{equation}
   \frac{2^{n-1}}{(n-1)!} C_{2n-2}^2 C_{2n-4}^2 \cdot \cdot\cdot C_2^2 = \frac{(2n-2)!}{(n-1)!}
\end{equation}
for the number of possible contract options.
Here, the factor $2^{n-1}$ indicates that one operator in a given pair is the creation operator. The factor $1/(n-1)!$ removes repeated options, as it does not make any difference to pick up one pair earlier or later.
Restoring the prefactor $1/(2n!)$ from the expansion, we obtain  for $|t|\gg \tau_0$:
\begin{equation}
\begin{aligned}
   \sum_{n=1}^\infty \left[ -\zeta_+ (\nu, \nu e V t)\frac{I_{A0} t}{\nu e}  \right]^{n-1} \frac{(2n-2)!}{(n-1)!} (2n-1) 2n \frac{1}{(2n)!}& =\sum_{n=1}^\infty  \frac{1}{(n-1)!}\left[ -\zeta_+ (\nu, \nu e V t)\frac{I_{A0} t}{\nu e}  \right]^{n-1} \\
   &= \exp \left[ -\zeta_+ (\nu, \nu e V t)\frac{I_{A0} t}{\nu e} \right].
\end{aligned}
\end{equation}
Upon this resummation, we arrive at the following correlation function for operators in channel $A$,
\begin{equation}
    \big\langle \Psi^\dagger_A (L,t^-) \Psi_A (L,0^+)\rangle = \sum_{n=0}^\infty D_{An} = \frac{\tau_0^{\frac{1}{\nu}-1}}{2\pi (\tau_0 + it)^{1/\nu}} \left\{ 1 + i \, c(\nu)\, e^{i\nu e V t} \frac{I_{A0} t}{(i \nu e V t)^{2\nu-1}} \exp\left[- \zeta_+ (\nu, \nu e V t)\frac{I_{A0} t}{\nu e} \right] \right\}.
\label{eq:dan_resum}
\end{equation}
Likewise, the corresponding correlation function in channel $B$ is evaluated as
\begin{equation}
    \big\langle \Psi_B (L,t^-) \Psi^\dagger_B (L,0^+)\rangle = \frac{\tau_0^{\frac{1}{\nu}-1}}{2\pi (\tau_0 + it)^{1/\nu}} \left\{ 1 + i\, c(\nu) \, e^{-i\nu e V t}  \frac{I_{B0} t}{(i \nu e V t)^{2\nu-1}} \exp\left[-\zeta_- (\nu, \nu e V t) \frac{I_{B0} t}{\nu e} \right] \right\}.
\label{eq:dan_resum_b}
\end{equation}
In addition to a trivial replacement $A\leftrightarrow B$, equation~\eqref{eq:dan_resum_b} differs from Eq.~\eqref{eq:dan_resum} by (i) an extra minus sign for the bias-dependent phase factor and (ii) by replacing $\xi_+$ with $\xi_-$.

\section{{\,\,\,\,\,\,\,}Anyon-quasihole braiding versus time-domain braiding:\\ 
{\phantom{....}\,}Braiding processes and corresponding correlation functions}
\label{sec:andreev_comparison}

In Eqs.~\eqref{eq:dan_resum} and \eqref{eq:dan_resum_b}, the exponential factor containing $\zeta_\pm$ is crucially important to establish the long-time decay of the non-equilibrium contribution to the correlation function in the Andreev-tunneling limit.
When $|t| \gg 1/\nu e V$, $\zeta_\pm$ reduces to the $|y| \gg 1$ limit of Eq.~\eqref{eq:zeta_definition},
with the phase factor $\pm 2 i \sin (\pi\nu) \exp (\mp i \pi\nu) = 1-\exp(\mp 2i\pi\nu)$ indicating the anyon-quasihole braiding phase between the quasihole reflected after an Andreev-like tunneling (triggered by an anyon arriving at the central collider) and other non-equilibrium anyons that bypass the central collider.
Noteworthy, in the opposite tunneling limit where the central collider allows anyons to tunnel (instead of electrons in the Andreev-like tunneling limit), a similar braiding process, known as the time-domain braiding, occurs~\cite{RosenowLevkivskyiHalperinPRL16S,SimNC16S,LeeSimNC22S,LeeNature23S,schillerPRL23S,LandscapePRL25S}.
To provide a better illustration of the difference and similarity of these two braiding processes, in this section, we focus on the long-time limit, $|t| \gg 1/\nu e V$.

\begin{figure}
  \includegraphics[width=1.0\linewidth]{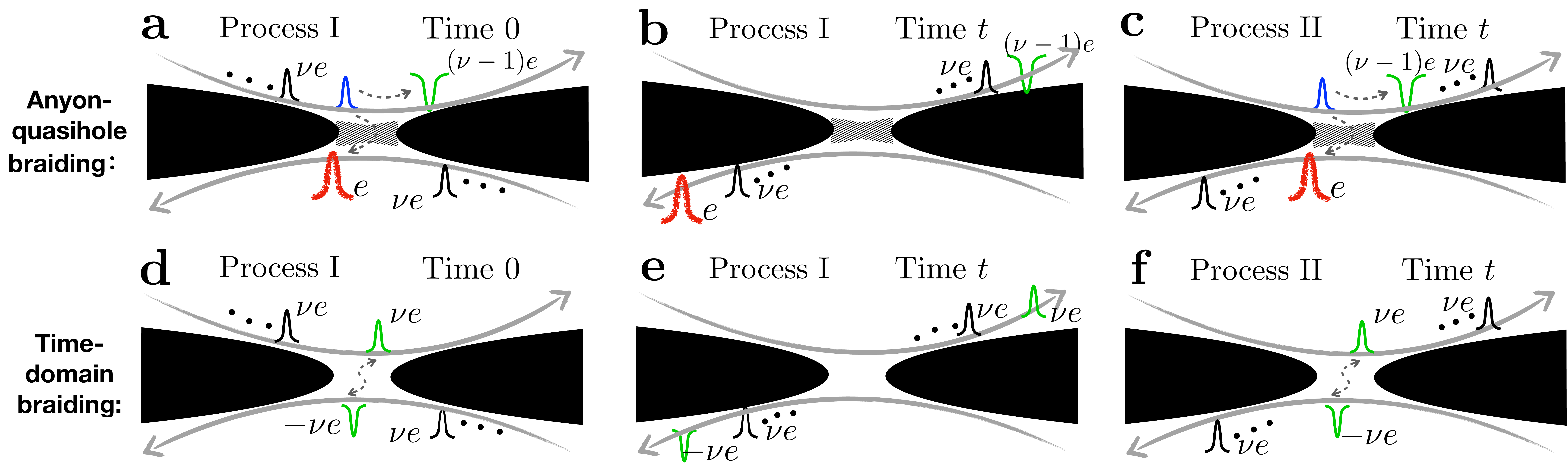}
  \caption{
  Comparison between anyon-quasihole braiding in Andreev-like tunneling and time-domain braiding in anyon-tunneling platforms. Edge chiralities are indicated by grey arrows. \textbf{Panels a, b, c}: Anyon-quasihole braiding processes that involve a single Andreev-like tunneling event. This tunneling process, triggered by an incoming non-equilibrium anyon (colored in blue), results in a transmitted fractional-charge hole (green), accompanied by an electron
  in the opposite edge (red).
  Non-equilibrium anyons bypassing the central collider are represented as black pulses (and dots). Panels \textbf{a} and \textbf{b} refer to configurations corresponding to Process I, where the Andreev-like tunneling occurs at time 0.
  Panel \textbf{c} refers to Process II, where the Andreev-like tunneling occurs at a later moment, $t > 0$.
  In panels \textbf{b} and \textbf{c}, non-equilibrium anyons in channel $A$ are upstream and downstream, respectively, of the reflected fractional-charge hole, thus generating the anyon-quasihole braiding in channel $A$ through the interference of Process I and Process II.
  \textbf{Panels d, e, f}: The interfering processes for the anyon-tunneling setup. Here, the green pulses refer to an anyon-hole pair generated at the collider.
  Panels \textbf{d} and \textbf{e} show configurations at moments 0 and $t$, respectively, for Process I, where anyonic tunneling occurs at time 0.
  Panel \textbf{f} instead presents the configuration of Process II at time $t>0$, when the anyonic tunneling occurs.
  Following panels \textbf{e} and \textbf{f}, time-domain braiding occurs in both channels $A$ and $B$, in sharp contrast to the Andreev-like tunneling situation, where anyon-quasihole braiding occurs only in the channel that contains a reflected fractional-charge quasihole.
  }
  \label{fig:disconnected_diagrams}
\end{figure}

In this limit, Eqs.~\eqref{eq:dan_resum} and \eqref{eq:dan_resum_b} for correlation functions in channels $A$ and $B$ become
\begin{equation}
\begin{aligned}
 & \left.\begin{array}{c}
 \big\langle\Psi^\dagger_{A}(t^-) \Psi_{A} (0^+) \big\rangle      \\
 \big\langle\Psi_{B}(t^-) \Psi_{B}^\dagger (0^+) \big\rangle     
\end{array}
\right\}
   =  \frac{\tau_0^{\frac{1}{\nu}-1}}{2\pi (\tau_0 + it)^{1/\nu}}
   \\
  & \qquad \times 
  \left[ 1 
   +  i\,  c(\nu)\, \mathrm{e}^{\pm i \nu e V t}\, \frac{I_{A0,B0}\, t }{e(i \nu e V t)^{2\nu - 1}}   
   \exp\left(-\frac{I_{A0,B0}}{\nu e} \Big\{ [1 - \cos (2\pi\nu)]  |t| \pm i \sin (2\pi\nu) t \Big\} \right) \right],
\end{aligned}
\label{eq:separate_correlations}
\end{equation}
where in the second line, the first (unity) and second terms correspond to the equilibrium and non-equilibrium contributions, respectively, and $c(\nu)$
is defined in Eq.~\eqref{eq:da0_da1}.
Two features of the non-equilibrium term are worth our special attention.

First, this term is proportional to the non-equilibrium current $I_{A0,B0}$ and, thus, the transmission probability of the corresponding diluter, which signifies that the generation of an Andreev-like tunneling requires the presence of a non-equilibrium anyon.
Second, the exponential factor, depending on the non-equilibrium currents $I_{A0}$ or $I_{B0}$, is generated by the anyon-quasihole braiding process, 
manifested via the appearance of the phase factor 
$\exp (\mp i 2\pi\nu)$. This factor is produced by braiding one Laughlin quasiparticle and one Laughlin quasihole, both with the same statistical angle $\pi \nu$.
As stated in the main text, in a given channel, the exponential factor in Eq.~\eqref{eq:separate_correlations} depends only on the non-equilibrium current ($I_{A0}$ or $I_{B0}$, for channels $A$ and $B$, respectively) in the corresponding channel, as anyon-quasihole braiding only occurs between fractional-charge quasihole and non-equilibrium anyons of the same channel (see  Fig.~\ref{fig:disconnected_diagrams}\textbf{c}).
The inclusion of the anyon-quasihole braiding factor distinguishes our work from Refs.~\cite{KaneFisherPRB03S,JonckheerePRB23S}.

For comparison, for a system where the central collider allows anyonic tunneling, the corresponding correlation functions read (see, e.g., Refs.~\cite{RosenowLevkivskyiHalperinPRL16S,LeeSimNC22S})
\begin{equation}
\begin{aligned}
 & \left.\begin{array}{c}
 \big\langle\psi^\dagger_{A}(t^-) \psi_{A} (0^+) \big\rangle      \\
 \big\langle\psi_{B}(t^-) \psi_{B}^\dagger (0^+) \big\rangle     
\end{array}
\right\}
   \approx  \frac{\tau_0^{\nu-1}}{2\pi (\tau_0 + it)^{\nu}} \\
   &\qquad \times \exp\left( -\frac{I_{A0}}{\nu e} \left\{ \left[1 - \cos (2\pi\nu)\right] |t| - i\sin(2\pi\nu) t \right\} \right)\, \exp\left( -\frac{I_{B0}}{\nu e} \left\{ \left[1 - \cos (2\pi\nu)\right] |t| + i\sin(2\pi\nu) t \right\} \right).
\end{aligned}
\label{eq:correlation_anyonic}
\end{equation}
Here, only time-domain braiding processes are taken into consideration for correlations of anyonic operators $\psi_A$, $\psi^\dagger_A$, $\psi_B$ and $\psi^\dagger_B$ (see Refs.~\cite{LeeSimNC22S, GuOneHalf24S, LandscapePRL25S} for results going beyond time-domain braiding).
Compared to the Andreev-like correlation functions [Eq.~\eqref{eq:separate_correlations}], correlations in the anyon-tunneling system [Eq.~\eqref{eq:correlation_anyonic}] have the following two unique features.

First, in Eq.~\eqref{eq:correlation_anyonic}, the correlation function of operators in either channel involves non-equilibrium currents of both channels ($I_{A0}$ and $I_{B0}$). This dependence on both currents originates from the fact that for anyonic tunneling, anyon-hole pair generated at the central collider braids with non-equilibrium anyons in both channels (see Fig.~\ref{fig:disconnected_diagrams}\textbf{f}).
This is in sharp contrast to Eq.~\eqref{eq:separate_correlations} in the Andreev-tunneling limit (see  Fig.~\ref{fig:disconnected_diagrams}\textbf{c}), where braiding only occurs between the fractional-charge quaihole and non-equilibrium anyons within the channel hosting the quasihole.
In addition, as the generation of anyon-quasihole braiding phase involves reflected holes (instead of transmitting anyons), it contains an extra minus sign, compared to the phase of time-domain braiding in the anyonic-tunneling limit.

Second, Eq.~\eqref{eq:separate_correlations} for Andreev-like tunneling contains an equilibrium contribution, i.e., the term ``1'' in the square brackets in the second line.
The non-equilibrium contribution [the second term of the second line of Eq.~\eqref{eq:separate_correlations}] is proportional to the non-equilibrium current ($I_{A0,B0}$) and, thus, to the transmission probability of the corresponding diluter. Physically, this means that the braiding process requires the presence of two anyons: one triggering Andreev tunneling and the other braiding with the resulting quasihole.
This is in great contrast to Eq.~\eqref{eq:correlation_anyonic} for anyonic tunneling, where the exponential factor induced by time-domain braiding factor directly multiplies the equilibrium correlation function, so that already a single non-equilibrium anyon is sufficient, as the other anyon is spontaneously generated (within an anyon-hole pair) at the collider. This implies that, for time-domain braiding, the separation between equilibrium and non-equilibrium contributions to the correlation functions becomes more involved than in the Andreev-tunneling limit.

\section{{\,\,\,\,\,\,\,}Evaluation of tunneling current and tunneling noise}

In Sec.~\ref{sec:correlation_derivations}, we have obtained the relevant correlation functions, Eqs.~\eqref{eq:dan_resum} and \eqref{eq:dan_resum_b}, for fermion tunneling operators in channels $A$ and $B$, respectively.
This section derives the tunneling current and the corresponding noise with these correlation functions.

\subsection{Integrals over time $t$}
\label{sec:integral_over_t}

When evaluating the tunneling current $I_T$ and its noise $S_T$, one needs to multiply the correlation functions Eqs.~\eqref{eq:dan_resum} and \eqref{eq:dan_resum_b}, and integrate the product over time $t$.
In this section, we describe the subtleties of calculating such an integral. We consider integrals of the general form,
\begin{equation}
    \int_{-\infty}^\infty dt \frac{e^{\xi(t)}}{(\tau_0 + it)^{n_0}},
    \label{eq:integral_target}
\end{equation}
where $\xi (t) \to - b|t| + c t$ in the long-time limit [cf. the first line of Eqs.~\eqref{eq:limit_positive} and \eqref{eq:limit_negative}], $|t| \gg 1/\nu e V$, with $b$ and $c$ two constant numbers related to either the non-equilibrium current and/or bias.
For instance, considering the case where only the upper diluter is on [corresponding to $\mathcal{T}_A$ finite, and $\mathcal{T}_B = 0$], $b = 2 I_{A0} \sin^2 (\pi\nu)/\nu e$ and $c = -I_{A0} \sin (2\pi\nu)/\nu e + \nu e V$, cf. Eq.~\eqref{eq:separate_correlations}.
Importantly, when $t \to 0$, the function $\xi(t)$ is analytic, so that the cusp $|t|$ is rounded around zero.
Based on Eq.~\eqref{eq:zeta_definition}, we see that this rounding does not occur at $t\sim 1/\nu e V$, as the small-$y$ asymptotics of the function $y\zeta_\pm(\nu,y)$ (obtained for $\tau_0\to 0$) is still a non-analytic function at $y=0$. As discussed below Eq.~\eqref{eq:limit_positive}, to cure this singularity, 
one needs to evaluate the integral keeping the ultraviolet cutoff, $\tau_0$, finite, which results in a parabolic dependence leading to the parabolic dependence $\xi (t) \approx t^2/(\mu \tau_0^2)^\nu$ around $t=0$.

The exponent $n_0$ in the power-law denominator in Eq.~\eqref{eq:integral_target} is determined by the scaling dimensions of the tunneling operators.
More specifically, $n_0 = \nu_\text{s} \equiv 2/\nu + 2\nu -2$ and $n_0 = \nu_\text{d} \equiv 2/\nu + 4\nu -4$ for integrals appearing in the single-source and double-source (collision-induced) quantities, respectively.
For Laughlin quasiparticles, where $\nu$ is the inverse of an odd integer, $\nu_\text{d}$ and $\nu_\text{s}$ are both larger than unity.
This feature, valid for the Andreev-like tunneling limit, however, greatly contrasts that in the opposite tunneling limit (when anyons are allowed to tunnel directly through the collider), where the $n_0<1$.
This feature of anyonic tunneling, importantly, is the reason why Refs.~\cite{LeeSimNC22S, MorelPRB22S, LeeNature23S, schillerPRL23S} focused on only the long-time ($|t| \gg 1/\nu e V$) limit, where one can simply replace $\xi (t)$ by $  - b|t| + c t$ (with $b$ and $c$ two constant numbers) for the entire range of integral).
Instead, for Andreev-like tunneling,  where $n_0 > 1$,   the integral with a non-analytic (at $t=0$) function $\xi(t)$ is dominated by $t\to 0$, see Eqs.~\eqref{eq:integral_double} and \eqref{eq:integral_single} below. For $\xi (t) = - b|t| + c t$, this happens because of the singularity generated by a second-order derivative of $|t|$, which produces $2 \delta (t)$. This unphysical result would imply that the integral depends explicitly on the ultraviolet scale $\tau_0$ and is, thus, non-universal. 
Moreover, such an ``ultraviolet'' result would completely miss the effects of quasiparticle braiding that determines the correlation functions at longer time $\nu e V |t|\gg 1$.

Therefore, in the Andreev-tunneling case, one has to exercise certain caution when dealing with the short-time limit of function $\xi(t)$ [cf. Eqs.~\eqref{eq:limit_positive} and \eqref{eq:limit_negative}] in the integrals of the type of Eq.~(\ref{eq:integral_target}). As we demonstrate below, the result of the integration in the Andreev-tunneling case will not depend on details of the correlation functions at $|t|\sim \tau_0$, once this function is analytic at $t=0$ (which is actually the case). As a result, one can still resort to the long-time asymptotics of the correlation function, which encode the information on anyon-quasihole braiding.

Before presenting the calculation of the integrals that determine the contributions of collision and single source to $I_T$ and $S_T$, we briefly analyze the general properties of the integral~(\ref{eq:integral_target}) for $n_0>1$. First, 
since the power-law function rapidly decays for $|t|>\tau_0$, it is tempting to use this fact to replace $\xi(t)$ with $\xi(0)$.
However, we then immediately get zero, since
\begin{equation}
    \int_{-\infty}^\infty dt \frac{1}{(\tau_0 + it)^{n_0}}=0.
    \label{eq:integral_0}
\end{equation}
Second, any power of $t$ in the numerator of such an integral
also gives zero, as long as the integral remains convergent.
For example, for $n_0>2$, 
\begin{align}
    \int_{-\infty}^\infty dt\, \frac{t}{(\tau_0 + it)^{n_0}}=i\tau_0\int_{-\infty}^\infty dt\, \frac{1}{(\tau_0 + it)^{n_0}}-i\int_{-\infty}^\infty dt\, \frac{1}{(\tau_0 + it)^{n_0-1}}=0,
    \label{eq:integral_00}
\end{align}
which can readily be generalized to higher powers of $t$ for larger $n_0$. Therefore, expanding the factor $e^{\xi(t)}$ in the numerator around $t=0$, the first terms of this expansion will produce zeros.
This suggests that the result of the integration is determined by the higher derivative of $e^{\xi(t)}$. To see this, it is convenient to perform integration over $t$ by parts. This strategy transforms the original integral into integrals that are dominated by the contribution of long times, such that the long-time (exponential)  asymptotics of the correlation functions are sufficient for evaluation of the tunneling current and its noise.

Indeed, for the collision processes [i.e., when we multiply the second parts of Eqs.~\eqref{eq:dan_resum} and \eqref{eq:dan_resum_b}, which describe the non-equilibrium contributions to the correlation functions], $\nu_\text{d} = 2/\nu + 4\nu -4$ is between 3 and 4 when $\nu = 1/3$. 
In this case, the integral \eqref{eq:integral_target} with $n_0=\nu_d$ is transformed as follows:
\begin{equation}
\begin{aligned}
    &\int_{-\infty}^\infty dt\, \frac{e^{\xi(t)}}{(\tau_0 + it)^{\nu_\text{d}}}  = -\frac{i}{(\nu_\text{d}-1) (\nu_\text{d}-2) (\nu_\text{d}-3)}\int_{-\infty}^\infty dt\, e^{\xi(t)} \frac{d^3}{dt^3} \frac{1}{(\tau_0 + i t)^{\nu_\text{d} - 3}}\\
    & = \frac{i}{(\nu_\text{d}-1) (\nu_\text{d}-2) (\nu_\text{d}-3)}\int_{-\infty}^\infty dt\,   \frac{1}{(\tau_0 + i t)^{\nu_d - 3}} \frac{d^3}{dt^3} e^{\xi(t)}\\
    & = \frac{i}{(\nu_\text{d}-1) (\nu_\text{d}-2) (\nu_\text{d}-3)}\int_{-\infty}^\infty dt   \frac{1}{(\tau_0 + i t)^{\nu_\text{d} - 3}} \left\{ [\xi'(t) ]^3 + 3 \xi'(t) \xi''(t) + \xi'''(t) \right\}\, e^{\xi(t)} .
    \end{aligned}
\label{eq:integral_double}
\end{equation}
Now that the exponent ($\nu_d-3$) in the power-law factor in the integrand is between 0 and 1, the transformed integral is no longer dominated by small $t$. We can thus use the long-time asymptotics of the function $\xi(t)$ to evaluate the resulting integrals: 
\begin{equation}
\xi(t)\to-b|t|+i c t, 
\label{xit-asympt}
\end{equation}
the exponential factor $e^{\xi(t)}$ guarantees the convergence of the integral at $|t|\to \infty$. At the same time, $\xi''(t)$ and $\xi'''(t)$ for the function \eqref{xit-asympt} have singularities at $t=0$. Recall that this point is beyond the validity range of the asymptotic expression for $\xi(t)$; the actual function is analytic near the origin. Ignoring this spurious contribution to the integral (we will return to the vicinity of $t=0$ shortly), we find 
\begin{equation}
\begin{aligned}
    \int_{-\infty}^\infty dt\, \frac{e^{\xi(t)}}{(\tau_0 + it)^{\nu_\text{d}}} & \approx \frac{i}{(\nu_\text{d}-1) (\nu_\text{d}-2) (\nu_\text{d}-3)}\int_{-\infty}^\infty dt   \frac{\left[ b(-b^2 + 3 c^2) \text{sgn} (t) + i c (3 b^2 - c^2) \right]}{(\tau_0 + i t)^{\nu_\text{d} - 3}} e^{-b|t| + i c t}\\
    & = 2 \text{Re} \left[ e^{-i\pi \nu_\text{d}/2} \left( b - ic \right)^{ \nu_\text{d} - 1} \right] \Gamma(1 - \nu_\text{d}).
\end{aligned}
\label{eq:integral_double-bc}
\end{equation}
Similar to integrals for the case of anyon tunneling~\cite{SimNC16S,LeeSimNC22S,LeeNature23S,schillerPRL23S,LandscapePRL25S}, the integral \eqref{eq:integral_double-bc} is dominated by long times $|t| > 1/\nu e V$.

\begin{figure}
\centering
  \includegraphics[width=0.8\linewidth]{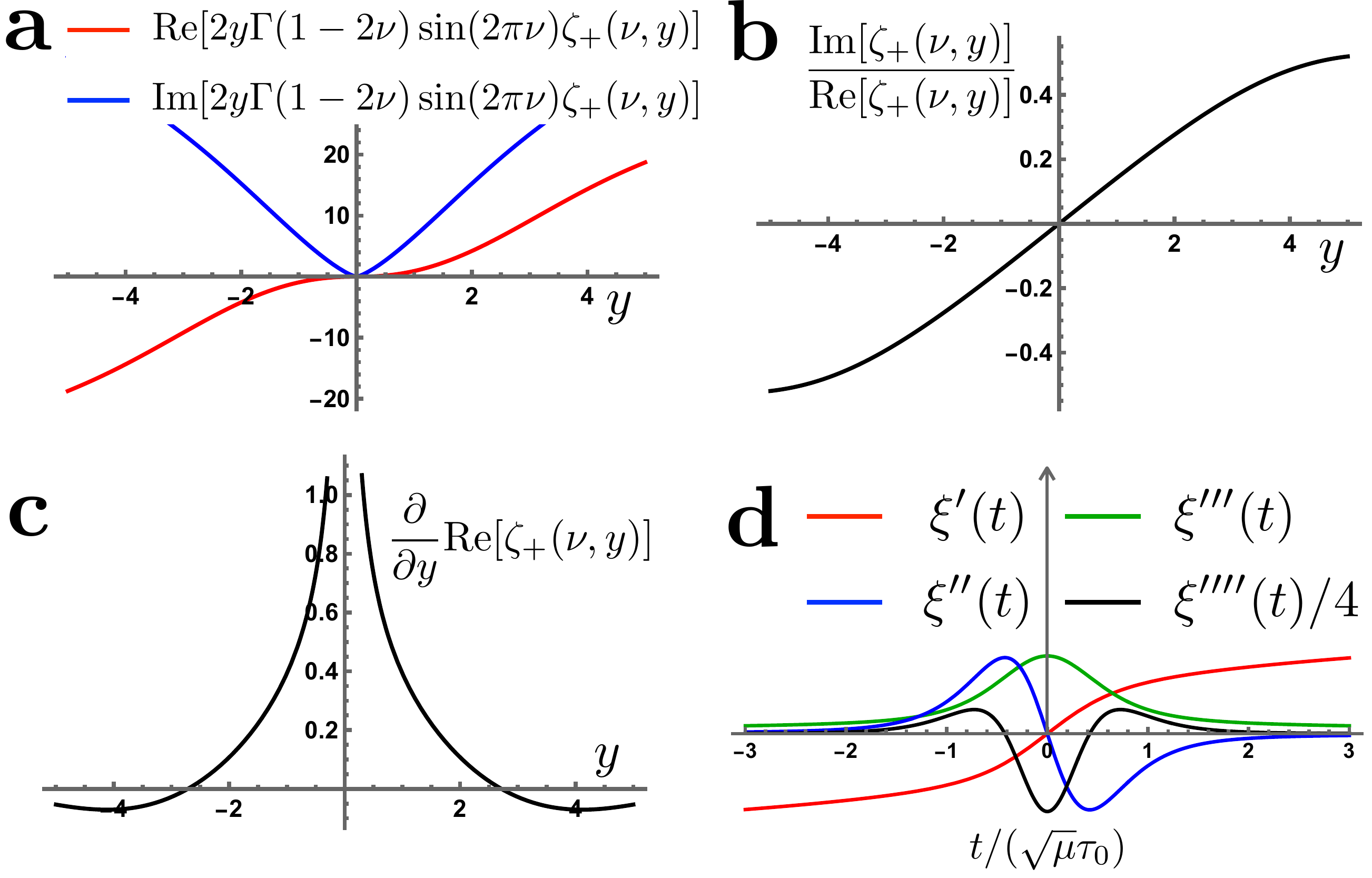}
  \caption{Plots of relevant functions, concerning the contribution to the integral in the small-$t$ limit.
  \textbf{Panel a}: The plots of the real and imaginary parts of the function $2 y \Gamma (1 - 2\nu) \sin (2\pi\nu) \zeta_+ (\nu,y)$ [corresponding to the last lines of Eq.~\eqref{eq:tdb_integral} and Eq.~\eqref{eq:tdb_integral_negative}] for $\nu =1/3$.
  Notice that both the real and imaginary parts of the plotted function vanish at $t=0$.
  Nevertheless, as shown by that in \textbf{Panel b}, the imaginary part vanishes faster than the real one, indicating that the function becomes approximately real in the $y\to 0$ limit.
  This fact implies $\xi (t)$ to be real in the $t\to 0$ limit, as $\xi (t)$ is basically the linear combination of $\xi_\pm$ functions.
  \textbf{Panel c}: The first-order derivative of the real part $\zeta_+ (\nu,y)$.
  The singularity at $y \to 0$ is removed after more carefully treating the integral in the $t\sim \tau_0$ limit; see the last lines of Eqs.~\eqref{eq:tdb_integral} and \eqref{eq:tdb_integral_negative}.
  \textbf{Panel d}:
  Plots of derivatives of $\xi(t)$ (being real in the leading order of $t$, following results of panels $\textbf{a}$ and \textbf{b}) for $|t| \ll 1/\nu e V$ [with expressions in this limit given by the second lines of Eqs.~\eqref{eq:limit_positive} and \eqref{eq:limit_negative}].
  All curves are normalized by a finite common factor.
  The red curve describes the value of the first derivative $\xi'(t)$. The green, blue, and black curves show absolute values of $\xi''(t)$, $\xi'''(t)$, and $\xi''''(t)$, respectively.
  The latter three are finite and decay for $|t|\gg \tau_0$, so that they become much smaller than $\xi'(t)$ away from the vicinity of $t=0$.
  }
  \label{fig:xi_differentiations}
\end{figure}

In Eq.~\eqref{eq:integral_double-bc}, we kept only the first term within the curly brackets of the third line of Eq.~\eqref{eq:integral_double}, i.e., $[\xi'(t)]^3$, since it dominates over other terms. The validity of this approximation is seen in Fig.~\ref{fig:xi_differentiations}, where we focus on expressions in the $|t| \ll 1/\nu e V$ limit [with corresponding expressions of the second lines of Eqs.~\eqref{eq:limit_positive} and \eqref{eq:limit_negative}].
Indeed, following this figure, when $t\to 0$, higher-order derivatives $\xi''(t)$, $\xi'''(t)$, and $\xi''''(t)$ are all finite. The widths of these curves are of the order of $\tau_0$. This is in great contrast to $\xi'(t)$, which instead grows slowly when $|t| \ll 1/\nu e V$, and then becomes a constant already when $|t| \gg 1/\nu e V$.
Importantly, this approximation used in Eq.~\eqref{eq:integral_double-bc} applies to other similar integrals of the type (\ref{eq:integral_target}) with the power-law factor in the denominator characterized by $n_0>1$.
Furthermore, as we see in Fig.~\ref{fig:xi_differentiations} for the full function $\xi(t)$, which is analytic at $t=0$, the contributions of the terms with higher derivatives to the integral
\eqref{eq:integral_double} are proportional to positive powers of the ultraviolet cutoff $\tau_0$ and, thus, can be sent to zero with $\tau_0\to 0$. This means that the result does not depend on the specific form of $\xi(t)$ near the origin, as long as this function is analytic.

When considering the single-source case [i.e., when combining the equilibrium contribution of either Eqs.~\eqref{eq:dan_resum} or \eqref{eq:dan_resum_b}, and the non-equilibrium contribution of the other one], the exponent of the power-law factor in the denominator of Eq.~\eqref{eq:integral_target} is $n_0=\nu_\text{s} = 2/\nu + 2\nu -2$, which has the value between 4 and 5 (thus larger than 1) for $\nu = 1/3$.
If this case, we take the same steps as in  Eq.~\eqref{eq:integral_double} but taking one more derivative in the partial integration:
\begin{equation}
\begin{aligned}
    &\int_{-\infty}^\infty dt\, \frac{e^{\xi(t)}}{(\tau_0 + it)^{\nu_\text{s}}}  = \frac{1}{(\nu_\text{s}-1) (\nu_\text{s}-2) (\nu_\text{s}-3) (\nu_\text{s} - 4)}\int_{-\infty}^\infty dt\, e^{\xi(t)} \frac{d^4}{dt^4} \frac{1}{(\tau_\text{s} + i t)^{\nu_\text{s} - 4}}\\
    & = \frac{1}{(\nu_\text{s}-1) (\nu_\text{s}-2) (\nu_\text{s}-3) (\nu_\text{s} - 4)} \int_{-\infty}^\infty dt\,   \frac{\left\{[\xi'(t)]^4 + 6 [\xi'(t)]^2 \xi''(t) + 3 [\xi''(t)]^2 + 4 \xi'(t) \xi'''(t) + \xi''''(t) \right\}}{(\tau_0 + i t)^{\nu_\text{s} - 4}} e^{\xi(t)}
     \\
    & \approx \frac{1}{(\nu_\text{s}-1) (\nu_\text{s}-2) (\nu_\text{s}-3) (\nu_\text{s} - 4)}\int_{-\infty}^\infty dt   \frac{ -4i bc (b^2 - c^2) \text{sgn}(t) + (b^4 - 6 b^2 c^2 + c^4) }{(\tau_0 + i t)^{\nu_s - 4}} e^{-b|t| + i c t}\\
    & = 2 \text{Re} \left[ e^{-i\pi \nu_\text{s}/2} \left( b - ic \right)^{ \nu_\text{s} - 1} \right] \Gamma(1 - \nu_\text{s}).
\end{aligned}
\label{eq:integral_single}
\end{equation}
The comparison of the last lines of Eqs.~\eqref{eq:integral_double} and \eqref{eq:integral_single} shows that the two results are related to each other by a simple replacement of the corresponding power-law exponents, i.e., $\nu_\text{d}$ and $\nu_\text{s}$.
This feature is not surprising, both physically and mathematically.
Physically, for both $n_0 > 1$ [e.g., $\nu_\text{d}$ and $\nu_\text{s}$ of Eqs.~\eqref{eq:integral_double} and \eqref{eq:integral_single}, respectively] and $n_0 < 1$ (corresponding to systems where anyons directly tunnel),
we are focusing on the braiding effects, for which $|t| \gg 1/\nu e V$ matters.
Thus, a universal result is expected, valid for all $n_0$ values, after removing the singularities $t\to 0$ with the integration-by-parts formalism.
Mathematically, as shown in Fig.~\ref{fig:xi_differentiations}, for all values of $n_0$, the integral is dominated by the product of first-order derivatives, thus leading to a uniform integral outcome. In other words, one can first calculate the integral for $n_0<1$, which converges at $t=0$ without any regularization, and the perform an analytical continuation to $n_0>0$ in the result.
As an important consequence, our analysis applies to general filling fractions $\nu$, although in Eqs.~\eqref{eq:integral_double} and \eqref{eq:integral_single}, we addressed $\nu = 1/3$.

\subsection{Tunneling current and tunneling current noises}

We are now ready to calculate tunneling current and tunneling noise, 
\begin{equation}
\begin{aligned}
        S_\text{T} & = e^2 \mathcal{T}^{(0)}_C \int dt \Big\langle \left\{ \Psi_B^\dagger (0) \Psi_A (0), \Psi_A^\dagger (t) \Psi_B (t) \right\} \Big\rangle_{\mathcal{T}^{(0)}_C = 0},\\
        I_\text{T} & = e \mathcal{T}^{(0)}_C \int dt \Big\langle \left[ \Psi_B^\dagger (0) \Psi_A (0), \Psi_A^\dagger (t) \Psi_B (t) \right] \Big\rangle_{\mathcal{T}^{(0)}_C = 0}.
    \end{aligned}
    \label{eq:it_st}
\end{equation}
Integrals in Eq.~\eqref{eq:it_st} can be evaluated with the correlation functions given by Eqs.~\eqref{eq:dan_resum} and \eqref{eq:dan_resum_b}, using the results of Sec.~\ref{sec:integral_over_t}.
Explicit expressions for the tunneling current and its noise are then given by (assuming $V_{sA} = V_{sB} = V$):
\begin{equation}
 I_\text{T}  = I_\text{T}^\text{single} + I_\text{T}^\text{collision},\quad S_\text{T}  = S_\text{T}^\text{single} + S_\text{T}^\text{collision},
\end{equation}
where
\begin{align}   
    I^\text{single}_\text{T} & = e \frac{\tau_0^{\nu_\text{s}-2}}{(2\pi)^2} \mathcal{T}_C^{(0)} 2\Gamma \left(1 - \nu_\text{s} \right) \left[ \mathcal{T}_A^{(0)} \text{Re}\left( e^{-i \pi\nu \nu_\text{s}/2} \left\{\frac{I_{A0}}{\nu e} \left[ 2i\sin(\pi\nu) e^{-i\pi\nu} -i \frac{\nu^2 e^2 V}{I_{A0}} \right]\right\}^{\nu_\text{s}-1} \right) \right. \notag \\
    & \left.- \mathcal{T}_B^{(0)} \text{Re}\left( e^{-i \pi\nu \nu_\text{s}/2} \left\{ \frac{I_{B0}}{\nu e} \left[ -2i\sin(\pi\nu) e^{i\pi\nu} + i \frac{\nu^2 e^2 V}{I_{B0}} \right]\right\}^{\nu_\text{s}-1} \right)\right], 
    \label{eq:tunneling_current_and_noise-Is}
    \\
    S^\text{single}_\text{T} & = e^2 \frac{\tau_0^{\nu_\text{s}-2}}{(2\pi)^2} \mathcal{T}_C^{(0)} 2\Gamma \left(1 - \nu_\text{s} \right) \left[\mathcal{T}_A^{(0)} \text{Re}\left( e^{-i \pi\nu \nu_\text{s}/2} \left\{\frac{I_{A0}}{\nu e} \left[ 2i\sin(\pi\nu) e^{-i\pi\nu} -i \frac{\nu^2 e^2 V}{I_{A0}} \right]\right\}^{\nu_\text{s}-1} \right) \right.\notag \\
    & \left.+ \mathcal{T}_B^{(0)}\text{Re}\left( e^{-i \pi\nu \nu_\text{s}/2} \left\{ \frac{I_{B0}}{\nu e} \left[ -2i\sin(\pi\nu) e^{i\pi\nu} + i \frac{\nu^2 e^2 V}{I_{B0}} \right]\right\}^{\nu_\text{s}-1} \right)\right],
    \label{eq:tunneling_current_and_noise-Ss}
    \end{align}
  and 
    \begin{align}
I_\text{T}^\text{collision} & = e \frac{2\tau_0^{\nu_\text{d}-2}}{ \pi } \mathcal{T}_A^{(0)} \mathcal{T}_B^{(0)} \mathcal{T}_C^{(0)}  \sin\left( \frac{\pi \nu_\text{d}}{2} \right) \Gamma (1 - \nu_\text{d})  \text{Im} \left\{ \left[\frac{I_{A0}}{\nu e} \left( 1 - e^{-2i\pi\nu} \right) + \frac{I_{B0}}{\nu e} \left( 1 - e^{2i\pi\nu} \right)\right]^{\nu_\text{d} - 1} \right\} ,
\label{eq:tunneling_current_and_noise-Ic}
\\
S_\text{T}^\text{collision} & = e^2 \frac{2\tau_0^{\nu_\text{d}-2}}{ \pi } \mathcal{T}_A^{(0)} \mathcal{T}_B^{(0)} \mathcal{T}_C^{(0)}   \cos\left( \frac{\pi \nu_\text{d}}{2} \right) \Gamma(1 - \nu_\text{d}) \text{Re} \left\{ \left[\frac{I_{A0}}{\nu e} \left( 1 - e^{-2i\pi\nu} \right) + \frac{I_{B0}}{\nu e} \left( 1 - e^{2i\pi\nu} \right)\right]^{\nu_\text{d} - 1} \right\}.
\label{eq:tunneling_current_and_noise-Sc}
\end{align}
Here, $I_\text{T}^\text{single}$ and $S_\text{T}^\text{single}$ are single-source tunneling current and corresponding noise, respectively, and 
$I_\text{T}^\text{collision}$ and $S_\text{T}^\text{collision}$ refer to extra contributions from two-particle collisions, when both sources are on.
In Eqs.~\eqref{eq:tunneling_current_and_noise-Is}-\eqref{eq:tunneling_current_and_noise-Sc}, the power-law exponents $\nu_\text{s}$ and $\nu_\text{d}$ are related to the tunneling scaling dimensions in the single-source and double-source collision contributions, respectively.

The single-source terms, Eqs.~\eqref{eq:tunneling_current_and_noise-Is} and \eqref{eq:tunneling_current_and_noise-Ss}, can be simplified in the strongly diluted limit  $I_{A0}, I_{B0} \ll e^2 V$:
\begin{align}
    I^\text{single}_\text{T} & \simeq e \frac{\tau_0^{\nu_\text{s}}}{(2\pi\tau_0)^2}  2\Gamma \left(1 - \nu_\text{s} \right) (\nu e V)^{\nu_\text{s} - 1}\,\mathcal{T}_C^{(0)} \notag \\
& \left(\left\{ \sin\left( \nu_\text{s} \pi \right) + 2 \nu (\nu_\text{s} - 1) \sin (\pi\nu) \left(\sin \left[ ( \nu_\text{s} - \nu ) \pi\right] + \sin (\pi\nu) \right) \frac{I_{A0}}{\nu^3 e^2 V} \right\} \mathcal{T}_A^{(0)} \right.\notag \\
&\left. -\left\{\sin\left( \nu_\text{s} \pi \right) + 2 \nu (\nu_\text{s} - 1) \sin (\pi\nu) \left(\sin \left[ ( \nu_\text{s} - \nu ) \pi\right] + \sin (\pi\nu) \right) \frac{I_{B0}}{\nu^3 e^2 V} \right\} \mathcal{T}_B^{(0)} \right) 
\label{eq:single_source_simplification-I}
\\
S^\text{single}_\text{T} & = e^2 \frac{\tau_0^{\nu_\text{s}}}{(2\pi\tau_0)^2}  2\Gamma \left(1 - \nu_\text{s} \right) (\nu e V)^{\nu_\text{s} - 1}\, \mathcal{T}_C^{(0)} \notag \\
& \left(\left\{ \sin\left( \nu_\text{s} \pi \right) + 2 \nu (\nu_\text{s} - 1) \sin (\pi\nu) \left(\sin \left[ ( \nu_\text{s} - \nu ) \pi\right] + \sin (\pi\nu) \right) \frac{I_{A0}}{\nu^3 e^2 V} \right\} \mathcal{T}_A^{(0)} \right.\notag \\
&\left. +\left\{\sin\left( \nu_\text{s} \pi \right) + 2 \nu (\nu_\text{s} - 1) \sin (\pi\nu) \left(\sin \left[ ( \nu_\text{s} - \nu ) \pi\right] + \sin (\pi\nu) \right) \frac{I_{B0}}{\nu^3 e^2 V} \right\} \mathcal{T}_B^{(0)} \right) .
\label{eq:single_source_simplification}
\end{align}
This simplification, however, does not apply to double-source quantities $I_\text{T}^\text{collision}$ and $S_\text{T}^\text{collision}$, as they depend only on non-equilibrium currents (but not explicitly on $V$).
Physically, this occurs since the frequency of two-particle collisions is determined by currents in two non-equilibrium channels.

Since the bare transmission probabilities are not accessible in experiment, we rewrite Eq.~\eqref{eq:single_source_simplification} in terms of the measurable transmission probabilities:
\begin{align}
    I^\text{single}_\text{T} & = \mathcal{T}_C \left\{\frac{I_{A0}}{\nu} \left[ 1- \frac{ f_1 (\nu)\mathcal{T}_A  }{\pi\nu \sin\left( \pi \nu_\text{s} \right) +2 f_1 (\nu) \mathcal{T}_A  }  \right] - \frac{I_{B0}}{\nu} \left[ 1- \frac{ f_1 (\nu)\mathcal{T}_B  }{\pi\nu^2 \sin\left( \pi \nu_\text{s} \right) +2 f_1 (\nu) \mathcal{T}_B  }  \right] \right\},
    \label{eq:single_leading_order2-I}\\
    S_\text{T}^\text{single} & = e\mathcal{T}_C \left\{\frac{I_{A0}}{\nu} \left[ 1- \frac{ f_1 (\nu)\mathcal{T}_A  }{\pi\nu \sin\left( \pi \nu_\text{s} \right) +2 f_1 (\nu) \mathcal{T}_A  }  \right] + \frac{I_{B0}}{\nu} \left[ 1- \frac{ f_1 (\nu)\mathcal{T}_B  }{\pi\nu^2 \sin\left( \pi \nu_\text{s} \right) +2 f_1 (\nu) \mathcal{T}_B  }  \right] \right\},
    \label{eq:single_leading_order2}
    \end{align}
where 
\begin{equation}
    f_1(\nu) \equiv (\nu_\text{s} - 1) \sin (\pi\nu) \big\{\sin \left[\pi (\nu_\text{s} -\nu)\right] + \sin (\pi\nu) \big\},
\end{equation}
and 
$\mathcal{T}_C = \nu \partial_{I_0} I_\text{T} (I_0,0) = - \nu \partial_{I_0} I_\text{T} (0, I_0)$ refers to the transmission probability through the central collider for the single-source case.
Again, Eqs.~\eqref{eq:single_leading_order2-I} and \eqref{eq:single_leading_order2} keep only terms to leading order of diluter transmissions $\mathcal{T}_A$ and $\mathcal{T}_B$.
As a reminder, these measurable transmission probabilities are related to tramission amplitude squares, i.e., $\mathcal{T}_A^{(0)}$ and $\mathcal{T}_B^{(0)}$, via the relation
\begin{equation}
\begin{aligned}
    \mathcal{T}_A & = \frac{I_{A0}}{\nu V} \frac{2\pi}{e^2} = \mathcal{T}^{(0)}_A \nu \tau_0^{2\nu-2} \sin(2\pi\nu) \Gamma (1-2\nu) (\nu e V)^{2\nu-2}/\pi \notag,\\
    \mathcal{T}_B & = \frac{I_{B0}}{\nu V} \frac{2\pi}{e^2} = \mathcal{T}^{(0)}_B \nu \tau_0^{2\nu-2} \sin(2\pi\nu) \Gamma (1-2\nu) (\nu e V)^{2\nu-2}/\pi \notag,
\end{aligned}
\end{equation}
such that $\mathcal{T}_A \propto \mathcal{T}^{(0)}_A$ and $\mathcal{T}_B \propto \mathcal{T}^{(0)}_B$, a fact that has been mentioned in Methods of the main text.
To obtain Eq.~\eqref{eq:single_leading_order2}, we have used the fact that to the leading order of transmission at diluters [cf. Eq.~\eqref{eq:single_source_simplification-I}], transmission through the central collider can be obtained via (assuming $\mathcal{T}_B^{(0)} = 0$, without loss of generality)
\begin{equation}
\begin{aligned}
    \mathcal{T}_C& \equiv \nu \partial_{I_{A0}} I_\text{T} (I_{A0},0) \\
    & \simeq e \frac{\tau_0^{\nu_\text{s}}}{(2\pi\tau_0)^2}  2\Gamma \left(1 - \nu_\text{s} \right) (\nu e V)^{\nu_\text{s} - 1}\,\mathcal{T}_C^{(0)}  \left[ \sin\left( \nu_\text{s} \pi \right) + 4 \nu f_1(\nu) \frac{I_{A0}}{\nu^3 e^2 V} \right] \frac{\partial}{\partial I_{A0}} \mathcal{T}_A^{(0)}\notag \\
    & = e \frac{\tau_0^{\nu_\text{s}}}{(2\pi\tau_0)^2}  2\Gamma \left(1 - \nu_\text{s} \right) (\nu e V)^{\nu_\text{s} - 1}\,\mathcal{T}_C^{(0)}  \left[ \sin\left( \nu_\text{s} \pi \right) +  f_1(\nu) \frac{2}{\pi \nu} \mathcal{T}_A \right] \frac{\nu e\tau_0^{2\nu-2} \sin(2\pi\nu) \Gamma (1-2\nu) (\nu e V)^{2\nu-1}}{2\pi^2}\notag,
\end{aligned}
\end{equation}
following which $\mathcal{T}_C\sim T_C^{(0)} \left[ \sin\left( \pi\nu_\text{s}  \right) +  \frac{2  f_1(\nu)}{\pi \nu} \mathcal{T}_A \right]$, and further simplifies into $\mathcal{T}_C \propto T_C^{(0)}$ in the strongly diluted limit where $\mathcal{T}_A \ll 1$: another conclusion mentioned in Methods of the main text.

Resummation of all orders in diluter transmissions [cf. Eqs.~\eqref{eq:tunneling_current_and_noise-Is}-\eqref{eq:tunneling_current_and_noise-Sc}], 
in the single-source contributions to the current and noise 
yields Eqs.~(10) and Eq.~(2) of the main text:
\begin{align}
    I_\text{T}^\text{single} \! = &\text{Re}\left\{\mathcal{T}_A \mathcal{T}_C \frac{\nu e^2 V}{2}\frac{ (2\pi \nu)^{1-\nu_\text{s}}\, e^{i\pi(1+\nu_\text{s}-\nu \nu_s)} }{ \pi\nu\sin (\pi\nu_\text{s}) + 2 f_1(\nu) \mathcal{T}_A}     
 \left[ 2i\pi \nu -  \mathcal{T}_A\left( 1 - e^{-2i\pi\nu} \right)  \right]^{\nu_\text{s}-1} \right\}
    - \left\{A\to B\right\},
    \label{eq:current_single_full}\\
    S_\text{T}^\text{single} &\! = \text{Re}\left\{\mathcal{T}_C\, \mathcal{T}_A \frac{\nu e^3 V}{2} \frac{ (2\pi \nu)^{1-\nu_\text{s}} e^{i\pi (1 + \nu_\text{s} -  \nu \nu_\text{s} )}}{ \pi\nu\sin (\pi\nu_\text{s}) + 2 f_1(\nu) \mathcal{T}_A}  
 \left[ 2i\pi \nu -  \mathcal{T}_A\left( 1 - e^{-2i\pi\nu} \right)  \right]^{\nu_\text{s}-1} \right\}  + \left\{A\to B\right\}.
  \label{eq:noise_single_full}
    \end{align}
The double-source collision contributions can be similarly written in terms of observables, leading to  Eqs.~(11) and Eq.~(3) of the main text:
\begin{align}
    I_\text{T}^\text{collision}& =  \frac{e^2 V \sqrt{\mathcal{T}_A \mathcal{T}_B}\, \mathcal{T}_C\, f_2(\nu) \sin \left(\pi \nu_\text{d}/{2} \right)}{\pi\nu\sin\left( \pi \nu_\text{s} \right) + 2 f_1 (\nu) \sqrt{\mathcal{T}_A \mathcal{T}_B}  }    \text{Im} \left\{ \left[\mathcal{T}_A \left( 1 - e^{-2i\pi\nu} \right) + \mathcal{T}_B \left( 1 - e^{2i\pi\nu} \right)\right]^{\nu_\text{d} - 1} \right\},
    \label{I-coll-nice}\\
    S_\text{T}^\text{collision}& =  \frac{e^3 V \sqrt{\mathcal{T}_A \mathcal{T}_B}\, \mathcal{T}_C\, f_2(\nu) \cos\left(\pi \nu_\text{d}/{2} \right) }{\pi\nu\sin\left( \pi \nu_\text{s} \right) + 2 f_1 (\nu) \sqrt{\mathcal{T}_A \mathcal{T}_B}  }    \text{Re} \left\{ \left[\mathcal{T}_A \left( 1 - e^{-2i\pi\nu} \right) + \mathcal{T}_B \left( 1 - e^{2i\pi\nu} \right)\right]^{\nu_\text{d} - 1} \right\},
    \label{S-coll-nice}
\end{align}
with 
\begin{equation}
    f_2 (\nu) \equiv \frac{4\pi^3\, (2\pi\nu)^{1-\nu_\text{d}} \Gamma(1 - \nu_\text{d}) }{\sin(2\pi\nu)\, \Gamma (1 - 2\nu)\, \Gamma\left(1-\nu_\text{s} \right)}.
\end{equation}

Before closing this section, we stress that $S_\text{T}^\text{collision} $ vanishes when $\nu = 1$, thus leading to a vanishing Andreev entanglement pointer in fermionic systems, following its definition, Eq.~(1) of the main text. It is thus different from the entanglement pointer introduced in Ref.~\cite{GuNC24S} for fermionic systems [see discussion below Eq.~\eqref{eq.S10}]. 
This fact, importantly, indicates that the entanglement pointer defined in the present work is in direct and close connection to the anyon-quasihole braiding that uniquely exists in anyonic systems in the Andreev-like tunneling limit. 
In addition, it unveils the topology of the system, considering the necessary connection between any braiding features and topology.

\section{{\,\,\,\,\,\,\,}Finite-temperature expressions}

In the main text, we assume that both sources are biased by the fixed voltage $V$ with respect to the two middle edges $A$ and $B$.
With this assumption, single-source measurement can be obtained by pinching off one of two diluters, which tunes either $\mathcal{T}^{(0)}_A$ or $\mathcal{T}^{(0)}_B$ to be zero.

In real experiments, the single-source correlation measurement can be alternatively performed by turning off either source, i.e., keeping the corresponding source grounded.
This option (i.e., taking zero voltage bias), however, cannot be described by Eqs.~\eqref{eq:dan_resum} and \eqref{eq:dan_resum_b}, because of the divergence of the non-equilibrium current [cf. Eq.~\eqref{eq:ia0_expression}] in the $V \to 0$ limit (as $\nu \le 1/3$ for Laughlin systems).
Consequently, when $V \to 0$, a finite temperature must be assumed to avoid the divergence of the current and to keep the diluter in the limit of weak (but finite) anyonic tunneling.
In this section, we consider the case of finite temperature $T$. We assume that (i) $T$ is much smaller than the corresponding bias if the source is on and (ii) $T$ is also large enough to keep the diluter in the anyonic tunneling limit, in which anyons are allowed to tunnel through the diluting QPC.
Notice that measurements at a finite temperature, or even with temperature difference, is capable of disclosing anyonic statistical feature~\cite{MartinDeltaT20S}, owing to the connection between delta-$T$ noise (noise induced by a temperature difference) and operator scaling dimension, which is a function of the filling factor, see Refs.~\cite{GuPRB22S, KyryloPRB22S, veillon2024S,schiller2024scalingS}.

\subsection{Modifications of the non-equilibrium current}

The inclusion of a finite temperature introduces two modifications: a modification of the non-equilibrium current through diluters and the modification of contour integrals.
At finite temperatures, the non-equilibrium current through a diluter involves the following integral:
\begin{equation}
    (\pi   T)^{2\nu} \int dt \frac{e^{i\frac{\nu e V}{\hbar} t}}{\sin^{2\nu} [\pi   T (\tau_0 + i t)  ]} = (2\pi   T)^{2\nu - 1} \frac{\hbar}{2\pi \Gamma (2\nu)} e^{\frac{\nu e V}{4\pi   T}} \Bigg| \Gamma \left(\nu + \frac{i\nu e V}{4\pi^2   T}\right) \Bigg|^2,
\end{equation}
where $2\nu < 1$ is assumed, as in the main text. This integral, which refers to the current from one source, was addressed, in particular, in Ref.~\cite{CampagnanoPRB16S}. With this integral, the leading-order currents that enter the two middle edges become
\begin{equation}
\begin{aligned}
    I_{A0,B0} (V_{sA,sB}, T) & = \frac{e^2}{\tau_0} \frac{\mathcal{T}^{(0)}_{A,B}}{4\pi^2}
\left(\frac{2\pi   T \tau_0}{\hbar}\right)^{2\nu-1} \frac{\nu }{\pi \Gamma (2\nu)} \sinh\left( \frac{\nu e V_{sA,sB}}{2   T} \right) \Bigg| \Gamma \left(\nu + \frac{i\nu e V_{sA,sB}}{2\pi   T}\right) \Bigg|^2 \equiv \nu \frac{e^2 V}{2\pi} \mathcal{T}_{A,B}  ,\\
\mathcal{T}_{A,B} & = \frac{1}{\tau_0} \frac{\mathcal{T}^{(0)}_{A,B}}{2\pi V}
\left(\frac{2\pi   T \tau_0}{\hbar}\right)^{2\nu-1} \frac{1}{\pi \Gamma (2\nu)} \sinh\left( \frac{\nu e V_{sA,sB}}{2   T} \right) \Bigg| \Gamma \left(\nu + \frac{i\nu e V_{sA,sB}}{2\pi   T}\right) \Bigg|^2 ,
\end{aligned}
\label{eq:finit-temperature_current}
\end{equation}
where $V_{sA}$ and $V_{sB}$ refer to the bias in sources $sA$ and $sB$, respectively, and the second line of Eq.~\eqref{eq:finit-temperature_current} refers to the experimentally accessible transmission probability at finite temperature.
We can briefly capture the current features by checking the asymptotic scaling of the function that depends on $\nu e V_{sA,sB}/  T$, i.e.,
\begin{align}
\sinh (x) \big|\Gamma (\nu + i x/\pi) \big|^2  \propto \begin{cases} \ \  x, & \text{if $x \ll 1$}, \\
\ \  x^{2\nu - 1}, & \text{if $x\gg 1$}. \end{cases}
\end{align}
Following the asymptotic features above, 
$$I_{A0,B0} \propto \mathcal{T}^{(0)}_{A,B} (e V_{sA,sB})^{2\nu - 1}$$
for $\nu e V_{sA,sB} \gg 2 T$, in agreement with Eqs.~(10) and (11) of the main text. In the opposite limit $\nu e V_{sA,sB} \ll 2   T$, we get $$I_{A0,B0} \propto \mathcal{T}^{(0)}_{A,B} (  T)^{2\nu - 2} e V_{sA,sB}.$$
In both limits, the current equals the product of $e V_{sA,sB}$, and the renormalization factor $[\max(\nu e V_{sA,sB}, 2   T)]^{2\nu - 2}$, in agreement with the scaling analysis for anyonic tunneling through a QPC.

\subsection{Finite-temperature contour integral}

The introduction of finite temperature also modifies the contour integral. In the present subsection, we focus, without loss of generality, on subsystem $\mathcal{A}$.
These constant factors will be included when showing the final results.
The finite-temperature situation involves two integrals below
\begin{equation}
\begin{aligned}
  &   \int ds_1 \frac{e^{-i\nu e V_{sA} s_1}}{\sinh \left\{ \pi T[i \tau_0 \chi_{+\eta_1} (-s_1) - (-s_1 - L) ]\right\}} \int ds_2 \frac{e^{i\nu e V_{sA} s_2}}{\sinh \left\{ \pi T [i \tau_0 \chi_{-\eta_2} (t-s_2) - (t-s_2 - L) ] \right\}}\\
 & =   \int ds_1 \frac{e^{i\nu e V_{sA} s_1}}{\sinh [ \pi T (s_1 - i \tau_0 \eta_1) ]} \int ds_2 \frac{e^{-i\nu e V_{sA} s_2}}{\sinh [ \pi T (s_2 - i \tau_0 \eta_2) ]},
\end{aligned}
\end{equation}
where we have shifted $s_1 \to - s_1 - L$, $s_2 \to - s_2 + t - L$, and taken the large-$L$ limit for the second line.
These integrals contain poles at $s_1 = i\tau_0\eta_1 + n_1/T$ and $s_2 = i\tau_0\eta_2 + n_2/T$, where $n_1$ and $n_2$ are integers, with their value ranges determined by $\eta_1$ and $\eta_2$.
Indeed, since $V_A > 0$, integrals over $s_1$ and $s_2$ include the upper and lower half-planes, respectively. As a consequence, $n_1 \ge 0$ if $\eta_1 = 1$, and $n_1 \ge 1$ otherwise; $n_2 \le 0$ if $\eta_2 = -1$, and $n_2 \le -1$ otherwise.
In contrast to the zero-temperature case, now $\eta_1$ and $\eta_2$ can take both values, as thermal fluctuations allow (exponentially suppressed) tunneling from $A$ to $sA$.

Since the poles $s_1$ and $s_2$ contain integer multiples of $1/T$, the contour integrals will be expressed in terms of series over $n_1$ and $n_2$.
Now, the correlator $D_{A1}$ is evaluated as
\begin{equation}
\begin{aligned}
    D_{A1} & = -\frac{\mathcal{T}^{(0)}_A}{(2\pi \tau_0)^3} \sum_{\eta_1\eta_2} \eta_1\eta_2 \iint ds_1 ds_2 \frac{(\pi T \tau_0)^{\frac{1}{\nu} + 2\nu} e^{-i\nu e V_{sA}(s_1 - s_2)} }{\sin^{1/\nu}[\pi T (\tau_0 + it)] \sin^{2\nu}\left\{ 2\pi^2 T [\tau_0 + i (s_1 - s_2) \chi_{\eta_1\eta_2} (s_1 - s_2)] \right\} }  \\
    & \times \frac{\sin\left\{ \pi T[\tau_0 + i (t - s_1 - L) \chi_{-\eta_1} (t-s_1)] \right\} \sin \left\{\pi T[\tau_0 + i (-s_2 - L) \chi_{+\eta_2} (-s_2)] \right\} }{ \sin \left\{ \pi T[\tau_0 + i (t - s_2 - L) \chi_{-\eta_2} (t-s_2)] \right\} \sin\left\{\pi T [\tau_0 + i (-s_1 - L) \chi_{+\eta_1} (-s_1)] \right\}}\\
    & = \frac{\mathcal{T}^{(0)}_A}{(2\pi \tau_0)^3} \sum_{\eta_1\eta_2} \eta_1\eta_2 \frac{(\pi T \tau_0)^{\frac{1}{\nu} + 2\nu}}{\sin^{\frac{1}\nu - 1} [\pi T (\tau_0 + it)] }e^{i\nu e V_{sA} t}\\
    & \times \iint ds_1 ds_2 \frac{e^{i\nu e V_{sA} ( s_1 - s_2)} \chi_{\eta_1\eta_2} (s_2 - s_1 - t) \sin^{1-2\nu} \left\{ [\pi T (\tau_0 + i (s_2-s_1 - t) \chi_{\eta_1\eta_2}(s_2 - s_1 - t)  ]\right\}}{\sinh [ \pi T (s_1 - i \tau_0 \eta_1) ]\sinh [ \pi T (s_2 - i \tau_0 \eta_2) ]}\\
    & = \frac{\mathcal{T}^{(0)}_A}{(2\pi \tau_0)^3} \sum_{\eta_1\eta_2} \eta_1\eta_2 \frac{(\pi T \tau_0)^{\frac{1}{\nu} + 2\nu}}{\sin^{\frac{1}\nu - 1} [\pi T (\tau_0 + it)] }e^{i\nu e V_{sA} t} \frac{4\pi^2}{(\pi T)^2} (-\eta_1) \sin^{1-2\nu} (i \pi T t \eta_1) \\
    & \times \sum_{n_1 = (1-\eta_1)/2}^\infty   \sum_{n_2 = (1 + \eta_2)/2}^\infty  e^{-\frac{\nu e V_{sA}}{ T} n_1} e^{-\frac{\nu e V_{sA}}{ T} n_2}\,  \theta^{1-2\nu} (n_1 + n_2) \\
    & = \frac{\mathcal{T}^{(0)}_A}{2\pi \tau_0} \frac{(\pi T \tau_0)^{\frac{1}{\nu} + 2\nu-2}}{\sin^{\frac{1}\nu + 2 \nu - 2} [\pi T (\tau_0 + it)] }e^{i\nu e V_{sA} t} \frac{\left[1 - \exp\left( -{\nu e V_{sA}}/{ T} \right)\right] \left[1 - (-1)^{-2\nu} \exp\left( -{\nu eV_{sA}}/{ T} \right)\right]}{ 1 + \exp\left( -{2\nu eV_{sA}}/{ T} \right)},
\end{aligned}
\label{eq:finite_t_da1}
\end{equation}
where $\theta (n) = 1 $ if $n$ is even, and equals $-1$ if $n$ becomes odd.
Equation~\eqref{eq:finite_t_da1} transforms into the zero-temperature expression when $V_{sA} \gg T$, and becomes proportional to $V_{sA}/T$ in the opposite limit.
This fact, in combination with Eq.~\eqref{eq:finit-temperature_current} for the non-equilibrium current, indicates that in the $V_{sA}\ll T$ limit (i.e., when source $sA$ is turned off),
one should replace the ratios $I_{A0,B0}/(\nu e V_{sA, sB})^{2\nu - 1}$ appearing in the zero-$T$ formulas with the modified expression $I_{A0,B0}/(2\pi T)^{2\nu - 1}$.

Based on the expressions above, we conclude that one can obtain the single-source contribution by tuning one of the source voltage biases ($V_{sA}$ or $V_{sB}$) to zero.
It is equivalent to pinching off the corresponding diluter (i.e., setting $\mathcal{T}^{(0)}_A$ or $\mathcal{T}^{(0)}_B$ to zero), as suggested in Eqs.~(2) and (3) in the main text.

\section{{\,\,\,\,\,\,\,}Assessing the role of interactions}
\label{Sec:SIII}

In the main text, we comment that the entanglement pointer $\mathcal{P}_\text{Andreev}$ has a strong resilience to interaction effects. Contrary to contributions to correlation functions due to intra-edge interactions among same-edge channels, such correlations are (to leading order) canceled out in the expression for the entanglement pointer.
In this section, we demonstrate the resilience to interaction effects by considering two types of screened Coulomb interaction, the one that couples the charge density in edge $A$ with the charge density in edge $B$ (i.e., inter-edge interaction); and that among charge densities of modes that belong to the same edge (i.e., intra-edge interaction).

\subsection{Influence of inter-edge interaction on correlation functions}

\begin{figure}  \includegraphics[width=0.9\linewidth]{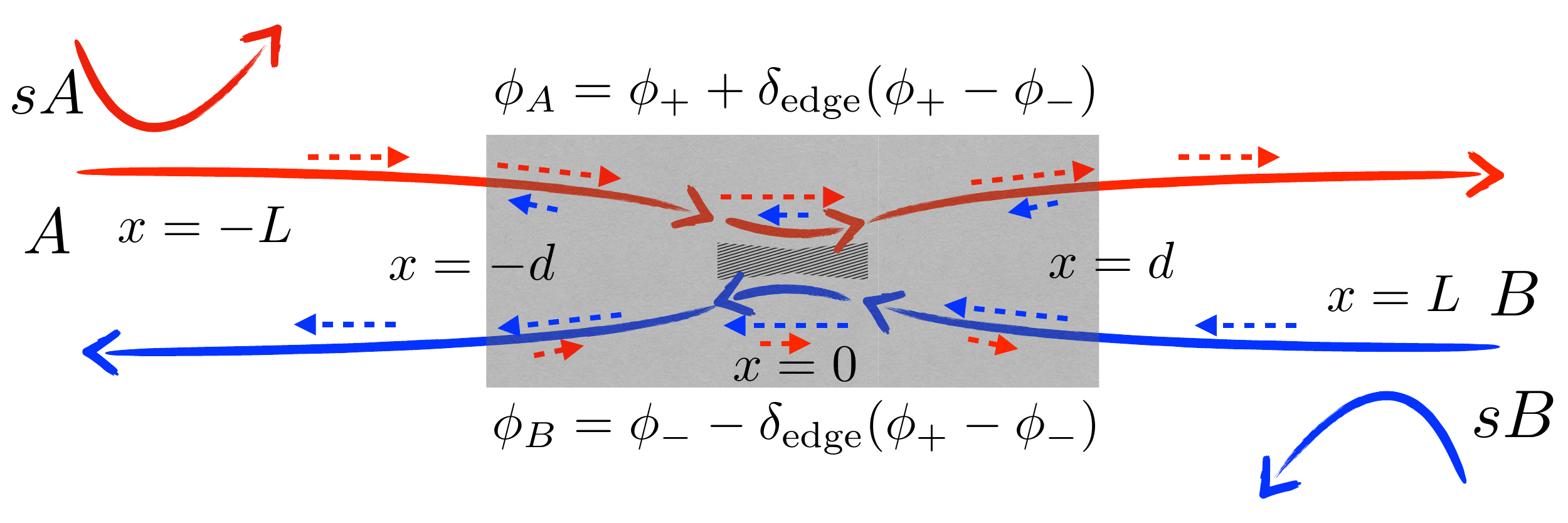}
  \caption{The schematics of the model with interaction between the two edges. We choose the convention of spatial coordinate such that $x$ increases from left to right. Two diluters and the central QPC are placed at $x = \pm L$ and $x = 0$, respectively.
  Within the area $-d \le x \le d$ (indicated by the shadowed gray box), particles in edge $A$ interact with particles in edge $B$, with the interaction strength quantified by the Luttinger parameter $K$.
  Outside the interacting area, the bosonic fields $\phi_A$ and $\phi_B$ equal the right-going and left-going chiral fields $\phi_+$ (the red dashed arrow) and $\phi_-$ (the blue dashed arrow), respectively.
  Within the interacting area, $\phi_A$ and $\phi_B$ are linear combinations of $\phi_+$ and $\phi_-$.
  }
  \label{fig:ll}
\end{figure}

In this section, we consider the model of Fig.~\ref{fig:ll}, where we introduce inter-edge Coulomb interaction, i.e., that between edges $A$ and $B$ only in the shadowed area (i.e., $-d \le x \le d$ in Fig.~\ref{fig:ll}, with $d < L$).
For simplicity, we further assume a constant interaction (quantified by the Luttinger liquid parameter $K$) within the interacting area ( in the experiment, this would correspond to the effect of screening of the long-range Coulomb repulsion by gates).

We point out that here we opt for a different convention of denoting the spatial coordinates: the two diluters are placed at $x = \pm L$, and the central QPC is located at $x = 0$.
Indeed, if we keep the convention employed in the other sections (i.e., x increases in the downstream direction of each mode), we end up with (formally)  
non-local interactions.

Within the interacting region, Luttinger-type interactions are easily incorporated within the bosonization approach.
For our later convenience, we follow Ref.~\cite{GiamarchiBookS}, and use canonical fields to bosonize \textit{fermionic} operators $\mathbf{\Psi}_A$ and $\mathbf{\Psi}_B$, 
\begin{equation}
\mathbf{\Psi}_A (x) = \frac{e^{i k_F x} e^{i [\theta (x ) - \phi (x) ]}}{\sqrt{2\pi a}} ,\ \quad \mathbf{\Psi}_B (x) = \frac{e^{-i k_F x} e^{i [\theta (x ) + \phi (x) ]}}{\sqrt{2\pi a}} ,
\end{equation}
where canonical phases follow the standard commutator $[ \phi(x), \partial_{x'}\theta (x') ] = i\pi \delta (x' - x)$, and are related to our original fields following $\phi_A = \theta - \phi$ and $\phi_B = \theta + \phi$.
Without interaction, $\phi_A$ and $\phi_B$ are the right-going and left-going modes in edges $A$ and $B$, respectively.
However, they are no longer chiral modes in the interacting area. Indeed, now the left-going and right-going chiral modes become $\phi_\pm = K \theta \mp \phi$, where $K$ refers to the Luttinger liquid parameter.
In the interacting area, edge fields and chiral fields are connected via
\begin{equation}
\phi_A (x) = \phi_+ (x) + \left( \frac{1}{2 K} -\frac{1}{2} \right) [\phi_+ (x) + \phi_- (x)],\ \quad \phi_B = \phi_- (x) + \left( \frac{1}{2 K} -\frac{1}{2} \right) [\phi_+ (x) + \phi_- (x)].
\label{eq:interacting_fields}
\end{equation}

For later convenience, we further define
\begin{equation}
\delta_\text{edge} \equiv \frac{1}{2 K } - \frac{1}{2},
\end{equation}
as the parameter that quantifies the effective difference from a non-interacting situation.
In addition to Eq.~\eqref{eq:interacting_fields} for the rotation of fields, interaction also modifies the quasiparticle velocity.
Indeed, following Ref.~\cite{GiamarchiBookS}, within the interacting area, the velocity $u$ and the Luttinger liquid parameter $K$ are related to the inter-edge interaction, following (after taking $v_F = 1$)
\begin{equation}
\begin{aligned}
    u = \sqrt{1 - (g_2/2)^2},\ \ K = \sqrt{\frac{1 - g_2/2}{1 + y_2/2}},
\end{aligned}
\end{equation}
where $g_2$ refers to the strength of the inter-edge Coulomb interaction (interaction between counterpropagating bare modes).
We can then express the ``plasmon'' velocity in the interacting area in terms of $K$,
\begin{equation}
    u = \frac{2K}{ 1 + K^2} \approx 1 - \frac{(K-1)^2}{2},
\end{equation}
where we have taken the weak-interaction assumption ($|K-1| \ll 1$) to expand $u$ to leading order in interaction.
In comparison to $\delta_\text{edge} \approx (1 - K )/2$ that is linear to $(1-K)$, 
the leading interaction-induced modification of the velocity is quadratic in $(1-K)$, underscoring a comparatively smaller correction from weak interactions.
In this section, we thus approximately take $u = 1$ in our calculation.

As we introduce sharp boundaries $x = \pm d$ that abruptly separate interacting and non-interacting areas, boundary conditions should be installed at these two boundaries.
These boundary conditions describe the Fresnel scattering of bosonic modes at the interfaces separating two media with different ``optical'' properties. This type scattering gives rise to fractionalization of the charge excitations at the interfaces~\cite{SafiSchulzPRB95S, Safi99S, ProtopopovGefenMirlinAoP17S, SpanslattPRB21S}.

Briefly, since edges $A$ and $B$ are spatially separated, we can enforce current conservation in each edge separately, at different boundaries.
For the boundary at $x = -d$, the incoming current in edge $A$ equals $\partial_x \phi_A/(2\pi)$.
This current should be equal to the current in edge $A$, right into the interacting area.
The current conservation requires the knowledge of current operators inside and outside of the interacting area.
More specifically, outside the interacting area, $\phi_{A,B} (x,t) = \phi_\pm (x \mp t)$, meaning that the current operator $$\hat{I}_{A,B} (|x| > d) = -\partial_t \phi_\pm (x\mp t)/2\pi = \pm \partial_x \phi_\pm/2\pi$$ 
(as a reminder, we take Fermi velocity $v_F = 1$ for simplicity in this work).
Inside the interacting area, instead $\phi_{A,B} (x,t) = (1 + \delta_\text{edge}) \phi_\pm (x \mp t) + \delta_\text{edge} \phi_\mp (x \pm t)$, leading to a modified current operator (we recall that we have neglected the difference of $u$ from 1, which is quadratic in the interaction strength)
\begin{equation}
\begin{aligned}
    \hat{I}_{A,B} (|x| < d) & = - \frac{1+\delta_\text{edge}}{2\pi}\, \partial_t \phi_\pm (x \mp t) - \frac{\delta_\text{edge}}{2\pi}\, \partial_t \phi_\mp (x \pm t)\\
    & =  \pm \frac{1+\delta_\text{edge}}{2\pi}\,  \partial_x \phi_\pm (x \mp t) \mp  \frac{\delta_\text{edge}}{2\pi}\, \partial_x \phi_\mp (x \pm t).
\end{aligned}
\end{equation}
Current conservation then leads to the following relations between the phases at the interfaces:
\begin{equation}
\begin{aligned}
& \phi_A (-d^-) = \phi_+ (-d^+) + \delta_\text{edge} [\phi_+ (-d^+) - \phi_- (-d^+) ],\\
& \phi_B (d^+) = \phi_- (d^-) + \delta_\text{edge} [ -\phi_+ (d^-) + \phi_- (d^-) ],
\label{eq:bcs}
\end{aligned}
\end{equation}
where superscript $\pm$ in $d^\pm$ labels the right and left sides, respectively, of a given boundary.
Since $\phi_{A,B}$ are free chiral fields in the non-interacting regions, with expressions of Eq.~\eqref{eq:bcs}, one can keep track of the positions of $\phi_\pm$ fields at earlier time moments, to express fields at diluters as
\begin{equation}
\begin{aligned}
\phi_A (-L, s) & = \phi_A (-d, s + L - d) = (1 + \delta_\text{edge}) \phi_+ ( -d,  s + L - d)  - \delta_\text{edge} \phi_- ( -d,  s + L - d),\\
\phi_B (L, s) & = \phi_B (d, s + L - d) = (1 + \delta_\text{edge}) \phi_- (d, s + L - d)  - \delta_\text{edge} \phi_+ (d, s + L - d).
\end{aligned}
\label{eq:diluter_fields}
\end{equation}

For further convenience, it is useful to imagine an auxiliary wire where the interaction region would be extended to the positions of diluters. In the interacting part of our setup, $|x|<d$, the chiral fields are equivalent to those in the auxiliary one: $\phi_\pm(x,t)=\tilde\phi_\pm(x,t)$. 
In the auxiliary system, we can further use  $\tilde\phi_+(-d,s+L-d)=\tilde\phi_+(L,s)$ and $\tilde\phi_-(-d,s+L-d)=\tilde\phi_-(-2d+L,s)$. Thus, the fields $\phi_{A,B}(-L,s)$
in the noninteracting parts of our setup near the diluters can be replaced by the combinations of the chiral fields $\tilde\phi_\pm$ of the virtual wire, where interaction is everywhere:  
\begin{align}
    \phi_A (-L, s)& \to (1 + \delta_\text{edge}) \tilde\phi_+ ( -L,  s )  - \delta_\text{edge} \tilde\phi_- ( -2d + L,  s ),\\
  \phi_B (L, s)   
& \to (1 + \delta_\text{edge}) \tilde\phi_- ( L,  s)  - \delta_\text{edge} \tilde\phi_+ (2d - L , s).
\end{align}  
Equations~\eqref{eq:diluter_fields} indicate that although field operators at two diluters are non-interacting, they can be written in terms of two counter-propagating fields of the auxiliary wire.
This substitution of fields is taken since $\phi_A$ and $\phi_B$ are not independent fields at the central QPC [see Eq.~\eqref{eq:expansion_qpc} below]. In what follows, for brevity, we will remove the tildes from fields in the virtual wire. 

With these expressions, we are ready to analyze the correlation function, in the presence of interaction.
To begin with, after including interactions, the correlator at the central QPC becomes
\begin{equation}
\begin{aligned}
& \mathcal{T}^{(0)}_C \big\langle \mathcal{T}_K \Psi^\dagger_A  (0, t^-) \Psi_B (0, t^-) \Psi^\dagger_B (0,0^+) \Psi_A (0,0^+) \big\rangle \\ 
= & \frac{\mathcal{T}^{(0)}_C}{(2\pi \tau_0)^2} \Big\langle \mathcal{T}_K  e^{-i\frac{1}{\sqrt{\nu}} [\phi_A (0,t^-) - \phi_B (0,t^-)] } e^{i\frac{1}{\sqrt{\nu}} [\phi_A (0,0^+) - \phi_B (0,0^+)] } \Big\rangle\\
= & \frac{\mathcal{T}^{(0)}_C}{(2\pi \tau_0)^2} \Big\langle \mathcal{T}_K  e^{-i\frac{1}{\sqrt{\nu}} \phi_+ (0,t^-)  } e^{i\frac{1}{\sqrt{\nu}} \phi_+ (0,0^+) } \Big\rangle \Big\langle \mathcal{T}_K  e^{i \frac{1}{\sqrt{\nu}} \phi_- (0,t^-)  } e^{-i \frac{1}{\sqrt{\nu}} \phi_- (0,0^+) } \Big\rangle,
\end{aligned} 
\label{eq:expansion_qpc}
\end{equation}
where now we need to evaluate the correlation function for $\pm$ modes, instead of $A$ and $B$ modes.
As a reminder, Eq.~\eqref{eq:expansion_qpc} can not capture the situation where edges $A$ and $B$ are biased at different voltages, as this situation requires the inclusion of another voltage-difference-dependent (and time-dependent) phase factor: in this non-equilibrium case the $\pm$ modes are not at equilibrium.
However, we can still use Eq.~\eqref{eq:expansion_qpc} for calculating perturbative expansions in diluter's transmissions even in the single-source case, since all the involved averages will be taken with respect to the equilibrium state.

The correlation functions determine the current and noises can not be factorized into the product of correlation functions for subsystems $\mathcal{A}$ and $\mathcal{B}$. 
For instance, when considering leading-order expansion at the upper diluter, the modified $D_{A2}$ of Eq.~\eqref{eq:da2} contains correlations like [following $\phi_A - \phi_B = \phi_+ - \phi-$, as given by Eq.~\eqref{eq:expansion_qpc}]:
\begin{equation}
\begin{aligned}
& \sum_{\eta_1\eta_2} \eta_1\eta_2 \iint ds_1 ds_2 \Big\langle \mathcal{T}_K  e^{-i\frac{\phi_+ (0,t^-) }{\sqrt{\nu}}  } e^{i\frac{\phi_+ (0,0^+)}{\sqrt{\nu}}  }  e^{i\frac{\phi_- (0,t^-)}{\sqrt{\nu}}   } e^{-i \frac{ \phi_- (0,0^+)}{\sqrt{\nu}} } e^{i\sqrt{\nu} \phi_A (-L, s_1^{\eta_1}) } e^{-i\sqrt{\nu} \phi_A (-L, s_2^{\eta_2}) } \Big\rangle \\
& \times  \Big\langle e^{-i\sqrt{\nu} \phi_{sA} (-L, s_1^{\eta_1}) } e^{i\sqrt{\nu} \phi_{sA} (-L, s_2^{\eta_2}) } \Big\rangle e^{i\nu e V (s_1 - s_2)}\\
& = \sum_{\eta_1\eta_2} \eta_1\eta_2 \iint ds_1 ds_2 \Big\langle \mathcal{T}_K  e^{-i\frac{1}{\sqrt{\nu}} \phi_+ (0,t^-)  } e^{i\frac{1}{\sqrt{\nu}} \phi_+ (0,0^+) }  e^{i\sqrt{\nu} (1 + \delta_\text{edge}) \phi_+ (-L, s_1^{\eta_1}) } e^{-i\sqrt{\nu} (1 + \delta_\text{edge}) \phi_+ (-L, s_2^{\eta_2}) } \Big\rangle\\
&\times \Big\langle \mathcal{T}_K e^{i\frac{\phi_- (0,t^-)}{\sqrt{\nu}}   } e^{-i\frac{\phi_- (0,0^+)}{\sqrt{\nu}}  } e^{- i\sqrt{\nu}  \delta_\text{edge} \phi_- (L - 2d, s_1^{\eta_1}) } e^{i\sqrt{\nu}  \delta_\text{edge} \phi_- (L - 2d, s_2^{\eta_2}) } \Big\rangle \frac{\tau_0^\nu e^{i\nu e V (s_1 - s_2)}}{[ \tau_0 + i (s_1 - s_2) \chi_{\eta_1\eta_2} (s_1 - s_2) ]^\nu} \\
& = \sum_{\eta_1\eta_2} \eta_1\eta_2 \iint ds_1 ds_2 \frac{e^{i\nu e V (s_1 - s_2)} \tau_0^{\frac{1}{2\nu} + 2\tilde{\nu}}}{(\tau_0 + it)^{\frac{1}{2\nu}} [ \tau_0 + i (s_1 - s_2) \chi_{\eta_1\eta_2} (s_1 - s_2) ]^{2\tilde{\nu} } }\\
& \times \left\{ \frac{[ \tau_0 + i ( t - s_2 - L + 2 d ) \chi_{-\eta_2} (t - s_2) ] [ \tau_0 + i (  - s_1 - L + 2 d ) \chi_{+\eta_1} ( - s_1) ]}{[ \tau_0 + i ( t - s_1 - L + 2 d ) \chi_{-\eta_1} (t - s_1) ] [ \tau_0 + i (  - s_2 - L + 2 d ) \chi_{+\eta_2} ( - s_2) ]} \right\}^{ \delta_\text{edge}}\\
& \times \left\{ \frac{[ \tau_0 + i ( t - s_2 - L ) \chi_{-\eta_2} (t - s_2) ] [ \tau_0 + i (  - s_1 - L ) \chi_{+\eta_1} ( - s_1) ]}{[ \tau_0 + i ( t - s_1 - L ) \chi_{-\eta_1} (t - s_1) ] [ \tau_0 + i (  - s_2 - L ) \chi_{+\eta_2} ( - s_2) ]} \right\}^{\frac{\tilde{\nu}}{\nu}},
\end{aligned}
\label{eq:single-source_correlation}
\end{equation}
where $\tilde{\nu} \equiv \nu (1 + \delta_\text{edge})$ is influenced by interaction.
Notice that bosonic operators with different chirality have different correlations: $\langle \phi_\pm (x,t) \phi_\pm (x',t') \rangle \propto \ln [(t \mp x) - (t' \mp x') ]$.
The last line of Eq.~\eqref{eq:single-source_correlation} represents the Coulomb-interaction-influenced ``tangling factor'' (cf. Ref.~\cite{GuOneHalf24S}) of the right-going field $\phi_+$.
It reduces to the last line of Eq.~\eqref{eq:leading1} for the non-interacting case, after taking $\delta_\text{edge} = 0$.
The last but one line instead refers to the field from the left-going field $\phi_-$.
This term is fully interaction-induced, as $\phi_-$ and $\phi_A$ are uncorrelated for the non-interacting situation.

We proceed by taking $s_1 \to s_2$ in the equation above to consider higher-order processes involving operators of non-equilibrium anyons.
We also perform shifts in time $s_1 \to s_1 - L $ and $s_2 \to s_2 - L$,
with which the last two lines of Eq.~\eqref{eq:single-source_correlation} equal
\begin{equation}
\begin{aligned}
& \left\{ \frac{[ \tau_0 + i ( t - s_1  + 2 d ) \eta_2 ] [ \tau_0 + i (  - s_1  + 2 d ) \eta_1 ]}{[ \tau_0 + i ( t - s_1  + 2 d ) \eta_1 ] [ \tau_0 + i (  - s_2 + 2 d ) \eta_2 ]} \right\}^{ \delta_\text{edge}} \left\{ \frac{[ \tau_0 + i ( t - s_1 ) \eta_2 ] [ \tau_0 + i (  - s_1  ) \eta_1]}{[ \tau_0 + i ( t - s_1  ) \eta_1 ] [ \tau_0 + i (  - s_1  ) \eta_2 ]} \right\}^{\frac{\tilde{\nu}}{\nu}}.
\end{aligned}
\label{eq:braiding_phases}
\end{equation}
The second part of Eq.~\eqref{eq:braiding_phases}, which corresponds to the braiding between the right-going $\phi_A$ mode (describing non-equilibrium anyons) and $\phi_+$ mode facilitated by the central QPC, equals $\exp[i\pi (\eta_2 - \eta_1) \tilde{\nu} / \nu]$.
In contrast to the non-interacting result, this phase is non-trivial and will not vanish after summations over Keldysh indices.
The first term of Eq.~\eqref{eq:braiding_phases} instead indicates the braiding between the $\phi_-$ mode at the central QPC, and counter-propagating non-equilibrium anyonic mode in the interacting area.
In this section, we assume the large-$d$ situation ($d>t$), with which this extra term equals one, a trivial value.
Notice that for a small value of $d$ (more specifically, when $2d < s_1 , s_2 < t$), this term equals $\exp[i\pi \delta_\text{edge}(\eta_2 - \eta_1)]$.
Combining this factor with  $\exp[i\pi (1 + \delta_\text{edge})(\eta_2 - \eta_1)]$, the total ``tangling'' part equals $\exp[i\pi (1 + 2\delta_\text{edge})(\eta_2 - \eta_1)]$ when considering pairs of non-equilibrium anyons whose operators have close time arguments, leading to an even stronger modification from interactions.

As another important piece of message, Eq.~\eqref{eq:braiding_phases} indicates that, when considering self-contracted anyonic pairs in the large-$d$ assumption, interactions mainly influence the correlation between $\phi_A$ at the diluter and $\phi_+$ at the central QPC.
The $\phi_A - \phi_-$ correlation instead remains negligible even with interaction involved.
Similarly, even for the interacting situation, we only need to worry about the correlation between $\phi_B$ at the diluter and $\phi_-$ at the central QPC.

In addition to contracted pairs of non-equilibrium anyons, the interaction between edges $A$ and $B$ also influences the leading-order Andreev-like tunneling processes.
Briefly, after choosing the contract option $s_1 \to t - L$ and $s_2 \to -L$, integrals over $s_1$ and $s_2$ of Eq.~\eqref{eq:single-source_correlation} can be rewritten as
\begin{equation}
\begin{aligned}
(\text{S39}) & = \frac{\tau_0^{\frac{1}{2\nu} + 2\tilde{\nu}}}{(\tau_0 + it)^{\frac{1}{2\nu}}} \sum_{\eta_1\eta_2} \eta_1\eta_2  \frac{( \tau_0 + i t \eta_2)^{\frac{\tilde{\nu}}{\nu}} [ \tau_0 + i (  -t ) \eta_1 ]^{\frac{\tilde{\nu}}{\nu}}}{ [ \tau_0 + i t \chi_{\eta_1\eta_2} (t) ]^{2\tilde{\nu} } } \times \left\{ \frac{[ \tau_0 + i ( t  + 2 d ) \eta_2 ] [ \tau_0 + i (  -t + 2 d ) \eta_1 ]}{( \tau_0 + i  2 d \eta_1 ) ( \tau_0 + i  2 d \eta_2 )} \right\}^{ \delta_\text{edge}} \\
& \times \iint ds_1 ds_2  \frac{e^{i\nu e V (s_1 - s_2)}}{\left\{[ \tau_0 + i ( t - s_1 - L ) \chi_{-\eta_1} (t - s_1) ] [ \tau_0 + i (  - s_2 - L ) \chi_{+\eta_2} ( - s_2) ]\right\}^{\frac{\tilde{\nu}}{\nu}}}\\
& = I_{\tilde{\nu} = \nu} \times \frac{ (it /\tau_0)^{(2-2\nu) \delta_\text{edge}} }{\Gamma^2\left( \frac{\tilde{\nu}}{\nu} \right)},
\end{aligned}
\end{equation}
where $I_{\tilde{\nu} = \nu}$ refers to the non-interacting result, where $\tilde{\nu} = \nu$.
The factor multiplying $I_{\tilde{\nu} = \nu}$ describes the modification from interactions.

With both modifications induced by the interaction taken into consideration, we arrive at modified correlation functions 
\begin{equation}
\begin{aligned}
    \big\langle\Psi^\dagger_{+}(t^-) \Psi_{+} (0^+) \big\rangle & = \frac{\tau_0^{\frac{1}{\nu}-1}}{2\pi (\tau_0 + it)^{\frac{1}{\nu}}} \left\{ e^{- \frac{I_{A0} }{e\nu} \zeta_- \left( \frac{\tilde{\nu}}{\nu},t \right) t } 
   + \frac{c(\nu) (it /\tau_0)^{(2-2\nu) \delta_\text{edge} }}{\Gamma^2\left( \frac{\tilde{\nu}}{\nu} \right)} \frac{ it \frac{I_{A0}}{e} e^{ i\nu e V t  } }{(i \nu e V t  )^{2\nu - 1}}   e^{- \frac{I_{A0} }{e\nu} \zeta_+ (\tilde{\nu}, t) t} \right\} \\
 &  \to  \frac{\tau_0^{\frac{1}{\nu}-1}}{2\pi (\tau_0 + it)^{\frac{1}{\nu}}} \left\{ e^{-\left( 1 - e^{ 2i\pi\frac{\tilde{\nu}}{\nu}} \right) \frac{I_{A0} t}{e\nu} } 
   + \frac{c(\nu) (it /\tau_0)^{(2-2\nu) \delta_\text{edge} }}{\Gamma^2\left( \frac{\tilde{\nu}}{\nu} \right)} \frac{ it \frac{I_{A0}}{e} e^{ i\nu e V t  } }{(i \nu e V t  )^{2\nu - 1}}   e^{-\left( 1 - e^{- 2i\pi\tilde{\nu}} \right) \frac{I_{A0} t}{e\nu}} \right\},\\
   \big\langle\Psi_{-}(t^-) \Psi^\dagger_{-} (0^+) \big\rangle & = \frac{\tau_0^{\frac{1}{\nu}-1}}{2\pi (\tau_0 + it)^{\frac{1}{\nu}}} \left\{ e^{- \frac{I_{B0}}{e\nu} \zeta_+ \left( \frac{\tilde{\nu}}{\nu},t \right) t } 
   + \frac{c(\nu) (it /\tau_0)^{(2-2\nu) \delta_\text{edge} }}{\Gamma^2\left( \frac{\tilde{\nu}}{\nu} \right)} \frac{ it \frac{I_{B0}}{e} e^{ -i\nu e V t  } }{(i \nu e V t  )^{2\nu - 1}}   e^{- \frac{I_{B0} }{e\nu} \zeta_- (\tilde{\nu}, t) t} \right\} \\
 &  \to  \frac{\tau_0^{\frac{1}{\nu}-1}}{2\pi (\tau_0 + it)^{\frac{1}{\nu}}} \left\{ e^{-\left( 1 - e^{- 2i\pi\frac{\tilde{\nu}}{\nu}} \right) \frac{I_{B0} t}{\nu}} 
    + \frac{c(\nu) (it /\tau_0)^{(2-2\nu) \delta_\text{edge}}}{\Gamma^2\left( \frac{\tilde{\nu}}{\nu} \right)} \frac{ it \frac{I_{B0}}{e} e^{ -i\nu e V t  } }{(i \nu e V t  )^{2\nu - 1}}  e^{-\left( 1 - e^{ 2i\pi\tilde{\nu}} \right) \frac{I_{B0} t}{e\nu}} \right\},
\end{aligned}
\label{eq:a_b_correlations}
\end{equation}
where the first term in each correlation function comes from braiding processes that are induced by interaction.
Notice that $I_{A0}$ and $I_{B0}$ are not influenced, as both diluters, where the non-equilibrium current values are emitted, are outside of the interacting area.

\subsection{Influence of inter-edge interaction on the tunneling current noise}
\label{sec:inter-edge-interaction}

With interaction-modified correlation functions, Eq.~\eqref{eq:a_b_correlations}, we are ready to calculate $S_T$ in the presence of inter-edge interactions.
For later convenience, we refer to the first and second terms within the curly brackets of Eq.~\eqref{eq:a_b_correlations} with subscripts $\alpha 1$ and $\alpha 2$, respectively, with $\alpha = \pm$ for the corresponding mode.
Combining different terms (1 or 2) of correlations in channels $+$ and $-$ leads to three types of contributions.

As the first type of contribution, we combine terms 1 for correlation functions of both the $+$ and $-$ modes.
In this case, both the tunneling current and its corresponding noise are associated with the integral (which becomes dimensionless due to the $\tau_0^5$ factor) below
\begin{equation}
\begin{aligned}
    \mathcal{I}^\text{int}_\text{1+,1-}&\equiv\, \tau_0^{\frac{2}{\nu}-1} \int dt \frac{1}{ (\tau_0 + it)^{2/\nu}}\, \exp\left[{- \frac{I_{A0} }{\nu e} \zeta_-\left(\frac{\tilde{\nu}}{\nu}, t\right) t - \frac{I_{B0}}{\nu e} \zeta_+\left(\frac{\tilde{\nu}}{\nu}, t\right)t}\right] \\
     &=  \Gamma(2/\nu)\, 
     i (-i)^{\frac{2}{\nu} - 4} \tau_0^{\frac{2}{\nu}-1} \int dt \ln (\tau_0 + i t) \frac{d^{2/\nu}}{d t^{2/\nu}} \exp\left[- \frac{I_{A0} }{\nu e} \zeta_-\left(\frac{\tilde{\nu}}{\nu}, t\right) t - \frac{I_{B0}}{\nu e} \zeta_+ \left(\frac{\tilde{\nu}}{\nu}, t\right) t\right]\\
     &\approx    \Gamma(2/\nu)\, 
     i (-i)^{\frac{2}{\nu} - 4} \tau_0^{\frac{2}{\nu}-1}  \int \!dt\, \ln (\tau_0 + i t) \, \exp\left\{-\frac{I_+}{\nu e} \left[1 - \cos\left(2\pi\frac{\tilde{\nu}}{\nu}\right)\right]|t| + i \frac{I_-}{\nu e} \,\sin \left(2\pi\frac{\tilde{\nu}}{\nu}\right) t \right\}
     \\
     &\qquad \qquad \times \left\{ -\frac{I_+}{\nu e} \left[1 - \cos\left(2\pi\frac{\tilde{\nu}}{\nu}\right)\right]\text{sgn}(t) + i\, \frac{I_-}{\nu e} \, \sin \left(2\pi\frac{\tilde{\nu}}{\nu}\right) \!\right\}^6.\\
     \end{aligned}
\label{eq:interaction_disconnected}
\end{equation}
Setting $\nu=1/3$, we obtain
\begin{align}
    \mathcal{I}^\text{int}_\text{1+,1-}& =  -5!\, i   
    \left\{ -2 b_0 (b_0^4 - 10b_0^2 c_0^2 + 5 c_0^4) \gamma_\text{E}  + c_0 (5b_0^4 - 10 b_0^2 c_0^2 + c_0^4) \pi
    \right. \notag 
    \\
    &- \left.(b_0 - i c_0)^5 \ln (b_0 - i c_0) - (b_0 + i c_0)^5 \ln (b_0+ i c_0) 
 \right\},
 \label{Iint1+1-13}
\end{align} 
where $\gamma_E$ is the Euler gamma constant 
and we have defined two current-dependent quantities: 
\begin{equation}
\begin{aligned}
    b_0& = \tau_0 \frac{I_+}{\nu e} \left[ 1 - \cos \left(2\pi\frac{\tilde{\nu}}{\nu}\right) \right],\\
    c_0 & = \tau_0 \frac{I_-}{\nu e} \sin \left(2\pi\frac{\tilde{\nu}}{\nu}\right).
\end{aligned}
\end{equation}
The subscript ``$1+,1-$'' in $\mathcal{I}^\text{int}_\text{1+,1-}$ signifies the combination of first terms in both Eqs.~\eqref{eq:dan_resum} and \eqref{eq:dan_resum_b}; 
the superscript ``int'' highlights the inclusion of interactions.
Notice that the imaginary and real parts of $\mathcal{I}^\text{int}_\text{1+,1-}$, being multiplied by  $\propto \mathcal{T}_C^{(0)}$, induce an extra double-source collision contribution to the tunneling current and the tunneling current noise, respectively.

In the weak-interacting limit ($|\delta_\text{edge}| = |\nu - \tilde{\nu}|/\nu \ll 1$), we have $b_0\propto \delta_\text{edge}^2$ and $c_0\propto\delta_\text{edge}$, to leading order of the interaction $\delta_\text{edge}$.
With Eq.~\eqref{eq:interaction_disconnected}, we obtain the following contribution to the tunneling current and its corresponding noise:
\begin{equation}
\begin{aligned}
    I_\text{T;1+,1-} & = 30 c_0 \Big(5 b_0^4 - 10 b_0^2 c_0^2 + c_0^4\Big)\,  \frac{e \mathcal{T}_C^{(0)}}{\pi \tau_0}\\
    S_{\text{T}; 1+,1-} & =  30 \left\{ 2 \Big(5 b_0^4 c_0 - 10b_0^2 c_0^3 + c_0^5\Big) \arctan \left( \frac{c_0}{b_0} \right) - b_0 \Big(b_0^4 - 10b_0^2 c_0^2 + 5c_0^4\Big) \left[2 \gamma_\text{E} + \ln \big(b_0^2 + c_0^2\big)\right] \right\} \frac{e^2 \mathcal{T}_C^{(0)}}{\pi^2 \tau_0},
\end{aligned}
\label{eq:it_st_11}
\end{equation}
which are proportional to $\delta_\text{edge}^5$ in the weak-interacting limit.
Due to this higher power factor in $\delta_\text{edge}$, modifications of Eq.~\eqref{eq:it_st_11} are negligible in the weak-tunneling limit.

As for the second type of contribution, we can combine the second terms (i.e., the terms with subscript 2) of correlation functions for both the $+$ and $-$ modes.
The relevant integral then becomes
\begin{equation}
\begin{aligned}
    \mathcal{I}_{2+,2-}^\text{int} & = \tau_0^{2/\nu + (4 \nu - 4) (1 + \delta_\text{edge}) - 1} \int dt \frac{1}{(\tau_0 + i t)^{(4 \nu - 4) (1 + \delta_\text{edge}) + 2 /\nu}}\, e^{-\frac{I_{A0}}{\nu e} \zeta_+(\tilde{\nu},t) t - \frac{I_{B0}}{\nu e} \zeta_- \left(\tilde{\nu},t\right) t }\\
    & \approx 2 \text{Re} \left( e^{-i\frac{\pi \nu_\text{d}^\text{int}}{2}} \left\{ \frac{\tau_0 I_+}{\nu e}\left[ 1 - \cos (2\pi\tilde{\nu}) \right]  -i \frac{\tau_0 I_-}{\nu e} \sin (2\pi\tilde{\nu})  \right\}^{-1 + \nu_\text{d}^\text{int}} \right) \Gamma (1 - \nu_\text{d}^\text{int})\\
    & \approx 2 \text{Re} \left( e^{-i\frac{\pi \nu_\text{d}}{2}} \left\{ \frac{\tau_0 I_+}{\nu e}\left[ 1 - \cos (2\pi\nu) \right]  -i \frac{\tau_0 I_-}{\nu e} \sin (2\pi\nu)  \right\}^{-1 + \nu_\text{d}} \right) \Gamma (1 - \nu_\text{d}) \left[ 1 + \gamma_\text{1,d} (I_+,I_-,\nu) \delta_\text{edge} \right]\\
    & + 2 \text{Im} \left( e^{-i\frac{\pi \nu_\text{d}}{2}} \left\{ \frac{\tau_0 I_+}{\nu e}\left[ 1 - \cos (2\pi\nu) \right]  -i \frac{\tau_0 I_-}{\nu e} \sin (2\pi\nu)  \right\}^{-1 + \nu_\text{d}} \right) \Gamma (1 - \nu_\text{d})  \gamma_\text{2,d} (I_+,I_-,\nu) \delta_\text{edge} ,
\end{aligned}
\label{eq:int_22}
\end{equation}
where $\nu_\text{d}^\text{int} \equiv 2/\nu + (4 \nu - 4) (1 + \delta_\text{edge})$ is the interaction-influenced power-law exponent in the denominator.
In great contrast to Eq.~\eqref{eq:interaction_disconnected}, Eq.~\eqref{eq:int_22} remains finite for the non-interacting case, i.e., when $\tilde{\nu} = \nu$.
Actually, the leading-order interaction-induced term in $\mathcal{I}_{2+,2-}^\text{int}$ is proportional to $\delta_\text{edge}$.
In the weakly interacting limit $|\delta_\text{edge}|\ll 1$, this effect of interaction is much stronger than that encoded in  $I_{1+,1-}^\text{int}$, as the latter is proportional to $\delta_\text{edge}^5$.

The functions $\gamma_{1,d}$ and $\gamma_{2,d}$ entering Eq.~\eqref{eq:int_22} have the following explicit expressions:
\begin{equation}
\begin{aligned}
    \gamma_\text{1,d} (I_+,I_-,\nu) & = \frac{ -2 I_+^2 + I_-^2 [1 - \cot^2 (\pi\nu)]}{I_+^2 + I_-^2 \cot^2 (\pi\nu) } \pi\nu\cot (\pi\nu) \left(\frac{2}{\nu} + 4\nu - 5\right) + 4 (1-\nu) \mathrm{\psi} \left( 5-\frac{2}{\nu} - 4\nu \right),\\
    & - 4 (1 - \nu) \ln \Big|  \frac{\tau_0 I_+}{\nu e}\left[ 1 - \cos (2\pi \nu) \right]  -i \frac{\tau_0 I_-}{\nu e} \sin (2\pi\nu) 
\Big| \\
    \gamma_\text{2,d} (I_+,I_-,\nu) & = \frac{I_+ I_-}{I_+^2 + I_-^2 \cot^2(\pi\nu)} \frac{\pi\nu}{\sin^2 (\pi\nu)} \left(\frac{2}{\nu} + 4\nu - 5\right) + 2 (1 - \nu) \pi  - 4 (1 - \nu)\arctan \left[ \frac{I_-}{I_+} \cot(\pi\nu)\right],
\end{aligned}
\label{eq:gamma_d}
\end{equation}
where $\mathrm{\psi}(x)$ is the polygamma function.
The values of $\gamma_\text{1,d}$ and $\gamma_\text{2,d}$ reflect the leading-order interaction influence on the tunneling current and tunneling current noise.
In agreement with our analysis, here leading-order corrections are proportional to $\delta_\text{edge}$, thus are much larger than interaction-induced corrections of $\mathcal{I}^\text{int}_\text{1+,1-}$ in Eq.~\eqref{eq:interaction_disconnected}.

Following Eq.~\eqref{eq:gamma_d}, the inclusion of interaction around the QPC leads to the modification of the tunneling-current noise, $\delta S_\text{T}^\text{collision} \equiv S_\text{T} (\delta_\text{edge}) - S_\text{T} (0) $, in the form of
\begin{equation}
\begin{aligned}
    \delta S_\text{T}^\text{collision} & \approx S_\text{T}^\text{collision} \delta_\text{edge} \left\{  \gamma_\text{1,d} (I_+,I_-,\nu) +  \gamma_\text{2,d} (I_+,I_-,\nu) \frac{\text{Im} \left[ \Pi_d (I_+,I_-,\nu) \right]}{\text{Re} \left[ \Pi_d (I_+,I_-,\nu) \right]} \right\},
\end{aligned}
\label{eq:st_double_modification}
\end{equation}
to the leading order of $\delta_\text{edge}$,
with the function
\begin{equation}
    \Pi_d (I_+,I_-,\nu)\equiv \left\{ \frac{\tau_0 I_+}{\nu e}\left[ 1 - \cos (2\pi\nu) \right]  -i \frac{\tau_0 I_-}{\nu e} \sin (2\pi\nu)  \right\}^{-1 + \nu_\text{d}}.
\end{equation}
Considering the experimentally relevant situation $\nu =1/3$, for the symmetric case, $I_- = 0$, these relevant parameters are as follows:
\begin{equation}
    \gamma_\text{1,d} (I_+,0,1/3) \approx 9.86,\qquad \gamma_\text{2,d} (I_+,0,1/3) \approx 4.19,\qquad \frac{\text{Im} \left[  \Pi_d (I_+,0,1/3) \right]}{\text{Re} \left[ \Pi_d (I_+,0,1/3) \right]} = 0,
\end{equation}
where we have taken $\tau_0 I_+/\nu e = 0.1$. 
Interaction-induced effect on the double-source collision contribution $\delta S_\text{T}^\text{collision}$,  Eq.~\eqref{eq:st_double_modification}, is thus linear in $\delta_\text{edge}$.

Finally, we can combine the first term from either the $+$ or $-$ mode with the second term from the other mode.
Without the loss of generality, here we present the result after combining the second term of mode $+$ and the first term of mode $-$. This option leads to the single-source contribution that is induced by the non-equilibrium current $I_{A0}$ in channel $A$.
The corresponding contribution involves the integral
\begin{equation}
\begin{aligned}
    \mathcal{I}^\text{int}_{2+,1-} & \equiv \tau_0^{\frac{2}{\nu} + (2 \nu - 2) (1 + \delta_\text{edge}) - 1} \int dt\, \frac{1}{(\tau_0 + i t)^{(2\nu-2) (1 + \delta_\text{edge})+ 2/\nu} }\, e^{-\frac{I_{A0}}{\nu e} \zeta_+(\tilde{\nu},t) t - \frac{I_{B0}}{\nu e} \zeta_+ \left(\frac{\tilde{\nu}}{\nu},t\right) t + i \nu e V t  }\\
    & \approx  2 \text{Re} \left( e^{-\frac{i\pi \nu_\text{s}^\text{int}}{2}} \left\{ \frac{\tau_0 I_{A0}}{\nu e} \left[ 1 - \cos (2\pi\tilde{\nu}) \right] + \frac{\tau_0 I_{B0}}{\nu e} \left[ 1 - \cos \left(2\pi\frac{\tilde{\nu}}{\nu}\right) \right] \right.\right.\\
    & \left.\left.  + i \left[ \frac{\tau_0 I_{A0}}{\nu e} \sin (2\pi\tilde{\nu}) + \frac{\tau_0 I_{B0}}{\nu e} \sin \left(2\pi\frac{\tilde{\nu}}{\nu}\right) - \frac{\nu e V}{\hbar} \right] \right\}^{-1 + \nu_\text{s}^\text{int}} \right) \Gamma (1 - \nu_\text{s}^\text{int})\\
    & \approx 2 \text{Re} \left\{ e^{-\frac{i\pi \nu_\text{s}}{2}}  \left[ \frac{\tau_0 I_{A0}}{\nu e} \left[ 1 - \cos (2\pi\nu) \right] + i \frac{\tau_0 I_{A0}}{\nu e} \sin (2\pi\nu) - i\frac{\tau_0 \nu e V}{\hbar} \right]^{-1 + \nu_\text{s}} \right\} \Gamma (1 - \nu_\text{s})\left( 1 + \gamma_\text{1,s}\ \right) \delta_\text{edge} \\
    & + 2 \text{Im} \left\{ e^{-\frac{i\pi \nu_\text{s}}{2}}  \left[ \frac{\tau_0 I_{A0}}{\nu e} \left[ 1 - \cos (2\pi\nu) \right] + i \frac{\tau_0 I_{A0}}{\nu e} \sin (2\pi\nu) - i\frac{\tau_0 \nu e V}{\hbar} \right]^{-1 + \nu_\text{s}} \right\} \Gamma (1 - \nu_\text{s})\gamma_\text{2,s} \delta_\text{edge} ,
\end{aligned}
\label{eq:i21}
\end{equation}
where, again, we keep only the leading-order contributions (the zeroth and the first, more specifically) of $\delta_\text{edge}$, and $\nu_\text{s}^\text{int} \equiv (2\nu-2) (1 + \delta_\text{edge}) + 2/\nu$ refers to the power-law exponent in the denominator of the single-source contribution. Equation~\eqref{eq:i21} contains two functions, with their explicit expressions given by
\begin{equation}
\begin{aligned}
    & \gamma_\text{1,s} (I_{A0},V,\nu) = \frac{2\pi\nu I_{A0} \left[ \nu^2 \cos(2\pi\nu) V - I_{A0} \sin(2\pi\nu) \right]}{\nu^2 \left[\nu^2 - \sin(2\pi\nu) \right] V^2 + 2 [1 - \cos(2\pi\nu)] I_{A0}^2 } \left(\frac{2}{\nu} + 2\nu -3 \right) + 2 (1 -\nu) \mathrm{\psi} \left( 3 - \frac{2}{\nu} - 2\nu \right)\\
    & -2 (1 - \nu) \ln \Big| \frac{\tau_0 I_{A0}}{\nu e} \left[ 1 - \cos (2\pi \nu) \right] + i \frac{\tau_0 I_{A0}}{\nu e} \sin (2\pi \nu) - i\frac{\tau_0 \nu e V}{\hbar}  \Big|,\\
    & \gamma_\text{2,s} (I_{A0},V,\nu) = \frac{4\pi\nu\sin(\pi\nu) I_{A0} \left[ \nu^2 \cos(\pi\nu) V - \sin(\pi\nu) I_{A0} \right]}{\nu^2 \left[\nu^2 - \sin(2\pi\nu) \right] V^2 + 2 [1 - \cos(2\pi\nu)] I_{A0}^2 } \left(\frac{2}{\nu} + 2\nu -3 \right) + (1 -\nu) \pi \\
    & - 2 (1-\nu) \arctan \left[ \frac{I_{A0} \sin(2\pi\nu) - \nu^2 e^2 V  }{2 I_{A0} \sin^2 (\pi\nu) } \right].
\end{aligned}
\end{equation}
With parameters defined above, interaction-induced correction on the single-source (due to $I_{A0}$ of channel $A$) tunneling current noise, $$\delta S_\text{T}^\text{single} (I_{A0},V,\nu,\delta_\text{edge}) \equiv S_\text{T}^\text{single} (I_{A0},V,\nu,\delta_\text{edge}) - S_\text{T}^\text{single} (I_{A0},V,\nu,0),$$ can be written as
\begin{equation}
\begin{aligned}
    \delta S_\text{T}^\text{single} (I_{A0},V,\nu,\delta_\text{edge}) & \approx S_\text{T}^\text{single} (I_{A0},V,\nu,0) \delta_\text{edge} \left\{  \gamma_\text{1,s} (I_{A0},V,\nu) +  \gamma_\text{2,s} (I_{A0},V,\nu) \frac{\text{Im} \left[ \Pi_s (I_{A0},V,\nu) \right]}{\text{Re} \left[ \Pi_s (I_{A0},V,\nu) \right]} \right\},
\end{aligned}
\label{eq:st_single_modification}
\end{equation}
with 
\begin{equation}
    \Pi_s (I_{A0},V,\nu) \equiv \left[ \frac{\tau_0 I_{A0}}{\nu e} \left[ 1 - \cos (2\pi\nu) \right] + i \frac{\tau_0 I_{A0}}{\nu e} \sin (2\pi\nu) - i\frac{\tau_0 \nu e V}{\hbar} \right]^{-1 + \nu_\text{s}}.
\end{equation}
For $\nu=1/3$, relevant single-source parameters are as follows
\begin{equation}
    \gamma_\text{1,s} (I_{A0},V,\nu) \approx -2.88,\qquad \gamma_\text{2,s} (I_{A0},V,\nu) \approx 4.15,\qquad \frac{\text{Im} \left[ \Pi_s (I_{A0},V,\nu) \right]}{\text{Re} \left[ \Pi_s (I_{A0},V,\nu) \right]} \approx 1.02,
\end{equation}
where we have again taken $\tau_0 I_+ /\nu e = 0.1 $, and assumed $\mathcal{T}_A = 0.1$ as the transmission probability through the diluter.

By combing interaction-induced noises for both the collision and single-source contributions [Eqs.~\eqref{eq:st_double_modification} and \eqref{eq:st_single_modification}, respectively] and using the definition of the entanglement pointer [Eq.~(1) of the main text], we have thus arrived at the expression of interaction-induced tunneling current noise and entanglement pointer, i.e.,
\begin{equation}
\begin{aligned}
    \delta S^\text{QPC}_\text{T} & = \left( S_\text{T}^\text{collision} \left\{  \gamma_\text{1,d} (I_+,I_-,\nu) +  \gamma_\text{2,d} (I_+,I_-,\nu) \frac{\text{Im} \left[ \Pi_d (I_+,I_-,\nu) \right]}{\text{Re} \left[ \Pi_d (I_+,I_-,\nu) \right]} \right\} \right.\\
    & \left. + S_\text{T}^\text{single} (I_{A0},V,\nu,0)  \left\{  \gamma_\text{1,s} (I_{A0},V,\nu) +  \gamma_\text{2,s} (I_{A0},V,\nu) \frac{\text{Im} \left[ \Pi_s (I_{A0},V,\nu) \right]}{\text{Re} \left[ \Pi_s (I_{A0},V,\nu) \right]} \right\} \right) \delta_\text{edge},\\
    \delta \mathcal{P}^\text{QPC}_\text{Andreev} & = \mathcal{P}_\text{Andreev}  \left\{  \gamma_\text{1,d} (I_+,I_-,\nu) +  \gamma_\text{2,d} (I_+,I_-,\nu) \frac{\text{Im} \left[ \Pi_d (I_+,I_-,\nu) \right]}{\text{Re} \left[ \Pi_d (I_+,I_-,\nu) \right]} \right\} \delta_\text{edge},
\end{aligned}
\end{equation}
where superscript ``QPC'' indicates that the interaction is introduced around the QPC.
Noticeably, both $\delta S^\text{QPC}_\text{T}$ and $\delta \mathcal{P}^\text{QPC}_\text{Andreev}$ are proportional to the interaction-induced factor $\delta_\text{edge}$.
We have thus arrived at the conclusion that the inter-edge interaction around the QPC introduces corrections to the tunneling noise and the entanglement pointer of the same order in the interaction strength.
This conclusion agrees with the observation reported for the integer quantum Hall setups~\cite{GuNC24S}, where the effects of interactions around the central QPC on the total noise and entanglement pointer were of the same order of magnitude. 
Importantly, for both $S_\text{T}$ and the entanglement pointer, corrections introduced by the inter-edge interactions are proportional to $\delta_\text{edge}$ and are thus negligible in the weak-interaction limit, which is realized in typical experimental settings~\cite{GuNC24S}.
However, as will be shown shortly in Sec.~\ref{sec:intra-edge-interaction}, the intra-edge interaction, in great contrast, can, in principle, induce a rather significant modification of correlation functions, even when the system interaction is very weak. This significant interaction effect is, however, avoided in the entanglement pointer.

\subsection{Influence of intra-edge interactions}
\label{sec:intra-edge-interaction}

The above analysis focused on the influence of inter-edge interaction, which is present only near the central collider.
The resilience of the entanglement pointer to interaction effects becomes manifest after including intra-edge interaction, i.e., the Coulomb interaction among channels of the same edge.
This is the situation for, e.g., the integer edges at $\nu>1$~\cite{GuNC24S} or high-order anyonic states that host multiple co-propagating channels at a single physical boundary.
Indeed, following Ref.~\cite{GuNC24S}, the intra-edge Coulomb interaction, inducing charge-fractionalization, leads to an extra correction to Eq.~\eqref{eq:a_b_correlations}, in the form of
\begin{equation}
\begin{aligned}
    & \big\langle \Psi_A (t^-) \Psi^\dagger_A (0^+)\big\rangle_\text{Intra} = \big\langle \Psi^\dagger_A (t^-) \Psi_A (0^+)\big\rangle_\text{Intra} = \big\langle \Psi_B (t^-) \Psi^\dagger_B (0^+)\big\rangle_\text{Intra} = \big\langle \Psi^\dagger_B (t^-) \Psi_B (0^+)\big\rangle_\text{Intra}\\
    &  = \delta_\text{coulomb} \frac{\tau_0^{n_\text{coulomb}-1}}{2 \pi (\tau_0 + i t)^{n_\text{coulomb}}},
\end{aligned}
    \label{eq:coulomb_interaction}
\end{equation}
where the subscript ``Intra'' highlights the fact that these extra terms are generated by intra-edge interactions, and $\delta_\text{coulomb}$ is a dimensionless quantity that depends on both the amplitude of Coulomb interaction and the channel filling fraction. 
Following similar setups of e.g., Ref.~\cite{WahlMartinPRL14S,GuNC24S}, Eq.~\eqref{eq:coulomb_interaction} can be generated via the charge fractionalization effect that is present with or without non-equilibrium anyons before the central colliders.
Importantly, when $n_\text{coulomb}$ is smaller than $1/\nu$, the interaction-induced current noise becomes much larger than the interaction-free single-source noise.
Indeed, in the presence of intra-edge interactions, an extra tunneling current noise $\delta S_\text{T}^\text{coulomb}$ is produced:
\begin{equation}
\begin{aligned}
    \delta S_\text{T}^\text{coulomb} & = \delta_\text{coulomb}  \frac{e^2\tau_0^{\nu_\text{C}}}{(2\pi \tau_0)^2}  2\Gamma \left(1 - \nu_\text{C} \right) (\nu e V)^{\nu_\text{C} - 1}\, \mathcal{T}_C^{(0)} \\
& \left(\left\{ \sin\left( \nu_\text{C} \pi \right) + 2 \nu (\nu_\text{C} - 1) \sin (\pi\nu) \left(\sin \left[ ( \nu_\text{C} - \nu ) \pi\right] + \sin (\pi\nu) \right) \frac{I_{A0}}{\nu^3 e V} \right\} \mathcal{T}_A^{(0)} \right.\\
&\left. +\left\{\sin\left( \nu_\text{C} \pi \right) + 2 \nu (\nu_\text{C} - 1) \sin (\pi\nu) \left(\sin \left[ ( \nu_\text{C} - \nu ) \pi\right] + \sin (\pi\nu) \right) \frac{I_{B0}}{\nu^3 e V} \right\} \mathcal{T}_B^{(0)} \right) ,
\end{aligned}
\label{eq:st_coulomb}
\end{equation}
where $\nu_\text{C} = n_\text{coulomb} + 2\nu - 2 + 1/\nu = \nu_\text{s} + n_\text{coulomb} - 1/\nu$.
The relative amplitude of the interaction-induced noise is then given by
$$\delta S_\text{T}^\text{coulomb}/S_\text{T}^\text{single} \sim \delta_\text{coulomb} (\tau_0 \nu e V)^{\nu_\text{C} - \nu_\text{s}}.$$

When $n_\text{coulomb} < 1/\nu$, we have $\nu_\text{C} < \nu_\text{s}$, such that $\delta S_\text{T}^\text{coulomb}$ can become of the same order, or even larger than the interaction-free noise, $S_\text{T}^\text{single}$ for sufficiently small $\tau_0 \nu e V$.
This is indeed the situation for interacting quantum Hall edges~\cite{GuNC24S}, where Coulomb interaction produces a term that is of a smaller order in the transmission probability of diluters, as a result of ``charge fractionalization''~\cite{WahlMartinPRL14S}.
In the weak-tunneling limit, this interaction-induced correction thus becomes more significant than the interaction-free result.
Luckily, for both our case and that of Ref.~\cite{GuNC24S}, such a correction is avoided when considering the entanglement pointer.
This relies on the fact that $S_\text{T,coulomb}$ does not contain the cross-source contributions, i.e., those proportional to $\mathcal{T}_A \mathcal{T}_B$. Because of this, $\delta S^\text{coulomb}_\text{T}$ from the double-source collision contribution is simply equal to the sum of two single-source contributions:
\begin{equation}
\begin{aligned}
    &\delta S^\text{coulomb}_\text{T}(\mathcal{T}_A,\mathcal{T}_B) - \delta S^\text{coulomb}_\text{T} (\mathcal{T}_A,0) - \delta S^\text{coulomb}_\text{T} (0,\mathcal{T}_B) = 0,\\
    & \delta \mathcal{P}^\text{coulomb}_\text{Andreev} = 0,
\end{aligned}
\end{equation}
where $\delta \mathcal{P}^\text{coulomb}_\text{Andreev}$ refers to the modification on the entanglement pointer [cf. the definition, Eq.~(1) of the main text] induced by the Coulomb interaction along the channel.

We conclude that the entanglement pointer defined in the present work is resilient to the influence from intra-edge Coulomb interactions within the edges.
This consideration applies to composite anyonic edges that host multiple co-propagating channels.
Importantly, it is also potentially applicable to other high-order anyonic edges (e.g., $\nu=2/3$, $2/5$) and, more interestingly, non-Abelian ones ($\nu=5/2$), hosting counter-propagating channels.
Considering the significance of these edges (non-Abelian $5/2$ edges) in the frontier of quantum research, the resilience of our entanglement pointer is highly important to relevant theoretical advances and practical applications.

\section{{\,\,\,\,\,\,\,}Single-particle and two-particle scattering}

In the main text, we provide analysis on the tunneling current noise and cross-correlation noise, within the scattering formalism \cite{MartinLandauerPRB92S,BlanterButtikerPhysRep00S}.
In this section, we detail how to arrive at Eq.~(5) of the main text.
Here, we consider cases where involved particles are either uncorrelated (i.e., where the particles are independent of each other) or correlated.
We introduce the probabilities of relevant processes $\mathbf{P}^{(s)}_{(N_A^0, N_B^0)}(N_A,N_B)$ and $\mathbf{P}^{(d)}(N_A,N_B)$, with the superscript $(s)$ and $(d)$ referring to the uncorrelated single-particle and the correlated double-particle system.
The arguments $N_A$ and $N_B$ refer to the final state that contains $N_A$ and $N_B$ non-equilibrium anyons in channels $A$ and $B$ after the central collider.
Notice that $e N_A$ and $e N_B$ are not equal to the non-equilibrium charges in channels $A$ and $B$, as charges that tunnel through the central collider ($e$, being fermionic) are different from those ($\nu e$) of transporting anyonic particles before the central collider.
For the single-particle process, the subscript $(N_A^0, N_B^0)$ marks the numbers of the particles involved, supplied from the sources $A$ and $B$.

This section addresses three major tasks.
Firstly, we express differential noises [see Eq.~\eqref{eq:correlation_in_energy} for the definition] in terms of probabilities $\mathbf{P}^{(s)}_{(N_A^0, N_B^0)}(N_A,N_B)$ and $\mathbf{P}^{(d)}(N_A,N_B)$.
Secondly, by comparing noises between correlated and uncorrelated situations, we identify and extract noises that are relevant to the entanglement pointer.
Finally, with the second task accomplished, we explicitly show that the entanglement pointer can be obtained by measuring auto- and cross-correlations in real experiments.
For later convenience, here we define the ``differential noise'',
\begin{equation}
\begin{aligned} 
       s^\text{type}(\hat{I}_1, \hat{I}_2)
    \equiv \partial_{e I_\text{neq}} \int dt \langle  \delta \hat{I}_1 (t) \delta \hat{I}_2 (0) \rangle_\text{type} +  \frac{\langle  \hat{I}_1 \rangle \langle \hat{I}_2 \rangle}{I_\text{neq}^2},
\end{aligned}
    \label{eq:correlation_in_energy}
\end{equation}
as the correlation function of current operators $\hat{I}_1$ and $\hat{I}_2$ (which can refer to $\hat{I}_\text{T}$, $\hat{I}_{A}$ and $\hat{I}_B$),
where $I_\text{neq} = \max (I_{A0},I_{B0})$ represent the amplitude of non-equilibrium current and
$\delta \hat{I}_1 \equiv  \hat{I}_1 - \langle \hat{I}_1\rangle$ refers to current fluctuations.
The subscript ``type'' refers to the type of particles involved; it is ``cl'' and ``anyon'', for uncorrelated (``classical'') and correlated anyons, respectively.
Notice that $s^\text{type}(\hat{I}_1, \hat{I}_2)$ is dimensionless.
Similarly, the correlation function of current fluctuations is defined as $s^\text{type}(\delta \hat{I}_1,\delta \hat{I}_2)
= s^\text{type}(\hat{I}_1, \hat{I}_2) - \langle \hat{I}_1\rangle\langle \hat{I}_2\rangle/I_\text{neq}^2$.
When ``type''$=$``anyon'', the correlator $\langle \ldots\rangle_\text{anyon}$ in Eq.~\eqref{eq:correlation_in_energy} can be straightforwardly evaluated via bosonization of vertex operators; When ``type''$=$``cl'', the correlator $\langle \ldots\rangle_\text{cl}$ equals the sum of two correlators for single-source cases. Alternatively, the correlation functions appearing in  Eq.~\eqref{eq:correlation_in_energy} can be expressed in terms of probabilities $\mathbf{P}^{(s)}_{(N_A^0, N_B^0)}(N_A,N_B)$ and $\mathbf{P}^{(d)}(N_A,N_B)$, following the scattering formalisms of, e.g., Ref.~\cite{BlanterButtikerPhysRep00S}.

The single-particle and two-particle scattering pictures apply to the analysis of non-interacting fermionic and bosonic systems, where non-equilibrium particles participate in scatterings at the tunneling QPC.
This tunneling of non-equilibrium particles at the QPC turns out to be crucial to the application of the scattering method.
Indeed, in  Refs.~\cite{RosenowLevkivskyiHalperinPRL16S,LeeSimNC22S,MorelPRB22S} where non-equilibrium particles do not tunnel at the central QPC, the obtained results are not described by the scattering picture. 
For the Andreev-tunneling situation, luckily, the leading contribution involves tunnelings of non-equilibrium particles, thus enabling the application of scattering theory.
Indeed, as a piece of evidence, now correlation functions Eqs.~\eqref{eq:dan_resum} and \eqref{eq:dan_resum_b} can be divided into the equilibrium (i.e., the factor ``1'') and non-equilibrium contributions, which is impossible for Refs.~\cite{RosenowLevkivskyiHalperinPRL16S,LeeSimNC22S,MorelPRB22S} that allow anyons to tunnel.

\subsection{Tunneling noise}

We first address the tunneling noise $S_\text{T}$.
As a benchmark, we begin by considering the reducible differential tunneling noise
\begin{equation}
\begin{aligned}
s^\text{cl}(\hat{I}_\text{T},\hat{I}_\text{T})
& = \mathbf{P}^{(s)}_{(0,1)}(1,0) + \mathbf{P}^{(s)}_{(1,0)}(0,1) + \mathbf{P}^{(d)} (2,0) + \mathbf{P}^{(d)} (0,2)\\
&= \mathcal{T}_A (1 - \mathcal{T}_B) \mathcal{T}_C + \mathcal{T}_B (1 - \mathcal{T}_A) \mathcal{T}_C + 2\mathcal{T}_A \mathcal{T}_B  \mathcal{T}_C (1 - \mathcal{T}_C) = (\mathcal{T}_A + \mathcal{T}_B) \mathcal{T}_C - 2 \mathcal{T}_C^2 \mathcal{T}_A \mathcal{T}_B ,
\end{aligned}
\label{eq:it_dis_reducible}
\end{equation}
that refers to noise generated by quasiparticles with the energy $2\pi I_\text{neq}/e$ [cf. Eq.~\eqref{eq:correlation_in_energy} for the definition], while
$\mathcal{T}_A$ and $\mathcal{T}_B$ refer to the probability to have a non-equilibrium particle from sources $A$ and $B$, respectively.
We notice that Eq.~\eqref{eq:it_dis_reducible} contains a term $\mathcal{T}_C^2 \mathcal{T}_A \mathcal{T}_B$ that is proportional to the two-particle scattering probability $\mathcal{T}_A \mathcal{T}_B$.
This term, however, disappears in the irreducible correlation, after the removal of the differential current average product $\langle \hat{I}_\text{T}/I_\text{neq}\rangle^2 = \mathcal{T}_C^2 (\mathcal{T}_A - \mathcal{T}_B)^2$. Indeed, now the irreducible correlation of the uncorrelated case becomes
\begin{equation}
s^\text{cl}(\delta \hat{I}_\text{T},\delta \hat{I}_\text{T})
= s^\text{cl}(\hat{I}_\text{T},\hat{I}_\text{T}) - \langle \hat{I}_\text{T}/I_\text{neq}\rangle^2 = (\mathcal{T}_A + \mathcal{T}_B) \mathcal{T}_C - (\mathcal{T}_A^2 + \mathcal{T}_B^2) \mathcal{T}_C^2,
\label{eq:it_dis}
\end{equation}
where $\delta \hat{I}_\text{T} \equiv \hat{I}_\text{T} - \langle \hat{I}_\text{T} \rangle$ is tunneling current fluctuation operator.
Here Eq.~\eqref{eq:it_dis} is irrelevant to the two-particle scattering probability $\propto \mathcal{T}_A \mathcal{T}_B$.
More specifically, it is equal to the sum of the probabilities of two independent single-particle tunneling processes: a solid benchmark for the absence of quantum statistics.
This fact, importantly, indicates that one can observe the statistical message from bilinear terms $\propto \mathcal{T}_A \mathcal{T}_B$.

Now we move to consider correlated particles.
When two anyons arrive at the central QPC simultaneously, the chance of having an Andreev-like tunneling is modified, in comparison to the uncorrelated case.
For simplicity, we call the probability of having the Andreev-like tunneling (when two anyons collide) as $P_\text{Andreev}^\text{anyon}$ and $1 - P_\text{Andreev}^\text{anyon}$ the probability of the scattering event without the Andreev-like tunneling.
The modification only exists in the reducible part, $\mathbf{P}^{(d)} (2,0) + \mathbf{P}^{(d)} (0,2)$, leading to 
\begin{equation}
\begin{aligned}
s_\text{T} \equiv s^\text{anyon}(\delta \hat{I}_\text{T},\delta \hat{I}_\text{T})
& = \mathcal{T}_A (1-\mathcal{T}_B) \mathcal{T}_C + \mathcal{T}_B (1 - \mathcal{T}_A) \mathcal{T}_C + \mathcal{T}_A \mathcal{T}_B P_\text{Andreev}^\text{anyon} - \mathcal{T}_C^2 (\mathcal{T}_A - \mathcal{T}_B)^2\\
& = (\mathcal{T}_A + \mathcal{T}_B) \mathcal{T}_C - ( \mathcal{T}_A^2 + \mathcal{T}_B^2 ) \mathcal{T}_C^2 + \mathcal{T}_A \mathcal{T}_B P_\text{Andreev}^\text{stat},
\end{aligned}
\label{eq:it_anyons}
\end{equation}
where $s_\text{T} \equiv \partial_{e I_+} S_\text{T}$ is defined in the main text, and $P_\text{Andreev}^\text{stat} = P_\text{Andreev}^\text{anyon} - P_\text{Andreev}^\text{cl}$ is the function that quantifies the Andreev-like tunneling probability from pure anyonic statistics.
Indeed, $P_\text{Andreev}^\text{stat}$ equals the difference between two functions: (i) the chance of Andreev-like tunneling where two involved anyons are uncorrelated, $P_\text{Andreev}^\text{cl} = 2 \mathcal{T}_C (1 - \mathcal{T}_C) $; and (ii) the function $P_\text{Andreev}^\text{anyon}$ that refers to the Andreev-like tunneling when all anyons are correlated.
After removing statistics-irrelevant contributions [first two terms of Eq.~\eqref{eq:it_anyons}, which perfectly equals that in Eq.~\eqref{eq:it_dis}], the rest noise discloses the influence of anyonic statistics on Andreev-like tunnelings.
Actually, by comparing Eq.~\eqref{eq:it_anyons} to Eqs.~(2) and (3) of the main text, 
one immediately notices that $S^\text{single}_\text{T}$ corresponds to the linear term (the first term) of Eq.~\eqref{eq:it_anyons}, in $\mathcal{T}_A$ or $\mathcal{T}_B$.
Its bilinear term (i.e., that is proportional to $\mathcal{T}_A \mathcal{T}_B$), on the other hand, is captured by the double-source collision contribution, $S_\text{T}^\text{collision}$, which yields $\mathcal{P}_\text{Andreev}$ following its definition.
With this message in mind, and the definition of $\mathcal{P}_\text{Andreev}$ in the main text, we can also express $\mathcal{P}_\text{Andreev}$ as
\begin{equation}
    \mathcal{P}_\text{Andreev} = -\frac{e}{2}\int dI_+ \mathcal{T}_A \mathcal{T}_B P_\text{Andreev}^\text{stat} (I_+).
\end{equation}

As another feature, all expressions for the tunneling-current noise do not contain fractional charge $e^*=\nu e$, as only electrons are allowed to tunnel through the central QPC.
As will be shown shortly, this feature greatly contrasts that for cross and auto correlations.

Before ending, we comment that following the differential form of $s_\text{T}$, Eq.~\eqref{eq:it_anyons}, and the comparison to Eqs.~(2) and (3) of the main text, $P_\text{Andreev}^\text{stat}$ can actually be described in terms of microscopic parameters, i.e.,
\begin{equation}
    P_\text{Andreev}^\text{stat} = \frac{4\pi}{\nu} \text{Re} \left\{\mathcal{T}_C\, \frac{ \, f_2(\nu) \cos\left(\pi \nu_\text{d}/{2} \right) }{\pi\nu\sin\left( \pi \nu_\text{s} \right) \sqrt{\mathcal{T}_A \mathcal{T}_B} + 2 f_1 (\nu) \mathcal{T}_A \mathcal{T}_B}  \,     \frac{\left[\mathcal{T}_A \left( 1 - e^{-2i\pi\nu} \right) + \mathcal{T}_B \left( 1 - e^{2i\pi\nu} \right)\right]^{\nu_\text{d} - 1}}{(\mathcal{T}_A + \mathcal{T}_B) } \right\}.
\end{equation}

\subsection{Cross-correlation noise}

Now we move on to consider the cross-correlation noise.
Once again, we start with the uncorrelated situation.
The reducible part of the cross-correlation differential noise then becomes
\begin{equation}
\begin{aligned}
s^\text{cl}( \hat{I}_{A}, \hat{I}_{B})
& = (\nu\! -\! 1) \left[\mathbf{P}^{(s)}_{(0,1)}(1,0)\! +\! \mathbf{P}^{(s)}_{(1,0)}(0,1)\right]\!  +\ (\nu^2\! -\! 1)\left[ \mathbf{P}^{(d)} (2,0)\! +\! \mathbf{P}^{(d)} (0,2)\right]\! +\! \nu^2 \left[ \mathbf{P}^{(d)} (1,1)\! +\! \mathbf{P}^{(d)} (1,1)\right]\\
& = (\nu\! -\! 1) \mathcal{T}_C \mathcal{T}_A (1-\mathcal{T}_B) \!+\! (\nu \! -\! 1) \mathcal{T}_C \mathcal{T}_B (1\! -\! \mathcal{T}_A) \!+\! \mathcal{T}_A \mathcal{T}_B \left[P_\text{Andreev}^\text{cl} (\nu^2 - 1) \!+\! (1 - P_\text{Andreev}^\text{cl}) \nu^2 \right]\\
& = (\nu\! -\! 1)\, \mathcal{T}_C\, [ \mathcal{T}_A (1-\mathcal{T}_B) + \mathcal{T}_B (1-\mathcal{T}_A) ] + \mathcal{T}_A \mathcal{T}_B \left(\nu^2 - P_\text{Andreev}^\text{cl}  \right),
\end{aligned}
\end{equation}
while the differential current average product now equals 
$$\frac{\langle \hat{I}_{A} \rangle \langle \hat{I}_{B} \rangle}{I_\text{neq}^2} = [\nu \mathcal{T}_A - \mathcal{T}_C (\mathcal{T}_A - \mathcal{T}_B)][\nu \mathcal{T}_B + \mathcal{T}_C (\mathcal{T}_A - \mathcal{T}_B)],$$
where the factor of $\nu$ refers to the fractional charge of an non-equilibrium anyon, and the factor of ``1'' in the other term (the term proportional to $\mathcal{T}_C$) refers to the transmission of a full electron across the central QPC.
With the reducible noise, and the current average product, we arrive at the irreducible differentiyl cross-correlation noise
\begin{equation}
\begin{aligned}
s^\text{cl}( \delta \hat{I}_{A}, \delta \hat{I}_{B})
& = -(1-\nu) \mathcal{T}_C (\mathcal{T}_A + \mathcal{T}_B) - \mathcal{T}_C ( \nu - \mathcal{T}_C ) (\mathcal{T}_A^2 + \mathcal{T}_B^2 ),
\end{aligned}
\label{eq:iaib_dis}
\end{equation}
which, similarly as the tunneling current noise Eq.~\eqref{eq:it_anyons}, does not contain the bilinear contribution $\propto \mathcal{T}_A \mathcal{T}_B$, and thus can be considered as the summation of two single-particle processes.
The second term of Eq.~\eqref{eq:iaib_dis}, corresponding to the current average product of single-source situation, has an apparent Andreev tunneling signature: the charge equals $\nu$ (corresponding to the non-equilibrium anyon) without charge tunneling, but becomes $\nu - 1$ (corresponding to the reflected hole) after the tunneling.

Now we move on to the situation where anyons in edges $A$ and $B$ are correlated, leading to the irreducible cross-correlation differential noise
\begin{equation}
\begin{aligned}
 s_{AB}& = s^\text{anyon}( \delta \hat{I}_{A}, \delta \hat{I}_{B})
 \\
& = (\nu  -  1) \mathcal{T}_C \mathcal{T}_A (1-\mathcal{T}_B)  +  (\nu  -  1) \mathcal{T}_C \mathcal{T}_B (1 - \mathcal{T}_A) + \mathcal{T}_A \mathcal{T}_B \left[(\nu^2 - 1)\,P_\text{Andreev}^\text{anyon}  +  \nu^2\,(1 - P_\text{Andreev}^\text{anyon}) \right]\\
& = (\nu - 1) \mathcal{T}_C \mathcal{T}_A (1-\mathcal{T}_B) + (\nu - 1) \mathcal{T}_C \mathcal{T}_B (1-\mathcal{T}_A) \\
& + \mathcal{T}_A \mathcal{T}_B \left\{ \left(\nu^2 - 1\right)( P_\text{Andreev}^\text{cl} + P_\text{Andreev}^\text{anyon} - P_\text{Andreev}^\text{cl})   +  \nu^2[1 - ( P_\text{Andreev}^\text{cl} + P_\text{Andreev}^\text{anyon} - P_\text{Andreev}^\text{cl})] \right\}\\
& = (\nu - 1) \mathcal{T}_C \mathcal{T}_A (1-\mathcal{T}_B) + (\nu - 1) \mathcal{T}_C \mathcal{T}_B (1-\mathcal{T}_A) + \mathcal{T}_A \mathcal{T}_B \left[  \left(\nu^2 - 1 \right)P_\text{Andreev}^\text{cl}  +  \nu^2 (1 - P_\text{Andreev}^\text{cl})  \right] 
\\
& 
- \mathcal{T}_A \mathcal{T}_B (P_\text{Andreev}^\text{anyon} - P_\text{Andreev}^\text{cl})\\
& = -(1-\nu) \mathcal{T}_C (\mathcal{T}_A + \mathcal{T}_B) - \mathcal{T}_C ( \nu - \mathcal{T}_C ) (\mathcal{T}_A^2 + \mathcal{T}_B^2 ) - \mathcal{T}_A \mathcal{T}_B P_\text{Andreev}^\text{stat}.
\end{aligned}
\label{eq:anyon_cc}
\end{equation}
Comparison of Eqs.~\eqref{eq:it_anyons} and \eqref{eq:anyon_cc} shows that for the leading contribution, i.e., terms linear in $\mathcal{T}_A$ or $\mathcal{T}_B$, the tunneling noise and the cross-correlation noise are proportional to each other.
This proportionality agrees with the experimental measurement of Ref.~\cite{PierreNC23S}.
More importantly, the bilinear term, i.e., the statistics-induced contribution, can be extracted via either tunneling current noise, or the cross-correlation: indeed, the obtained statistics-induced noise has only a difference in sign.
This result indicates that one can obtain the entanglement pointer of Andreev-like tunnelings, through either tunneling current, or cross-correlation noise measurements, whichever is more convenient.

\subsection{Auto-correlation}

Finally, we move to consider two auto-correlations, $\langle \delta \hat{I}_{A}^2 \rangle_\text{anyon}$ and $\langle \delta \hat{I}_{B}^2 \rangle_\text{anyon}$.
Once again, we start with the benchmark scenario where anyons in $A$ are uncorrelated from those in $B$.
In this case, the reducible differential correlations are given by
\begin{equation}
\begin{aligned}
   & s^\text{cl}( \hat{I}_{A}, \hat{I}_{A})\\
    & =  \nu^2\mathbf{P}^{(s)}_{(1,0)}(1,0) + \mathbf{P}^{(s)}_{(0,1)}(1,0) + (1-\nu)^2 \mathbf{P}^{(s)}_{(1,0)}(0,1) + (1-\nu)^2\mathbf{P}^{(d)} (0,2) + (1+\nu)^2\mathbf{P}^{(d)} (2,0) + \nu^2 \mathbf{P}^{(d)} (1,1) \\
    & = \nu^2 (1 - \mathcal{T}_C ) \mathcal{T}_A ( 1- \mathcal{T}_B ) + \mathcal{T}_C \mathcal{T}_B (1 - \mathcal{T}_A) + \mathcal{T}_C \mathcal{T}_A ( 1- \mathcal{T}_B) (1 - \nu)^2 \\
    & + \mathcal{T}_A \mathcal{T}_B \left[ \frac{P_\text{Andreev}^\text{cl}}{2} (\nu - 1)^2 + \frac{P_\text{Andreev}^\text{cl}}{2} (\nu + 1)^2 + (1 - P_\text{Andreev}^\text{cl}) \nu^2 \right]\\
    & = \nu^2  \mathcal{T}_A ( 1- \mathcal{T}_B ) + \mathcal{T}_C [\mathcal{T}_B (1 - \mathcal{T}_A) + (1 - 2\nu) \mathcal{T}_A (1 - \mathcal{T}_B)]  + \mathcal{T}_A \mathcal{T}_B (P_\text{Andreev}^\text{cl} + \nu^2),
    \end{aligned}
\end{equation}
\begin{equation}
\begin{aligned}
      & s^\text{cl}( \hat{I}_{B}, \hat{I}_{B})\\
    & =  \nu^2\mathbf{P}^{(s)}_{(0,1)}(0,1) + \mathbf{P}^{(s)}_{(1,0)}(0,1) + (1-\nu)^2 \mathbf{P}^{(s)}_{(0,1)}(1,0) + (1-\nu)^2\mathbf{P}^{(d)} (2,0) + (1+\nu)^2\mathbf{P}^{(d)} (0,2) + \nu^2 \mathbf{P}^{(d)} (1,1) \\
    & = \nu^2 (1 - \mathcal{T}_C ) \mathcal{T}_B ( 1- \mathcal{T}_A ) + \mathcal{T}_C \mathcal{T}_A (1 - \mathcal{T}_B) + \mathcal{T}_C \mathcal{T}_B ( 1- \mathcal{T}_A) (1 - \nu)^2 \\
    & + \mathcal{T}_A \mathcal{T}_B \left[ \frac{P_\text{Andreev}^\text{cl}}{2} (\nu - 1)^2 + \frac{P_\text{Andreev}^\text{cl}}{2} (\nu + 1)^2 + (1 - P_\text{Andreev}^\text{cl}) \nu^2 \right]\\
    & = \nu^2  \mathcal{T}_B ( 1- \mathcal{T}_A ) + \mathcal{T}_C [\mathcal{T}_A (1 - \mathcal{T}_B) + (1 - 2\nu) \mathcal{T}_B (1 - \mathcal{T}_A)]  + \mathcal{T}_A \mathcal{T}_B (P_\text{Andreev}^\text{cl} + \nu^2).
\end{aligned}
\end{equation}
We can again use the current averages $\langle \hat{I}_{A} \rangle/I_\text{neq} = \nu \mathcal{T}_A - \mathcal{T}_C (\mathcal{T}_A - \mathcal{T}_B)$ and $\langle \hat{I}_{B} \rangle/I_\text{neq} = \nu \mathcal{T}_B + \mathcal{T}_C (\mathcal{T}_A - \mathcal{T}_B)$, to rewrite the reducible differential correlations into irreducible ones,
\begin{equation}
\begin{aligned} 
    s^\text{cl}( \delta \hat{I}_{A}, \delta \hat{I}_{A})& = \mathcal{T}_A [ \mathcal{T}_C + \nu (1 - \mathcal{T}_A) (\nu - 2\mathcal{T}_C) ] + \mathcal{T}_B \mathcal{T}_C - (\mathcal{T}_A^2 + \mathcal{T}_B^2) \mathcal{T}_C^2,\\
     s^\text{cl}( \delta \hat{I}_{B}, \delta \hat{I}_{B})& = \mathcal{T}_B [ \mathcal{T}_C + \nu (1 - \mathcal{T}_B) (\nu - 2\mathcal{T}_C) ] + \mathcal{T}_A \mathcal{T}_C - (\mathcal{T}_A^2 + \mathcal{T}_B^2) \mathcal{T}_C^2,
\end{aligned}
\end{equation}
which also display the separation of contributions from different edges.

For the situation where all anyons are perfectly indistinguishable (being correlated), once again only the value of $P_\text{Andreev}^\text{cl}$ is replaced by $P_\text{Andreev}^\text{anyon}$, leading to modified differential auto-correlations
\begin{equation}
\begin{aligned}
   s_{AA}\equiv s^\text{anyon}( \delta \hat{I}_{A}, \delta \hat{I}_{A})
   & = \mathcal{T}_A [ \mathcal{T}_C + \nu (1 - \mathcal{T}_A) (\nu - 2\mathcal{T}_C) ] + \mathcal{T}_B \mathcal{T}_C - (\mathcal{T}_A^2 + \mathcal{T}_B^2) \mathcal{T}_C^2 + \mathcal{T}_A \mathcal{T}_B   P_\text{Andreev}^\text{stat} ,\\
   s_{BB}\equiv s^\text{anyon}( \delta \hat{I}_{B}, \delta \hat{I}_{B})
   & = \mathcal{T}_B [ \mathcal{T}_C + \nu (1 - \mathcal{T}_B) (\nu - 2\mathcal{T}_C) ] + \mathcal{T}_A \mathcal{T}_C - (\mathcal{T}_A^2 + \mathcal{T}_B^2) \mathcal{T}_C^2 + \mathcal{T}_A \mathcal{T}_B P_\text{Andreev}^\text{stat},
\end{aligned}
\end{equation}
which contains the same form of the statistical term $\mathcal{T}_A \mathcal{T}_B P_\text{Andreev}^\text{stat}$, as that in Eqs.~\eqref{eq:it_anyons} and \eqref{eq:anyon_cc}, for tunneling current noise and cross-correlation, respectively.
In addition, the same as cross-correlation and tunneling current noise, the only bilinear term of auto correlations equal $\mathcal{T}_A \mathcal{T}_B P_\text{Andreev}^\text{stat}$, meaning that $\mathcal{P}_\text{Andreev}$ can also be measured with the auto-correlation:
\begin{equation}
\begin{aligned}
    \mathcal{P}_\text{Andreev} & = S_\text{T} (\mathcal{T}^{(0)}_A,0) + S_\text{T} (0,\mathcal{T}^{(0)}_B) - S_\text{T} (\mathcal{T}^{(0)}_A, \mathcal{T}^{(0)}_B) \\
    & =- [S_{AB} (\mathcal{T}^{(0)}_A,0) + S_{AB} (0,\mathcal{T}^{(0)}_B) - S_{AB} (\mathcal{T}^{(0)}_A, \mathcal{T}^{(0)}_B) ]\\
    & = S_{AA} (\mathcal{T}^{(0)}_A,0) + S_{AA} (0,\mathcal{T}^{(0)}_B) - S_{AA} (\mathcal{T}^{(0)}_A, \mathcal{T}^{(0)}_B) \\
    & = S_{BB} (\mathcal{T}^{(0)}_A,0) + S_{BB} (0,\mathcal{T}^{(0)}_B) - S_{BB} (\mathcal{T}^{(0)}_A, \mathcal{T}^{(0)}_B) ,
\end{aligned}
\label{eq:ep_different_forms}
\end{equation}
with similar definitions, by removing single-source contributions.
Here, $$S_{AA} \equiv \int dt \langle \delta \hat{I}_{A} (t) \delta \hat{I}_{A} (0) \rangle \quad \text{and} \quad S_{BB} \equiv \int dt \langle \delta \hat{I}_B (t) \delta \hat{I}_B (0) \rangle$$ refer to the auto-correlations.

\section{{\,\,\,\,\,\,\,}Experiment}

In this section, we briefly describe the experimental setup of Ref.~\cite{PierreNC23S} used to obtain the data, which we analyze in the main text in the context of the theory of entanglement pointer. 
\begin{figure}
  \includegraphics[width=0.7\linewidth]{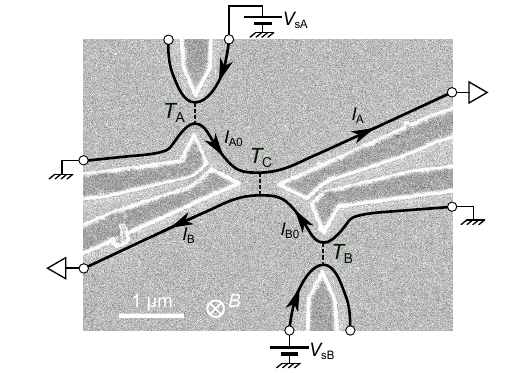}
  \caption{E-beam micrograph of the experimental device (cf. Ref.~\cite{PierreNC23S}), which is schematically shown in Fig.~1 of the main text.
  The QPCs are formed by applying negative voltages to metallic gates deposited at the surface (darker with bright edges). The chiral edge channels are displayed as continuous black lines with arrows. The tunneling processes take place along the dashed lines. The source QPCs in subsystems $\mathcal{A}$ and $\mathcal{B}$ are set for $e/3$ quasiparticle tunnelings, whereas the central QPC is tuned to $e$ quasielectron tunnelings. The tunneling quasiparticles are ascertained from shot noise measurements of the tunneling charge, in the presence of a direct voltage bias applied to the considered QPC.
    }
  \label{fig:device}
\end{figure}

The experiment is performed on the Ga(Al)As  device shown in Fig.~\ref{fig:device} (see Ref.~\cite{PierreNC23S} for details). 
It is cooled at an electronic temperature of 35\,mK and set at the center of the $\nu=1/3$ fractional quantum Hall plateau.
The spectral density of the current auto- and cross-correlations $\big<\delta \hat{I}_{A}^2\big>$, $\big<\delta \hat{I}_{B}^2\big>$ and $\big<\delta \hat{I}_{A} \delta \hat{I}_{B}\big>$ are simultaneously measured around a frequency of $0.86$\,MHz.
The dc currents $I_{A,B,\text{T}}$ are obtained by integrating the differential conductances $\partial I_{A,B,\text{T}}/\partial V_{sA,sB}$ directly measured by standard lock-in techniques at frequencies below 100\,Hz.

Importantly, the present data-theory comparison is performed on a specific data set, which was measured following a protocol optimized to limit as much as possible any changes between the different configurations of the sources.
This is essential for the entanglement pointer, which is obtained from the small difference of large signals.
Note that the data shown in the main text of Ref.~\cite{PierreNC23S} do not fully follow the procedure described below:\\
\noindent
First, the source QPCs are activated not by changing the gate voltage controlling their transmission parameter $\mathcal{T}_{A,B}$ but instead by setting the dc bias voltage $V_{sA,sB}$ to $V$.
Indeed, changing the gate voltage controlling one source (e.g. in branch $A$) would also change the other transmissions ($\mathcal{T}_{B}$ and $\mathcal{T}_{C}$) by capacitive crosstalk, and thereby introduce unwanted artifacts in $P_\mathrm{Andreev}$.
Note that the applied dc bias voltage itself also acts electrostatically on the QPCs. 
This can play a role, as further discussed in the experiment-theory comparison, yet it is a smaller effect since the bias voltage changes ($V_{sA,sB}\lesssim0.1$\,mV) are much smaller than the gate voltage changes to open or close a QPC ($\sim1\,$V).\\
\noindent
Second, the necessary averaging time is split in several sequences alternating between the following successive configurations: \textit{(i)} source $sA$ is ON and $sB$ is OFF ($V_{sA}=V$, $ V_{sB}=0$), \textit{(ii)} source $sA$ is OFF and $sB$ is ON ($V_{sA}=0$, $V_{sB}=V$), \textit{(iii)} The central QPC is directly voltage biased for tunneling charge characterization, and \textit{(iv)} sources $sA$ and $sB$ are both ON ($V_{sA}=V_{sB}=V$).
This allows us to effectively cancel out in $P_\text{Andreev}$ the small drifts of the QPCs with time, which could otherwise have a noticeable impact.

\section{{\,\,\,\,\,\,\,}Details on the experiment-theory comparison}
\label{ssec:comparison}

We now compare our theoretical results with the experimental data.
This comparison involves $P_\text{Andreev}$ obtained with two methods.
Within the first method, $P_\text{Andreev}$ is obtained directly following its definition, Eq.~(1) of the main text, and Eq.~\eqref{eq:tunneling_current_and_noise-Sc}.
The latter equation allows us to obtain $P_\text{Andreev}$ alternatively, with the explicit current and noise expressions provided.
The tunneling noise is, however, not easily accessible experimentally.
To proceed, we thus need to establish the relation between $S_\text{T}^\text{collision}$ and the quantities measured in the experiment.
In short, following Eq.~\eqref{eq:tunneling_current_and_noise-Ic} [alternatively, Eq.~\eqref{eq:noise_single_full}], the double-source collision contribution to the tunneling current, $I^\text{collision}_\text{T}$, is related to the corresponding contribution to the noise $S^\text{collision}_\text{T} $, Eq.~\eqref{S-coll-nice}, via 
\begin{equation}
\begin{aligned}
    &\frac{\partial}{\partial I_-} I_\text{T}^\text{collision} \Big|_{I_- = 0} 
    \\
    &=  
    e \frac{\tau_0^{\nu_\text{d}}}{(2 \pi \tau_0^2)} \mathcal{T}_A^{(0)} \mathcal{T}_B^{(0)} \mathcal{T}_C^{(0)}  4 \sin\left( \frac{\pi \nu_\text{d}}{2} \right) \Gamma (1 - \nu_\text{d})  \frac{\partial}{\partial I_-} \text{Im} \left\{ \left[\frac{I_{A0}}{\nu e} \left( 1 - e^{-2i\pi\nu} \right) + \frac{I_{B0}}{\nu e} \left( 1 - e^{2i\pi\nu} \right)\right]^{\nu_\text{d} - 1} \right\} \Bigg|_{I_- = 0}
    \\
    & = \frac{e^2 V \sqrt{\mathcal{T}_A \mathcal{T}_B}\, \mathcal{T}_C\, f_2(\nu) \sin \left(\pi \nu_\text{d}/{2} \right)}{\pi\nu\sin\left( \pi \nu_\text{s} \right) + 2 f_1 (\nu) \sqrt{\mathcal{T}_A \mathcal{T}_B}  }  \frac{\partial}{\partial I_-
    }  \text{Im} \left\{ \left[\mathcal{T}_A \left( 1 - e^{-2i\pi\nu} \right) + \mathcal{T}_B \left( 1 - e^{2i\pi\nu} \right)\right]^{\nu_\text{d} - 1} \right\}\\
    &=  (\nu_\text{d} - 1) \frac{\tan\left( \frac{\pi \nu_\text{d}}{2} \right)}{\tan (\pi\nu)} \frac{S_\text{T}^\text{collision}}{ e I_+} \Big|_{I_- = 0}.
\end{aligned}
\label{eq:it_st_relation}
\end{equation}

With Eq.~\eqref{eq:it_st_relation}, we can further obtain $S_\text{T}^\text{collision}$ by using 
$$I_\text{T} (I_{A0}, I_{B0}) = I^\text{collision}_\text{T} (I_{A0}, I_{B0}) +  I^\text{single}_\text{T} (I_{A0}, 0) +  I^\text{single}_\text{T} (0,I_{B0}),$$ 
with which we split the tunneling current into single-source and double-source collision contributions, yielding
\begin{equation}
\begin{aligned}
    &\frac{\partial}{\partial I_-} I_\text{T} (\mathcal{T}_A^{(0)}, \mathcal{T}_B^{(0)}) \Big|_{I_- = 0}  = \left\{ \frac{\partial}{\partial I_-} I_\text{T}^\text{collision} (I_{A0}, I_{B0})  + \frac{1}{2} \left( \frac{\partial}{\partial I_{A0}} - \frac{\partial}{\partial I_{B0}}  \right) \left[ I^\text{single}_\text{T} (I_{A0}, 0) +  I^\text{single}_\text{T} (0,I_{B0}) \right] \right\} \Big|_{I_- = 0} \\
    & = \left\{ (\nu_\text{d} - 1) \frac{\tan\left( \frac{\pi \nu_\text{d}}{2} \right)}{\tan (\pi\nu)} \frac{S_\text{T}^\text{collision}}{ e I_+}  + \frac{1}{2} \left( \frac{\partial}{\partial I_{A0}} - \frac{\partial}{\partial I_{B0}}  \right) \left[ I^\text{single}_\text{T} (I_{A0}, 0) +  I^\text{single}_\text{T} (0,I_{B0}) \right] \right\} \Big|_{I_- = 0}.
\end{aligned}
\end{equation}
This leads to
\begin{equation}
\begin{aligned}
    S_\text{T}^\text{collision} & = \frac{e I_+ \tan (\pi\nu)}{(\nu_\text{d} - 1) \tan\left( \frac{\pi \nu_\text{d}}{2} \right) } \left\{ \frac{\partial}{\partial I_-} I_\text{T} (I_{A0}, I_{B0}) -\frac{1}{2} \left( \frac{\partial}{\partial I_{A0}} - \frac{\partial}{\partial I_{B0}}  \right) \left[ I^\text{single}_\text{T} (I_{A0}, 0) +  I^\text{single}_\text{T} (0, I_{B0}) \right]    \right\}\Bigg|_{I_- = 0}\\
    & = \frac{e I_+ \tan (\pi\nu)}{(\nu_\text{d} - 1) \tan\left( \frac{\pi \nu_\text{d}}{2} \right) } \left\{ \frac{\partial}{\partial I_-} I_\text{T} (I_{A0}, I_{B0}) - \frac{1}{2} \left( \frac{\partial}{\partial I_{A0}} - \frac{\partial}{\partial I_{B0}}  \right) \left[ I_\text{T} (I_{A0}, 0) +  I_\text{T} (0, I_{B0}) \right]  \right\}\Bigg|_{I_- = 0},
\end{aligned}
\label{eq:st_with_it_conductance}
\end{equation}
where in the second line we write $I^\text{single}_\text{T} (I_{A0}, 0)$ and $I^\text{single}_\text{T} (0,I_{B0})$ as $I_\text{T} (I_{A0}, 0)$ and $I_\text{T} (0, \mathcal{T}_B^{(0)})$, respectively, as for the single-source case the tunneling current is totally from the single-source contribution.
Notice that Eq.~\eqref{eq:st_with_it_conductance} requires knowing differential conductances, defined as the response of tunneling current $I_\text{T}$ at the central collider to non-equilibrium currents $I_{A0}$ and $I_{B0}$:
\begin{equation}
    \mathcal{T}_C (I_{A0},I_{B0}) \equiv \nu \frac{\partial}{\partial I_-} I_\text{T} (I_{A0},I_{B0}),
    \label{eq:central_QPC_transmissions}
\end{equation}
which is actually the definition of $\mathcal{T}_C$ in the main text.
This quantity was measured experimentally (cf. Fig.~\ref{fig:transmissions}). 

\begin{figure}
  \includegraphics[width=0.99\linewidth]{fig_transmissions}
  \caption{Transmission data of three quantum point contacts (including two diluters and the central collider). \textbf{Panel a}: Transmission probabilities of diluters $A$ and $B$, for both double-source and single-source situations.
  \textbf{Panel b}: Conductances at the central collider for the single-source situation. Here, conductance is defined as the response of tunneling current ($I_\text{T}$, through the central collider) to the modification of non-equilibrium current $I_A$ and $I_B$. \textbf{Panel c}: Corresponding conductances of the double-source scenario.
  \textbf{Panel d}: The rescaling factor $\chi(I_{A0},I_{B0})$ that depends on the total current $I_+$.
  \textbf{Panel e}: The ratio between the single-source and double-source $\mathcal{T}_C$. Either ratio is close to unity for the entire range of current.
  \textbf{Panel f}: The value of $-\nu \partial_{I_{B0}} I_\text{T} (I_{A0},0)$, from the data of \textbf{Panel b}, and that obtained indirectly following the tunneling current expressions of Eqs.~\eqref{eq:tunneling_current_and_noise-Is} and \eqref{eq:tunneling_current_and_noise-Ic}.}
  \label{fig:transmissions}
\end{figure}

Ideally, Eqs.~\eqref{eq:st_with_it_conductance} and \eqref{eq:central_QPC_transmissions} then provide an alternative (indirect) method [in addition to that defined by Eq.~(1) of the main text] to obtain $S_\text{T}^\text{collision}$.
In real experiments, however, $\mathcal{T}_C^{(0)}$ may depend on the system details, e.g., effects of electrostatic landscape in the given geometry (see Fig.~5 of the main text).
Rescaling is thus necessary to avoid corresponding distortions.
Within this work, we perform the rescaling by approximately taking
\begin{equation}
    \mathcal{T}_C (I_{A0},0) + \mathcal{T}_C (0,I_{B0}) = 2 \chi(I_{A0},I_{B0}) \mathcal{T}_C (I_{A0},I_{B0}),
    \label{eq:rescaling_requirement}
\end{equation}
with $\chi(I_{A0},I_{B0})$ the rescaling factor that depends on non-equilibrium currents.
For the ideal case where $\mathcal{T}_C^{(0)}$ is a constant number, $\chi(I_{A0},I_{B0})$ simply equals one,
since $I^\text{single}_\text{T}$ of Eq.~\eqref{eq:noise_single_full} does not contain any term that depends on both currents $I_{A0}$ and $I_{B0}$, leading to $\partial_{I_{A0}} I_\text{T}^\text{single} (I_{A0},I_{B0}) = \partial_{I_-} I_\text{T}^\text{single} (I_{A0},0)$, and $ \partial_{I_{B0}} I_\text{T}^\text{single} (I_{A0},I_{B0}) = -\partial_{I_-} I_\text{T}^\text{single} (0,I_{B0})$.

As discussed in Sec. 7.2 of the main text, the electrostatic landscape is likely different for single-source and double-source situations. The bare transmission coefficient of the central QPC is a function of all (gate and bias) voltages in the sample, including both $V_A$ and $V_B$ (in particular, since the geometric distance between the edges at the collider can be slightly different for the single- and double-source cases). This bias-voltage dependence of the collider transmission then leads to $\chi\neq 1$ in the experiment.
Strictly speaking, the tunneling conductance in the presence of the double-source collisions also receives a contribution from $I_\text{T}^\text{collision}$ [Eq.~\eqref{S-coll-nice}].
This contribution is, however, neglected, as it is much smaller than $I_\text{T}^\text{single}$ in the strongly diluted limit.
Based on the experimental data shown in Fig.~\ref{fig:transmissions}\text{e}, the value of $\chi (I_{A0}, I_{B0})$, however, slightly deviates from one.
We thus rescale the collision-induced noise $S_\text{T}^\text{collision}$ into $\chi S_\text{T}^\text{collision}$. We stress that, since $\chi (I_{A0}, I_{B0})$ is close to one, the effects leading to this rescaling are rather minor.

With Eqs.~\eqref{eq:st_with_it_conductance}, \eqref{eq:central_QPC_transmissions} and \eqref{eq:rescaling_requirement}, we are ready to obtain $S_\text{T}^\text{collision}$ from the measured conductances.
The results, together with Eq.~\eqref{eq:ep_different_forms}, are shown in Fig.~4 of the main text.
In addition to the final result, here we further show, in Fig.~\ref{fig:transmissions}, extra messages concerning transmissions at different quantum point contacts.
Here, \textbf{Panel a} presents tunneling probabilities of non-equilibrium anyons for the two diluters.
Clearly, for both the single-source and double-source contributions, both diluters are in the weak-tunneling regime,  such that each tunneling of non-equilibrium anyons is an independent event.
The next two panels, \textbf{b} and \textbf{c} present differential conductances of the central collider.
Differential conductances in both panels are again small, indicating scarce occurrence of Andreev-like tunnelings at the central collider.
In panel \textbf{d}, we show the rescaling factor $\chi(I_{A0},I_{B0})$, which turns out to be close to unity, indicating a rather small difference between the single-source and double-source cases.
With Eq.~(1) of the main text, Eq.~\eqref{eq:st_with_it_conductance}, as well as the data shown in Fig.~\ref{fig:transmissions}, we plot Fig.~4 of the main text.
The ratio of single-source and double-source $\mathcal{T}_C$, when turning on either the upper or lower diluter, is further shown in panel \textbf{e}.
Like the results presented in panel \textbf{d}, each ratio is close to one, indicating a minor influence from electrostatic interactions induced by gates and contacts.
Finally, in the last panel, \textbf{f}, we validate expressions for the tunneling current, Eqs.~\eqref{eq:tunneling_current_and_noise-Ic} and \eqref{eq:tunneling_current_and_noise-Is}, by comparing the value of $-\nu \partial_{I_{A0}} I_\text{T}(0,I_\text{B0})$ of panel \textbf{b}, and that obtained indirectly from Eqs.~\eqref{eq:tunneling_current_and_noise-Ic} and \eqref{eq:tunneling_current_and_noise-Is}.

\section{{\,\,\,\,\,\,\,}Prospective direction: Application to non-Abelian systems}

Here we present a sketchy outline, indicating how our approach might be generalizable to non-Abelian platforms.

Entanglement in non-Abelian states remains an outstanding challenge, given the crucial role of exotic non-Abelian statistics in the realization of topological quantum information processing; see, e.g., Refs.~\cite{ZhangYiPRB15S, HaegemanNC15S, BondersonAoP17S, CornfeldPRB19S, YangPRB19S, ManiPRA20S, WuCPB2022S, SreedharX2022S}.
However, generation of entanglement between initially unentangled non-Abelian quasiparticles has not been the focus of that line of research. In this section, we briefly discuss the applicability of our approach for the realization of entanglement-by-braiding  (at the collider) of non-Abelian quasiparticles.
We further show that the entanglement, generated via Andreev-like tunnelings, can be quantified following methods similar to our present study.

\begin{figure}
  \includegraphics[width=1\linewidth]{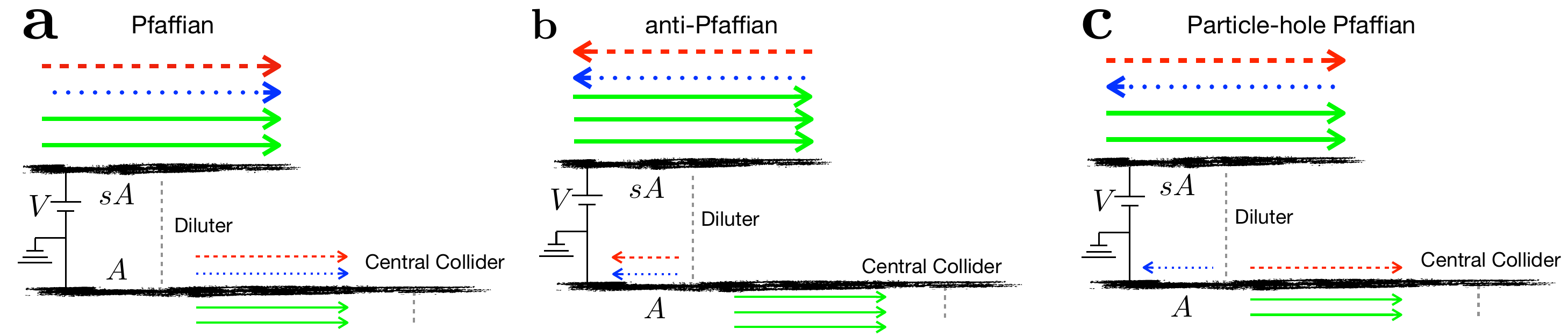}
  \caption{Andreev-like tunneling setup for the candidate $\nu=5/2$ states: (a) Pfaffian, (b) anti-Pfaffian, (c) particle-hole Pfaffian. 
  Black thick lines refer to the physical edges $sA$ and $A$ that are connected by the diluter.
  Following the convention used in the main text, the source channel $sA$ is biased at voltage $V$ with respect to the other channel $A$. The edge structure of each candidate state is shown by colored arrows. Particles from edge $sA$  enter edge $A$ through the diluter (tunneling through the quantum Hall bulk). In edge $A$, arrows indicate the nonequilibrium diluted states that arrive at the collider (which transmits only fermions in this setup). Each green arrow refers to an edge mode with the filling factor one. Red dashed and blue dotted lines represent the two nontrivial edge modes, i.e., the charged $\nu = 1/2$, and the Majorana edge states, respectively. 
  (a) For the Pfaffian state, both nontrivial edge states arrive at the collider. (b) For the anti-Pfaffian state, nontrivial state will not arrive at the collider. (c) For the particle-hole Pfaffian state, only the charge $\nu=1/2$ state arrives at the central collider.
  These three cases lead to different anyonic-statistics-induced entanglement.
    }
  \label{fig:non_abelian}
\end{figure}

As discussed in Refs.~\cite{ImuraSSC98S,OhashiJPS22S,MaX2022S}, Andreev-like tunneling can be realized in $\nu = 5/2$ non-Abelian systems~\cite{MooreRead91S,LevinHalperinRosenowPRL07S,LeeRyuNayakFisherPRL07S,FidkowskiChenVishwanathPRX13S,ZuckerFeldmanPRL16S,AntoniPRB18S}.
It is thus natural to extend our method, which quantifies entanglement induced by Andreev-like tunneling of $\nu = 1/3$ anyons, to non-Abelian systems.
Actually, all the prominent candidate states for filling factor $\nu=5/2$ contain a $1/2$ channel that carries quasiparticles with fractional charge $e/2$
(red dashed arrows in  Fig.~\ref{fig:non_abelian}).
Entanglement of these fractional-charge quasiparticles generated by Andreev-like tunneling can be quantified by extending our method, in particular, by taking $\nu=1/2$ in the general-$\nu$ formulas for the corresponding correlation functions for fractional chiral modes [cf. Eqs.~\eqref{eq:da0_da1} and \eqref{eq:S14}]. 
These correlation functions for fractional chiral modes will appear in the general expression for the entanglement pointer, where they will be multiplied by the correlation functions of other modes (yet to be calculated) in the integrand of the time integral determining the entanglement pointer. These latter correlation functions will account for the intricacy of the particles involved being non-Abelian.

In addition to the fractional chiral modes, there are also modes with the integer filling factor $\nu=1$, represented by green arrows in Fig.~\ref{fig:non_abelian}.
The generation of entanglement between the integer modes can be described similarly to the case of the integer quantum Hall effect addressed in Ref.~\cite{GuNC24S}. Importantly, there is no braiding between the quasiparticles of the fractional and integer modes (see the discussion of anyon-quasihole braiding for Andreev-like tunneling in the main text). This is expected to simplify the consideration of non-Abelian Andreev tunneling and could allow one to single out the contributions to the entanglement pointer stemming from the collisions of fractional quasiparticles. Furthermore, the Coulomb interaction between different ($\nu=1/2$ and $\nu=1$) ``bare'' modes, which results in the rearrangement of the complex edge structure akin to the fractionalization discussed in Sec.~\ref{Sec:SIII}, should also be taken into account in the non-Abelian edges. Given the relative resilience of the entanglement pointer to interaction effects in fermionic (Ref.~\cite{GuNC24S}) and Abelian-anyonic (main text) cases, we expect that, also in the non-Abelian case, interaction among the bare modes will not essentially affect entanglement generated in the Andreev-Hong-Ou-Mandel setting. 

Further, tunneling processes are affected by the presence of Majorana modes (blue dotted arrows in Fig.~\ref{fig:non_abelian}), which encode the non-Abelian character of the emergent quasiparticles. The influence of Majorana modes on the entanglement pointer in our case can be addressed within the bosonization technique (see, e.g., Ref.~\cite{Asasi2020S}).
The ``dressing'' of the tunneling events at the QPCs by Majorana modes will affect the contribution of the fractional and integer bare modes to the entanglement pointer through the additional correlation functions appearing in the time integrals for the noise contributions.
This is akin to the dressing of Abelian anyon tunneling by neutralon modes, cf. Ref.~\cite{ParkPRB15S}.
The combination of the above ingredients, each of which can be addressed by extending the theory developed here, should provide an understanding of the signatures of non-Abelian statistics on the generation of entanglement by Andreev-like tunneling between complex edges.

From another perspective, the challenge of identifying the proper state from a set of Pfaffian candidate states has attracted a lot of efforts, both on the experimental side ~\cite{IulianaScience08S,LinPRB12S,LinDuXieNSR14S,BanerjeeNature18S,BivasScience22S,BivasWenminScience22S,ArupX23S} and in theory~\cite{Lai2013S,Park2020S,Simon2020S,Asasi2020S,Hein2023S,Yutushui2023S,manna2023classificationS,park2024fingerprintsS}.
Most of those efforts have resorted to noise (both charge and thermal) measurements, to indirectly read out the chirality of $\nu = \pm 1/2$ mode and the Majorana mode.
Nevertheless, identifying Pfaffian nature remains a challenging task. 
Our approach provides promising direct diagnostics of the structure of the edge states.
Indeed, in our setup (Fig.~\ref{fig:non_abelian}), only edge states with the ``correct'' chirality (i.e., $\nu=1/2$ and Majorana edge modes for Pfaffian, and $\nu=1/2$ edge mode for particle-hole Pfaffian) arrive at the central collider in the diluted edge $A$, to influence both the tunneling current through the central collider and the corresponding statistics-induced entanglement pointer.
Measurements of these quantities would thus provide another option for the identification of composite edge states of a $\nu=5/2$ non-Abelian state.

Finally, it is important to explore the effects of intra-edge equilibration (mediated by the interplay of disorder-induced and Coulomb couplings between the edge modes of the complex edge) on the entanglement pointer in sufficiently long edges. Based on the considerations of Abelian quasiparticles, one can again expect that the entanglement pointer would be more robust with respect to such processes than ordinary current-current correlators.

\end{document}